\documentclass[fleqn,usenatbib]{mnras}
\usepackage{newtxtext,newtxmath}
\usepackage[T1]{fontenc}
\DeclareRobustCommand{\VAN}[3]{#2}
\let\VANthebibliography\thebibliography
\def\thebibliography{\DeclareRobustCommand{\VAN}[3]{##3}\VANthebibliography}
\usepackage{graphicx}	
\usepackage{amsmath}	
\usepackage{longtable}
\usepackage{multicol}
\usepackage{orcidlink}
\usepackage{utfsym}
\usepackage[none]{hyphenat}
\usepackage{soul,xcolor}
\usepackage[caption=false]{subfig}
\usepackage{float}
\usepackage{multirow}

\title[Morphology, SFH, and environment in the local universe]{The relation between morphology, star formation history, and environment in local universe galaxies}

\author[D. Pérez-Millán et al.]{
D. Pérez-Millán,$^{1}$\thanks{E-mail: astrodapm@gmail.com}\orcidlink{0000-0002-4507-9571}
Jacopo Fritz,$^{1}$\orcidlink{0000-0002-7042-1965}
Rosa A. González-Lópezlira,$^{1}$\orcidlink{0000-0003-1557-4931}
Alessia Moretti,$^{2}$\orcidlink{0000-0002-1688-482X}
\newauthor 
Bernardo Cervantes Sodi,$^{1}$\orcidlink{0000-0002-2897-9121}
Benedetta Vulcani,$^{2}$\orcidlink{0000-0003-0980-1499}
Marco Gullieuszik,$^{2}$\orcidlink{0000-0002-7296-9780}
Gustavo Bruzual,$^{1}$\orcidlink{0000-0002-6971-5755}
\newauthor 
Stéphane Charlot,$^{3}$\orcidlink{0000-0003-3458-2275}
and Daniela Bettoni$^{2}$\orcidlink{0000-0002-4158-6496}
\\
$^{1}$Instituto de Radioastronomia y Astrofisica, UNAM, Campus Morelia, A.P. 3-72, C.P. 58089, Morelia, Mexico \\
$^{2}$INAF-Osservatorio Astronomico di Padova, Vicolo Osservatorio 5, I-35122 Padova, Italy\\
$^{3}$Sorbonne Université, UPMC-CNRS, UMR7095, Institut d’Astrophysique de Paris, F-75014, Paris, France\\
}

\date{Accepted 2023 February 09. Received 2023 February 09; in original form 2022 June 09} 

\pubyear{2023}

\begin{document}
\label{firstpage}
\pagerange{\pageref{firstpage}--\pageref{lastpage}}
\maketitle

\begin{abstract}
The observed properties of galaxies are strongly dependent on both their total stellar mass and their morphology. Furthermore, the environment is known to play a strong role in shaping them. The galaxy population in the local universe that is located in virialized clusters is found to be red, poorly star-forming, and mostly composed of early morphological types. Towards a holistic understanding of the mechanisms that drive galaxy evolution, we exploit the spectrophotometric data from the WINGS and OmegaWINGS local galaxy cluster surveys, and study the role of both the local and the large-scale environments. We attempt to disentangle their effects from the intrinsic characteristics of the galaxies, in shaping the star formation activity at fixed morphological type and stellar mass. Using a sample of field galaxies from the same surveys for comparison, we analyse the effects of the environment, embodied by the local density, clustercentric distance, and close neighbours, respectively, on the star formation histories of cluster galaxies. We find that local effects have a more relevant impact on galaxy stellar properties than the large-scale environment, and that morphology needs to be taken into account to pinpoint the mechanisms that are driving the influence of clusters in galaxy evolution. 
\end{abstract}



\section{Introduction}

Observations in the local universe have shown that galaxies in clusters are significantly different from their field counterparts, in terms of both stellar content and morphology. These differences point to a quite rapid evolution of cluster galaxies, compared to field ones, and to the well-known morphology-density (MD) relation  \citep[e.g.][]{Dressler1980,Fasano2015}.

It is in dense environments such as clusters or compact groups that we can directly observe the effects of repeated interactions between galaxies and with the intergalactic and intracluster medium, and the effects of a very high spatial density of galaxies. As a consequence, galaxy clusters constitute an ideal place to trace the evolution and investigate the effects of the environment on galaxies. To explain the evolutionary paths that galaxies follow in clusters, different environment-dependent processes have been identified and proposed, such as harassment \citep{Moore1996}, starvation or strangulation \citep{Larson1980}, ram-pressure stripping \citep[RPS;][]{Gunn&Gott1972}, thermal evaporation \citep{Cowie&Songaila1977}, major and minor mergers \citep{Toomre1977}; and the overall tidal influence of the cluster \citep{Byrd&Valtonen1990}. 

Environment impacts not only the morphology of galaxies but also their gas content. Since the star formation history (SFH) of a galaxy crucially depends on the amount of gas available, any process that removes, adds, or even perturbs the gas is ultimately determining the evolution and the fate of a galaxy, at least as far as its stellar content is concerned. Galaxies are likely to have their star-formation activity quenched if they are massive, or located in dense environments \citep{Kauffmann2003a}, and the vast majority of quenched galaxies are early-types, suggesting that morphological type and quenching of star formation are related. This interdependence between morphology/star formation/environment complicates disentangling the processes affecting galaxies and their evolution. 

In an attempt to address this issue, \cite{Schawinski2014} studied the relation between morphology and SFH in low-redshift galaxies, without considering the environment, and found two evolutionary pathways towards quenching: the slow quenching of late-types through secular evolution, and the fast quenching of early-types with star formation, probably driven by major mergers. A third pathway could be due to weaker interactions that cause an intermediate quenching \citep{Smethurst2015}. Other studies \citep[e.g.,][]{Liu2015, Contini2019} have also suggested that the environment may not be the fundamental parameter in the quenching of star formation, but rather stellar mass, on which the SFR is strongly dependent. But at least at $z \leq 0.8$, local density is more important than large-scale environment in determining the stellar mass distribution of galaxies, suggesting that galaxy properties are more strongly dependent on local processes \citep{Vulcani2012}. \citet{Guglielmo2015} and \citet{Liu2019} also find that both morphological transformation and quenching of star formation are mainly driven by stellar mass, at least for massive galaxies. Concerning the SFH, on average it depends on galaxy mass but, at fixed mass, it depends on the environment and is almost independent of present morphology \citep{Guglielmo2015}.

Both the value of the stellar mass and the colours of galaxies are the result of their star formation history: the former mostly depends on the older SFR, while the latter is sensitive to the more recent one. Analysing differences in the SFR as a function of time can give us clues about the processes regulating these properties. Several investigations (e.g., \citealp{Wijesinghe2012}) have failed to find a relation between SFR, stellar mass, and environmental properties, such as the local density, hence hinting at a predominance of nature over nurture. Conversely, \cite{Calvi2018}, studying these relations for field galaxies in the local universe, found that, at a given stellar mass, galaxies in low-density environments have systematically higher SFRs than those in denser environments, with a wide spread, which can be explained in terms of morphology. Accordingly, they concluded that processes acting at local scales have a larger effect on galaxy properties than the large-scale environment. 

In particular, late-type galaxies are significantly different in the cluster and field environments \citep{Boselli&Gavazzi2006}; in clusters, they are more gas deficient, redder, and with a lower SFR \citep{Guglielmo2015, Paccagnella2016}. The mechanisms that drive changes in these galaxies might depend not only on cluster properties but also on galaxy stellar mass \citep{Fraser-McKelvie2018}. Low-mass spirals are quenched via gas stripping and heating processes operating in rich clusters, while for high-mass spirals there is no privileged mechanism, but rather a mixture of all processes \citep{Fraser-McKelvie2018b}. \cite{Cava2017} analysed a large sample of galaxies in clusters, and concluded that late-types are a recently accreted population; in time, they will change not only their morphology but their phase-space distribution as well, approaching that of cluster earlier types. 

Several investigations have shown, both through observations \citep[e.g.,][]{Park-Hwang2009, Cao2016} and simulations, \citep[e.g.,][]{Hwang2018, Patton2020} that hydrodynamical/tidal galaxy interactions with nearby companions can enhance (for late-late pairs) or suppress (late-early pairs) the SFR. Minor mergers may induce a similar effect \citep[e.g.,][]{Lambas2012, Kaviraj2014}. Interactions among galaxies may be a key piece to study the star formation (SF) activity at present time, and  to understand how it was in earlier epochs.

Leveraging current facilities, in this work we exploit one of the largest, most complete, and most homogeneous databases of cluster galaxies in the local universe: the WIde-field Nearby Galaxy-cluster Survey (WINGS; \citealp{Fasano2006}), and its follow-up OmegaWINGS \citep{Gullieuszik2015}. The ultimate goal of this paper is to establish causal connections between the galactic stellar population properties and the characteristics of the environment in which they are found, and possibly disentangle their importance. At the same time, we attempt to qualitatively and quantitatively assess the evolutionary characteristics driven by clusters (e.g., colour, SFR, morphological fraction, star-forming fraction),  and determine which galaxy types are the most affected.

To this end, we try to perform a ``holistic'' analysis of the stellar population properties of galaxies in clusters, taking simultaneously into account both the main features of the galaxies themselves, i.e., their stellar mass and morphology, and all the information we have at our disposal about their environment, both local and large scale. In order to ensure self-consistency and homogeneity, this work repeats analyses that have already been partially published, given the differences in the samples between this and previous papers.

The WINGS and OmegaWINGS projects, together with their products, have been formerly presented in several papers. We briefly recap all the information on the data, as well as the criteria that have been used to define the final sample of cluster and field galaxies we are exploiting, in  \autoref{sec:Dataset}. As this work heavily relies on the spectrophotometric code \texttt{SINOPSIS}, in \autoref{sec:Sinopsis} we summarise the main features of the modelling and its limitations, and we also list its main products. \autoref{sec:Sample_properties} is devoted to the presentation of the main characteristics of the galaxies in both the cluster and the field samples: the mass distribution, the star formation rates (SFRs), and the colour--mass relations. Total stellar mass is one of the most important galactic physical properties, both in general and in the context of scaling relations, while colours and SFRs are fundamental for the study of galaxy evolution.
In \autoref{sec:SFH_environment}, we start diving into the relations and possible correlations between SFH and environment, through its various parameterisations, simultaneously --when possible-- taking into account  galaxy stellar mass and morphology. As a first step, we characterise the SFH of galaxies in cluster and field environments, while keeping mass and morphology fixed. This comparison allows us to study environmentally--driven differences in the stellar content, focusing on different parameterisations of the environment: projected clustercentric distance, local density, and closest neighbour morphology. Here, we also try to understand which one of these parameters affects more the ability to form stars of the recently infalling galaxies. The ensuing \autoref{sec:global} approaches the issue from another --more global-- point of view, checking if and how total cluster mass drives differences in SFH of galaxies in comparison to those in the field. 
Since local and large-scale environments are not independent from one another, and neither are the other expressions of the environment that we have considered, we also attempt to disentangle their effects. All the aspects that we have studied somewhat separately in the previous three sections are then combined and analysed in \autoref{sec:discussion}, where we try to give a self-consistent picture of the different findings. The most important results are highlighted in \autoref{sec:conclusions}.

Throughout this work, we make a distinction between ``local'' and  ``large-scale'' environment. By local, we refer to the space immediately surrounding the galaxies, and we quantify it through the galaxy number density. Conversely, the large-scale environment denotes the structure to which a galaxy belongs, such as a cluster, a group, or the field; we characterise it by the cluster mass, and by the cluster membership or lack thereof. 

In this paper, we assume a $\Lambda$CDM cosmological model with $\Omega_{\text{m}} = 0.3$, $\Omega_\Lambda = 0.7$, and $ H_0 = 70 \ \text{km} \ \text{s}^{-1} \ \text{Mpc}^{-1}$, and use a \cite{Chabrier2003} initial mass function (IMF).

\section{The datasets and observations}
\label{sec:Dataset}
Here, we briefly recall the properties of the data we have used for this work. A more complete and detailed description can be found in previously published works, which we refer to in the following sections. 

WINGS and OmegaWINGS observed galaxies in local  $(0.04 < z < 0.07)$  clusters, obtaining both images and optical spectra. Photometry for the central parts ($\sim$ 30$^\prime$) of 77 clusters was obtained by WINGS \citep{Fasano2006}, using the 2.5 m Isaac Newton Telescope (INT/WFC) for the northern clusters, and the MPG/ESO-2.2 m telescope for the southern clusters. The spectroscopic follow-up of 46 of them \citep{Cava2009} was taken through the AF2/WYFFOS multifiber spectrograph, mounted on the $4.2$ m William Herschel Telescope (WHT) for the northern clusters, while the southern clusters were observed with the 2dF multifiber spectrograph on the $3.9$ m Anglo Australian Telescope (AAT).

OmegaWINGS represents the extension to a wider field ($\sim 1^\circ$, up to about 2.5 virial radii) for 46 clusters, randomly selected from the 57 clusters that can be observed with the Very Large Telescope (VLT) Survey Telescope (VST; $\delta < 20$ deg) to span a wide range in X-ray luminosity and hence probe a complete range in halo masses. OmegaWINGS was spectroscopically followed up as well; 33 of the 46 clusters were observed with the VST fibre spectrograph. 
WINGS spectra cover a range of $\sim 3800 - 7000$ \AA,  with an intermediate resolution of $6-9$ \AA, and a fibre aperture of $1\farcs6$ for northern clusters and $2''$ for southern clusters. OmegaWINGS spectra cover $\sim 3800 - 9000$ \AA, with a resolution of $3.5-6$ \AA, and a fibre diameter of $2\farcs16$.

For this work, we use aperture spectra, which only cover the central part of the galaxies (aperture diameter $\sim 2-3$ kpc), as many other authors have done (see, e.g., \citealp{Kauffmann2003a, Kauffmann2003b, Kauffmann2004, Gallazzi2005, Casado2015}, for the SDSS; \citealp{Guglielmo2015, Paccagnella2016}, for WINGS and OmegaWINGS).

The reader can refer to \cite{Fasano2006} for a full description of the cluster sample, to \cite{Varela2009} for details about the WINGS photometric survey, and to \cite{Cava2009, Moretti2014} for facts about the spectroscopic one. For the OmegaWINGS dataset, \cite{Gullieuszik2015} and \cite{Moretti2017} are the references for information on photometry and spectroscopy, respectively.

\subsection{Previously derived properties}
\label{subsec:Galaxy-properties}
Redshift measurements were made by  \cite{Cava2009} for WINGS, and by \cite{Moretti2017} for OmegaWINGS; we refer the reader to those works for more details.  From these, cluster velocity dispersions were calculated, and they span a range of $\sigma_{\rm cl} \simeq 400 - 1300$ km  $\text{s}^{-1}$ (see \autoref{tab:WINGS_clusters} in \autoref{sec:Clusters_table}). Four clusters in OmegaWINGS (A1069, A2382, A3158, and A4059) have a second group of galaxies outside the main group (with radial velocities $\pm 3 \times \sigma_{\rm cl}$ different from the systemic velocity of the cluster). In these cases, such galaxies were also included as cluster members. We use the updated $R_{200}$ and $\sigma_\text{cl}$ values given by \cite{Biviano2017} and \cite{Gullieuszik2020}.

Another important piece of information used throughout this work is galaxy morphology. This property was assessed through \texttt{MORPHOT} \citep[see][for details]{Fasano2012}, a tool specifically designed to maximise the ability to distinguish between elliptical and lenticular galaxies. This classification is based on the $T$ parameter \citep{deVaucouleurs1974} which, for the purposes of this work, we have smoothed in order to deal with 4 broader morphological classes: ellipticals (E; $-5.5 < T_M < -4.25$), lenticulars (S0; $-4.25 \leq T_M \leq 0$), early spirals (SpE; $0 < T_M \leq 4$), and late spirals including irregulars (SpL; $4 < T_M \leq 11$). In the following, when necessary, we will also refer to E and S0 as ``early-types'', and to SpEs and SpLs as ``late-types''.

To statistically account for the fact that not all galaxies detected in the images have a spectroscopic counterpart, throughout this work we have corrected for both radial (geometrical) $C(r)$ and magnitude $C(m)$ completeness (see \citealp{Cava2009} for details on their calculation). This is done by weighting the properties of each galaxy with the product of the inverse of the two completeness values.
\subsection{Sample selection}
\label{subsec:Sample_selection}
Stellar masses calculated as described in \autoref{sec:Sinopsis} were used to select galaxies in the final sample, whose characteristics should satisfy the following requirements:
\begin{enumerate}
    \item Stellar mass $\mathcal{M}_* > 3 \times 10^9 \ \text{M}_\odot$, roughly corresponding to an absolute $V$ magnitude brighter than $M_V = -18.5$ mag.
    \item An acceptable spectral fit from \texttt{SINOPSIS}, represented by a value of $\chi^2 \leq 5$.
\end{enumerate}

The limit imposed on the stellar mass is related to the absolute magnitudes of the galaxies, given the photometric threshold reached by the observations, whereby spectroscopy has been obtained down to $V = 20.0$ with a satisfactory enough completeness.

The $\chi^2$ limit was empirically chosen, by comparing the model spectra to the observed ones. This allowed us to account for high $\chi^2$ still providing a good fit in those cases for which some of the observational constraints were severely affected by locally bad signal-to-noise ratio (SNR). Good results were obtained for values equal to or lower than the limit. We also excluded cD galaxies in clusters, and neglected the AGN contribution, since the fraction of AGNs in WINGS/OmegaWINGS clusters is approximately $3\%$ \citep{Marziani2017}. By taking the imposed limit in EW(H$\alpha)$ (see end of \autoref{sec:Sinopsis}), low-luminosity AGN are excluded, while type-1 AGNs still present cannot be fitted through stellar populations synthesis, and hence the resulting  $\chi^2$ value will be above the threshold we have defined as acceptable.

Regarding the control sample, we have adopted the same approach as in \cite{Paccagnella2016}. The sample is assembled with galaxies located in the fields of view of the WINGS/OmegaWINGS clusters with $0.02 < z < 0.09$, and whose relevant physical properties were measured/calculated, but that do not satisfy the membership condition in velocity. 

A quality check assessment, based on the SNR, made us discard data from all but 7 (A376, A1795, A1983, A2457, A2626,  Z8338, and Z8852) WINGS clusters, because of weather problems during the observations. In this work, we have used 3,097 individual galaxy spectra from WINGS and all 17,985 spectra from OmegaWINGS, in a total of 43 clusters (see \autoref{tab:WINGS_clusters}). From these, 2,711 WINGS and 7,627 OmegaWINGS objects are cluster member galaxies. 

The final sample from WINGS and OmegaWINGS, taking into account the criteria defined above, contains 4,598 members (8,845 after weighting) in 43 clusters. Regarding the morphological classes E, S0, SpE, and SpL, their percentages are, respectively: $0.261 \pm 0.006$, $0.425 \pm 0.007$, $0.251 \pm 0.006$, and $0.063 \pm 0.004$, where the uncertainties correspond to binomial errors. For the field galaxy sample, we have 676 galaxies (1,071 after weighting); their percentages of the same morphological classes are: $0.167 \pm 0.014$, $0.266 \pm 0.017$, $0.374 \pm 0.019$, and $0.192 \pm 0.015$.
We have compared the stellar mass distributions between the cluster and field samples as a function of morphology, in order to guarantee that the comparison is statistically significant. The results are discussed in \autoref{subsec:Mass_distribution} for spiral galaxies, and in \autoref{sec:Stellar-mass_distribution} for all the other types.

\subsection{Environmental tracers}
\label{subsec:environment-tracers}
We used three parameterisations of the cluster environment for cluster members: the projected distance to the cluster centre, the projected local density (LD), and the projected distance to the nearest galaxy. 

The clustercentric distance is calculated in projection to the centre of the cluster in the X-ray emission (the updated coordinates in \citealp{Biviano2017} have been used). These values are given in units of the virial radius of each cluster ($R_{200}$; see \autoref{subsec:Galaxy-properties}). 
\begin{figure}
 \centering 
  \subfloat{ \hspace{-10px}
    \includegraphics[trim={0 0 0 0}, clip, height=4.4cm]{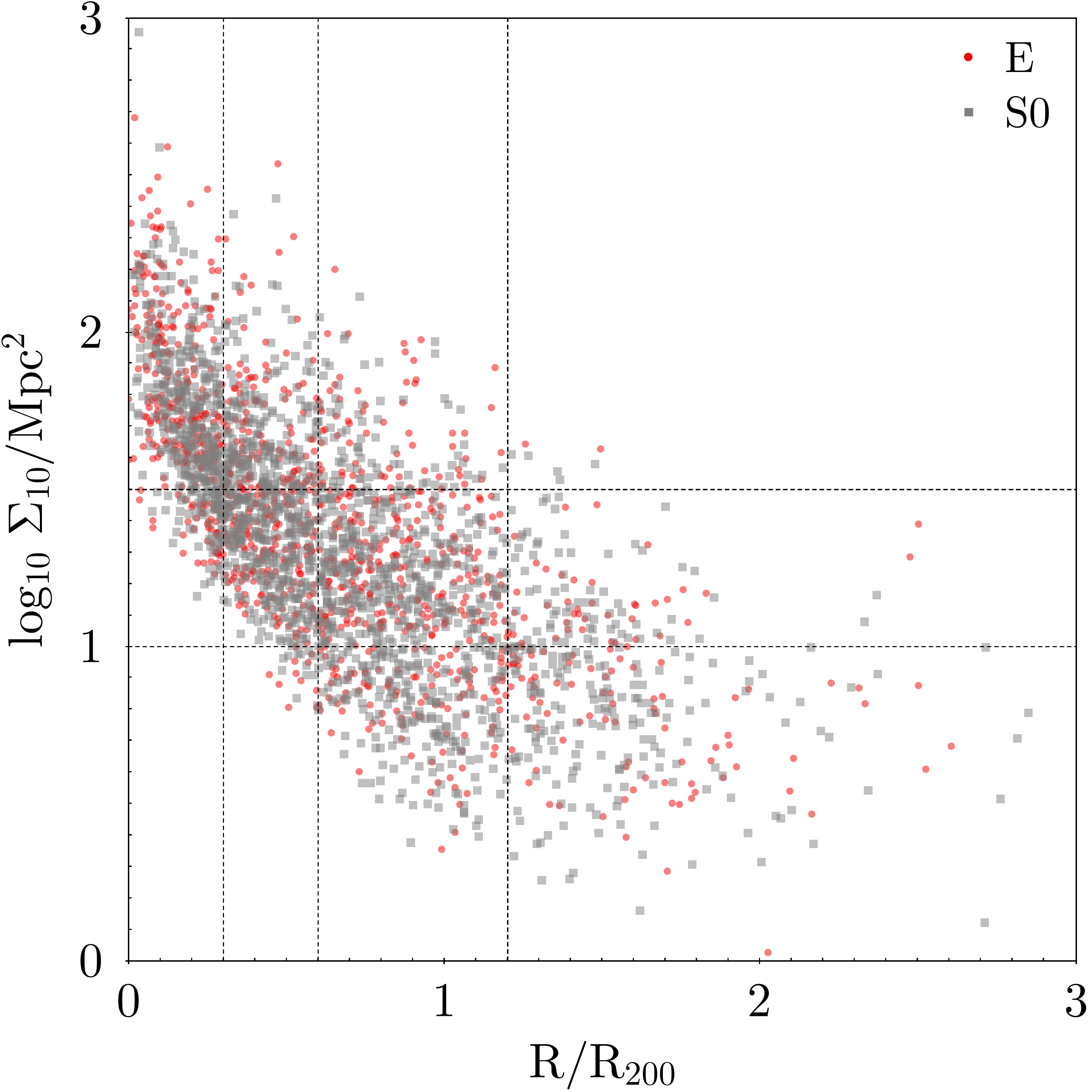}}
  \subfloat{
    \includegraphics[trim={95 0 0 0}, clip, height=4.4cm]{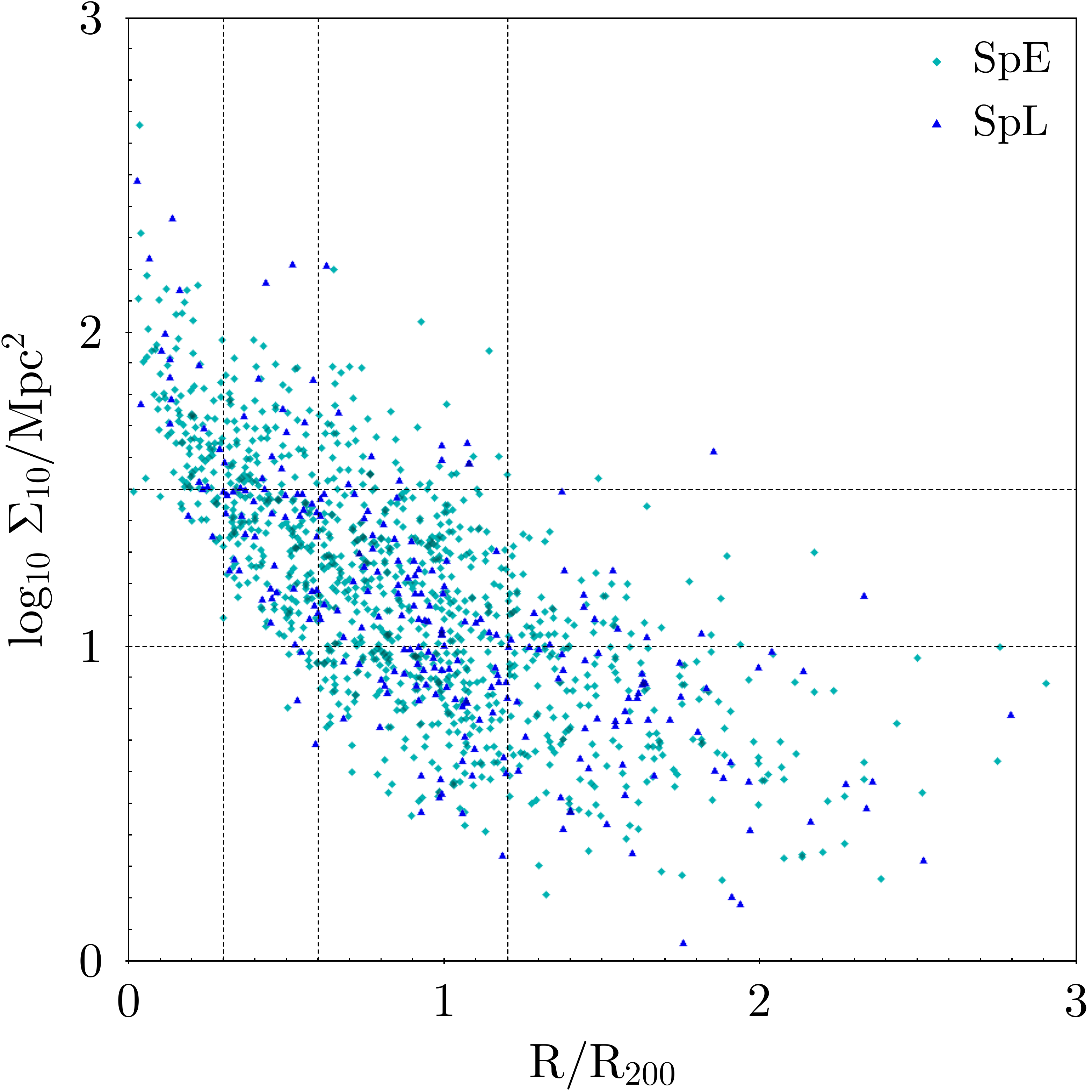}}
    \vspace{1px}\hspace{-3px}
    
    \subfloat{
    \includegraphics[trim={0 0 0 0}, clip, height=4.4cm]{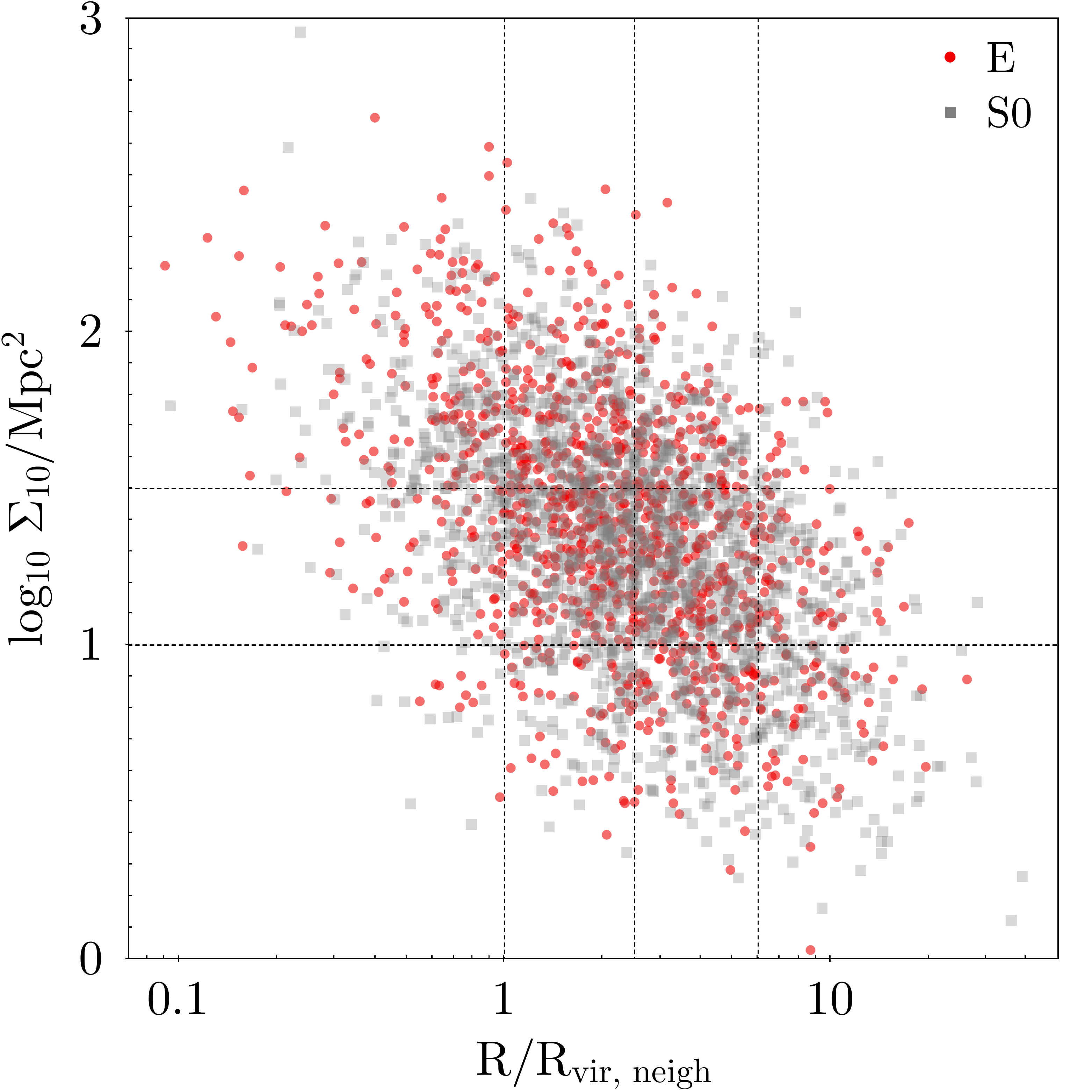}}
  \subfloat{
    \includegraphics[trim={100 0 0 0}, clip, height=4.4cm]{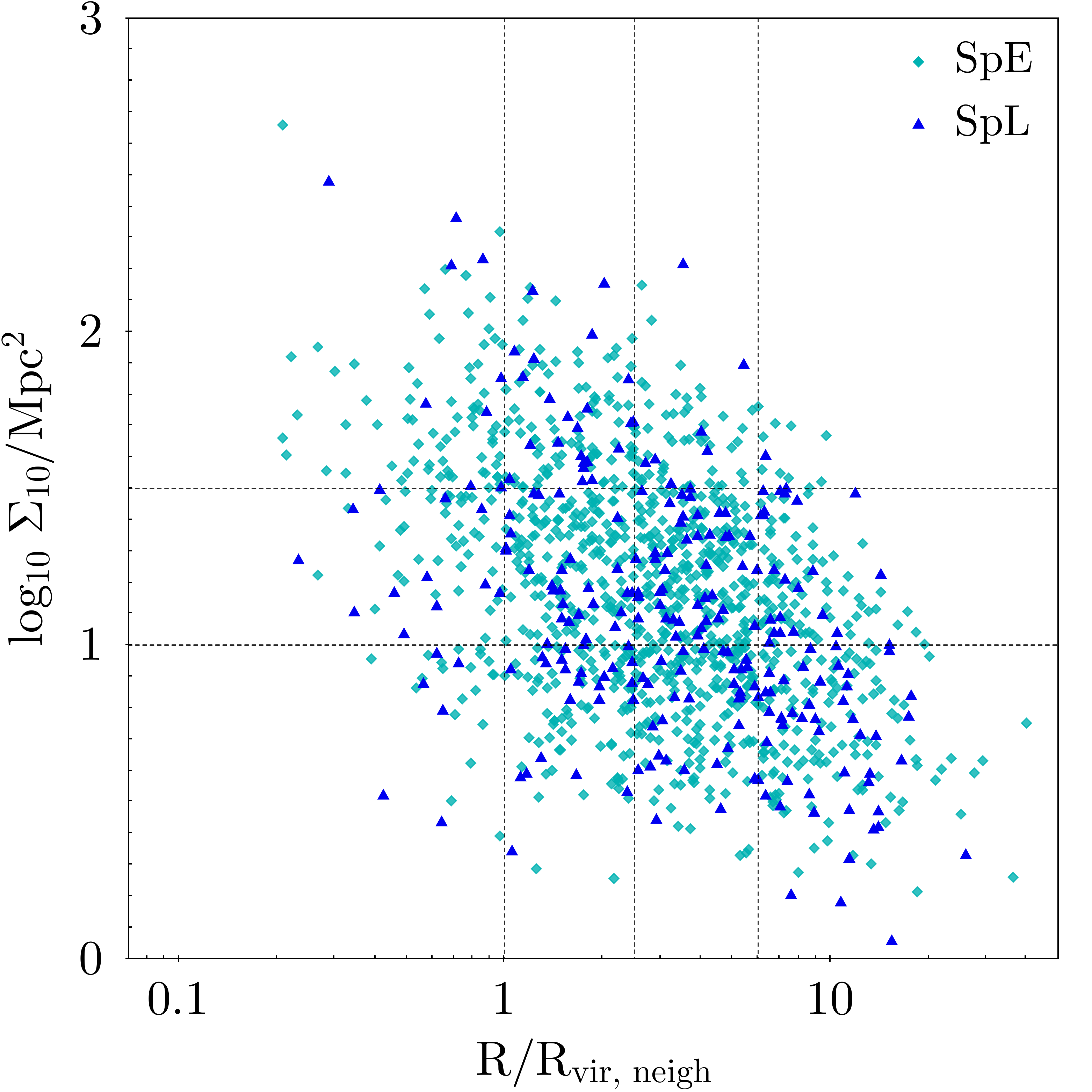}}
    
  \caption{Local density vs. clustercentric projected distance ({\it top row}), and vs. projected distance to the nearest neighbour galaxy ({\it bottom row}) for early- (\emph{left}) and late- (\emph{right}) types in the cluster galaxy sample. Horizontal and vertical dotted black lines indicate the bins used to divide the sample.}
 \label{fig:LD-R_R-neigh}
\end{figure}

The local density for a galaxy is calculated from the circular area projected in the sky, enclosing the $N$ nearest projected neighbours brighter than a certain limit, corrected for spectroscopic incompleteness. In this case, $N=10$. Given the lack of spectroscopy for all galaxies, a limit of M$_V = -19.5$ was imposed to eliminate background galaxies. LD values for WINGS are taken from \cite{Vulcani2012}, and for OmegaWINGS from \cite{Vulcani2023}. 

The distance to the closest neighbour galaxy is the projected distance, normalised by the virial radius of the neighbour ($R_{\rm vir, \ neigh}$), calculated as the radius of a sphere whose density of baryonic matter (approximated by the stellar mass obtained through \texttt{SINOPSIS}) is equal to 200 times the critical density of the universe at that redshift:
\begin{equation}
    R_{\textrm{vir}}=\left(\frac{3}{4\pi}\cdot \frac{M_\star}{200 \rho_{\rm c}}\right)^{1/3}.
\end{equation}
In \autoref{fig:LD-R_R-neigh}, we show the distribution of LD as a function of projected distance to the cluster centre, and projected distance to the closest neighbour galaxy, separated into the four main morphological classes of the cluster galaxy sample. We also represent, as black dotted lines, the bins used to divide galaxies according to the three environment tracers used throughout the text.

\section{Stellar population properties}
\label{sec:Sinopsis}
The stellar population properties are derived with the spectrophotometric code \texttt{SINOPSIS}\footnote{\url{http://www.irya.unam.mx/gente/j.fritz/JFhp/SINOPSIS.html}} (SImulatiNg OPtical Spectra wIth Stellar population models). The code is described in detail in \cite{Fritz2007, Fritz2017}, and here we will only summarize the setup used to fit the spectra in this sample. 

To reproduce an observed spectrum, \texttt{SINOPSIS} uses the theoretical spectra of simple stellar populations (SSP) with 12 different ages, from $10^6$ yr up to the age of the universe at the galaxy redshift, and four metallicity values: sub-solar ($Z = 0.004$), solar ($Z = 0.017$), and super-solar ($Z = 0.03$, $0.04$). SSP in the 4 youngest bins, up to $2\times 10^7$ yr, also display nebular emission lines from ionised gas in the typical physical conditions found in {\sc Hii} regions, which have been calculated in \cite{Fritz2017}. SSP models are those from Charlot \& Bruzual (in prep.), with a \cite{Chabrier2003} IMF with masses in the range $0.1 - 100\  M_\odot$. The adopted SFH is non--parametric, i.e., the SFR is free to vary for each stellar age independently. Only the 4 youngest SSP are constrained to have the same SFR, in order to be consistent with the assumption of a constant SFR over $\sim 10^7$, which is conventionally used to convert nebular lines luminosities into SFR values \citep[e.g.][]{Kennicutt1998}. 

Dust extinction is modelled as a uniform screen in front of the stars, adopting the \cite{Cardelli1989} extinction curve, and fully embracing the selective extinction hypothesis, whereby extinction is allowed to be a function of stellar age: generally speaking, the younger the stellar population is, the more likely it is located in a region with higher obscuration (i.e., in the molecular clouds where they were born; see, e.g., \citealt{Charlot&Fall2000,Poggianti2001}).

Once the best fit is obtained, the solution is far from unique. Thus, a further binning in age is implemented, such that the final resolution is lowered to four age bins. The derived SFRs for these four ages constitute the final SFH. The ages are listed in \autoref{tab:stellar_ages}, and were selected based on the presence and intensity of stellar population features \citep{Fritz2007}, namely:
\begin{table}
\caption{\label{tab:stellar_ages} Age intervals used in \texttt{SINOPSIS} to obtain the SFH. $t_u$ represents the age of the universe at the cluster redshift.}
\begin{center}
\begin{tabular}{ c c c } 
\hline
$\rm{SFR}_i$ & Age range & Age bin \\ \hline \hline
SFR$_1$ & $0 - 19.95 $ Myr    & 19.95 Myr  \\
SFR$_2$ & $19.95 - 571.5$ Myr  & 551.55 Myr \\
SFR$_3$ & $0.5715 - 5.754$ Gyr & 5.183 Gyr  \\
SFR$_4$ & $5.754-t_u$ Gyr      & $\Delta t_u$ Gyr \\ \hline
\end{tabular}
\end{center}
\end{table}

\begin{description}
    \item[SFR$_1$:] Characterised by emission lines and the strongest ultraviolet emission.
    \item[SFR$_2$:] Hydrogen lines from the Balmer series reach their maximum intensity {\em in absorption}, while the Ca{\sc k,h} UV lines still have low equivalent width.
    \item[SFR$_3$:] The intensities of Balmer absorption lines decrease as stellar ages increase, while the Ca absorption lines reach their maximum intensity.
    \item[SFR$_4$:] Stellar populations in this age bin are reddest, and display the highest 4000 \AA \ Break (D4000) values. Other spectral characteristics reach an asymptotic behaviour at these ages.
\end{description}

SSPs models for which the nebular emission has been self-consistently calculated eliminate the need to mask parts of the spectra that display emission lines, and allow the calculation of the current SFR (i.e., in the last $\sim 20$ Myr, or SFR$_1$, in the notation here adopted). For this calculation, a fit is performed to the equivalent widths (EWs) of the emission lines, when present. Mainly H$\alpha$, H$\beta$, and {\sc [Oii]} are used, but other Balmer lines are employed as well. This approach has the advantage that the correction for dust absorption is automatically taken into account (dust extinction is, in fact, one of the output parameters). Furthermore, since we fit both emission and absorption lines simultaneously, the effects of a possible absorption component originating in the photosphere of intermediate-age stars are naturally taken into consideration as well. 

Ionisation mechanisms are entirely attributed to star formation processes. Nevertheless, other mechanisms, like evolved stars, shocks, and AGNs, can produce ionising photons and, hence, emission lines. If this is not properly taken into account, the current SFR could be overestimated. As a conservative approach, we have assumed that a low EW of the H$\alpha$ line (i.e., $0 <$ EW(H$\alpha) \le 6$ \AA; \citealp{Sanchez2014, Cid-Fernande2013}) is not due to star formation. Therefore, for these low values, we always impose that the present-day SFR, as calculated by \texttt{SINOPSIS}, is 0.

\section{Physical properties of the sample}
\label{sec:Sample_properties}

Observed properties of galaxies such as colour, SFR, and internal structure are known to depend on the stellar mass \citep[e.g.,][]{Kauffmann2003a, Kauffmann2004, Mannucci2010, Kelvin2014}. It is hence of paramount importance to characterise this property in the galaxies of our sample. 

\subsection{Mass distribution}
\label{subsec:Mass_distribution}
\begin{figure}
\centering
   \subfloat{\label{fig:Mass-dist_SpE-Cl-Field}
     \includegraphics[height=4.3cm]{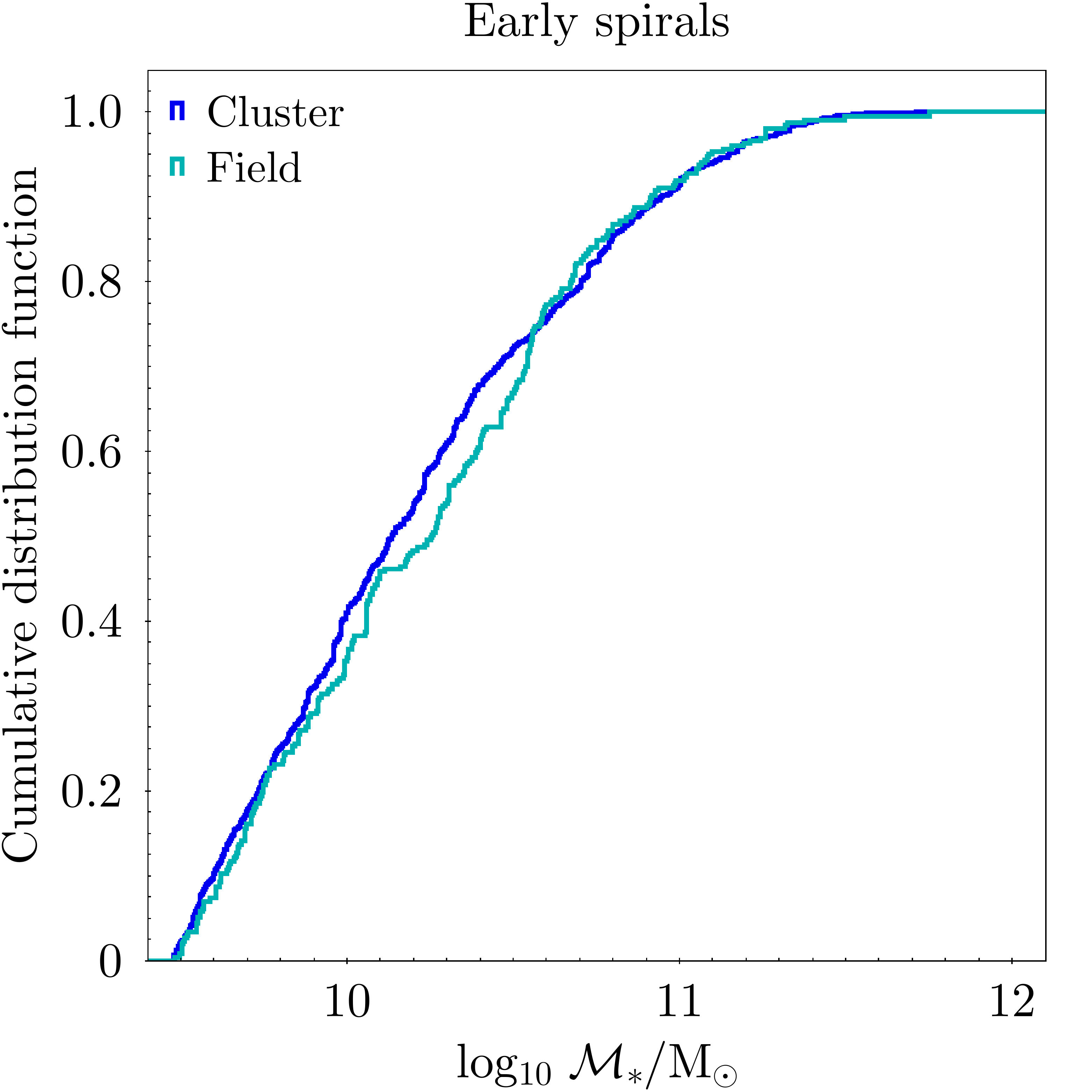}}
   \subfloat{\label{fig:Mass-dist_SpL-Cl-Field}
     \includegraphics[trim={120 0 0 0}, clip,  height=4.3cm]{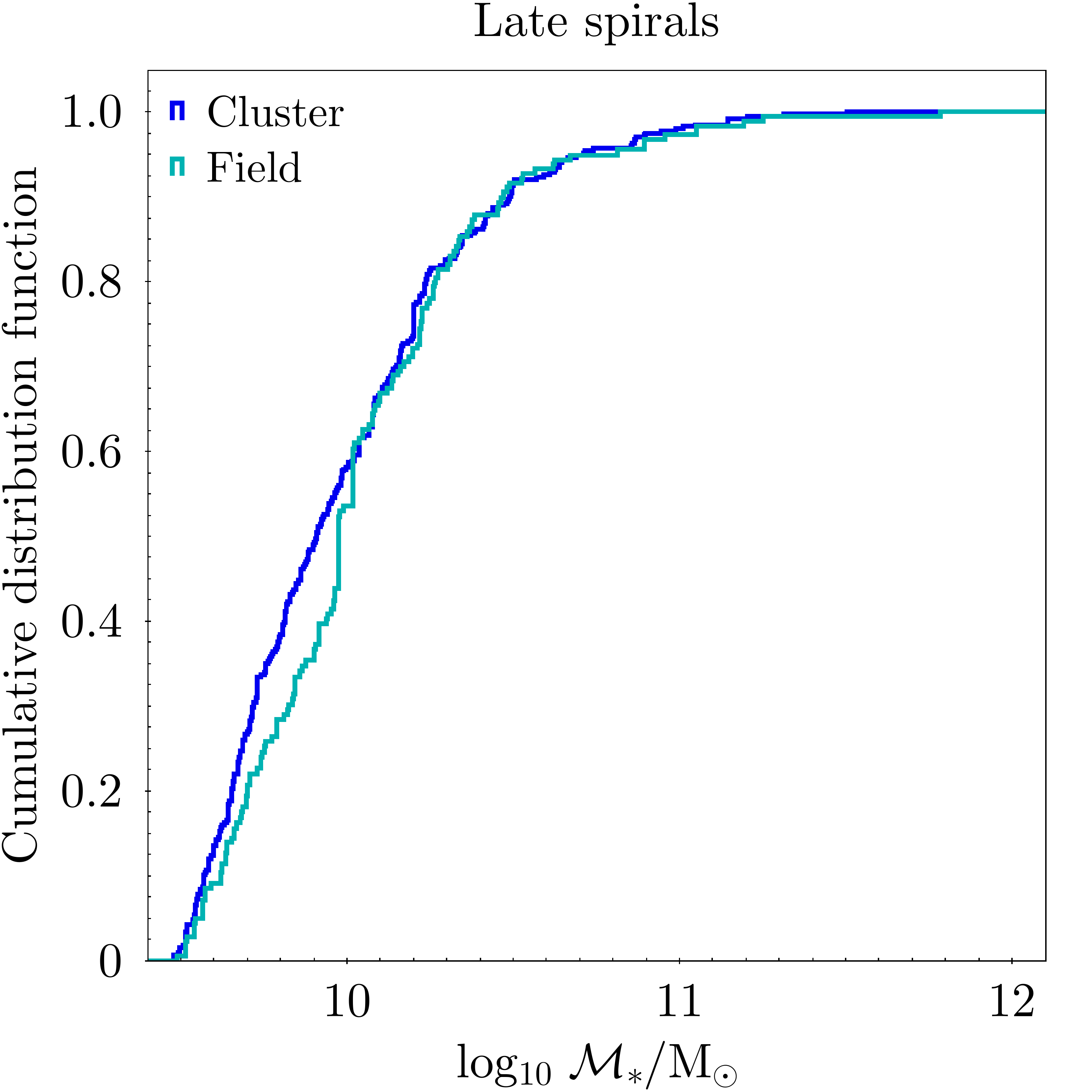}}

\caption{Cumulative distribution function of weighted total stellar mass for early- ({\it left}) and late- ({\it right}) spiral galaxies in the cluster ({\it blue lines}) and field ({\it cyan lines}) samples.}
 \label{fig:Mass_distr_Sp}
\end{figure}

The stellar mass function of WINGS galaxies was presented by \cite{Vulcani2011}. However, we cannot compare their results with ours in a straightforward manner, since we have used different parameters and theoretical SSP models in \texttt{SINOPSIS}, plus we include here also OmegaWINGS galaxies. 

In \autoref{fig:Mass_distr_Sp}, we present the weighted cumulative distribution of the stellar mass for early- and late-spiral galaxies, distinguishing between the cluster and field samples. We focus here on the spiral population because in clusters it is very likely dominated by recently infalling and acquired galaxies, and to better understand the differences with field galaxies. 
From a Kolmogorov-Smirnoff (K-S) test (see \autoref{sec:Stellar-mass_distribution}), the mass distributions of SpE and SpE+SpL are statistically the same in clusters and field, whereas the mass distributions of SpL are statistically different. Besides, as expected, SpL are, on average, less massive than SpE. Regarding early-type galaxies, the K-S test reveals that the mass distributions of ellipticals are different in clusters and the field, while the mass distributions of S0s are indistinguishable. The result for ellipticals does not exactly match the one found by \cite{Calvi2013}, for the WINGS and PM2GC (Padova-Millennium Galaxy and Group Catalogue; \citealp{Calvi2011}) samples. In \autoref{sec:Stellar-mass_distribution} we dig into the issue and propose possible explanations.

\subsection{SFR-mass relation}
\label{subsec:SFR-Mass_relation}
\begin{figure}
\centering
   \subfloat{\label{fig:SFR1-M_SpE-Cl}
     \includegraphics[width=0.23\textwidth]{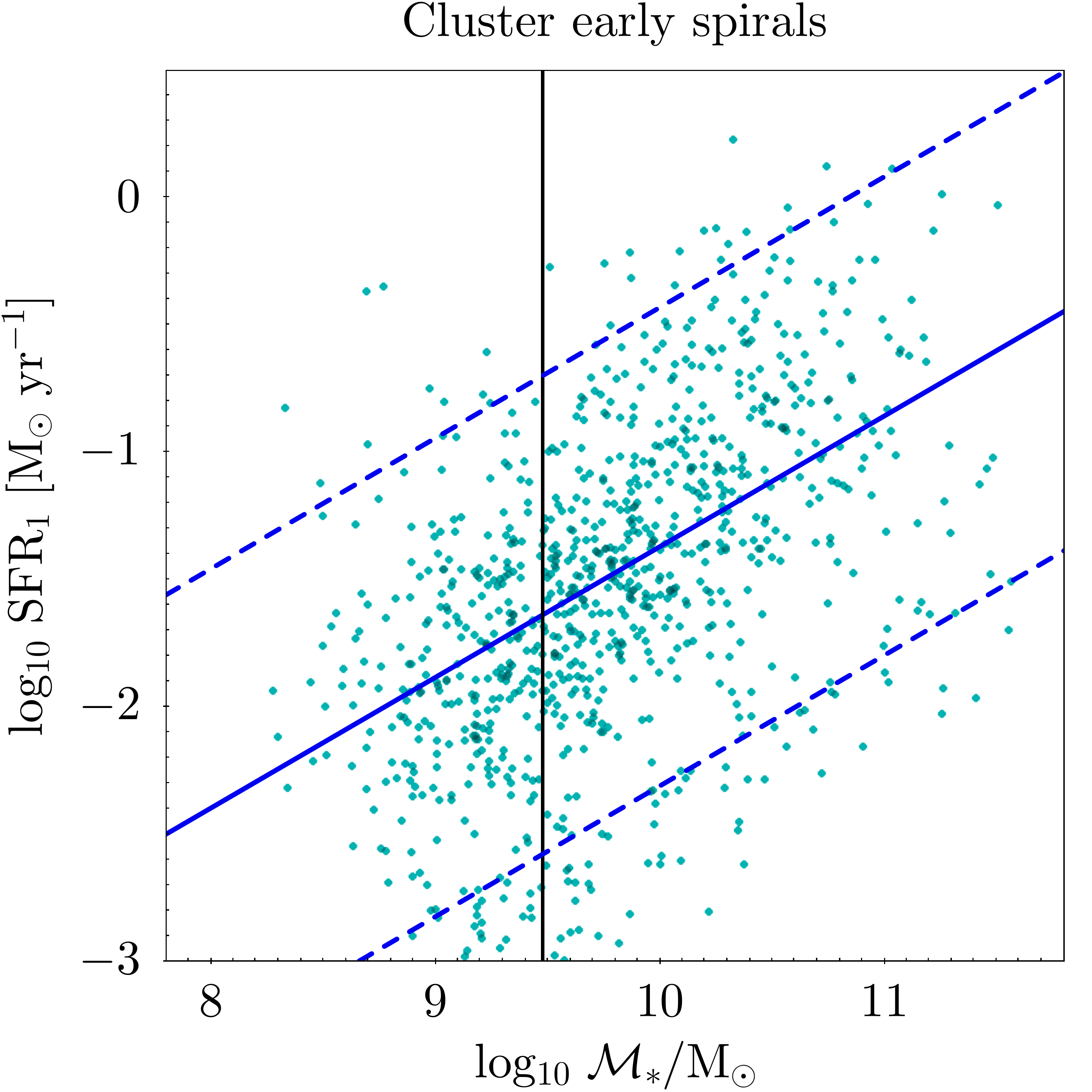}}
   \subfloat{\label{fig:SFR1-M_SpE-Field}
     \includegraphics[width=0.23\textwidth]{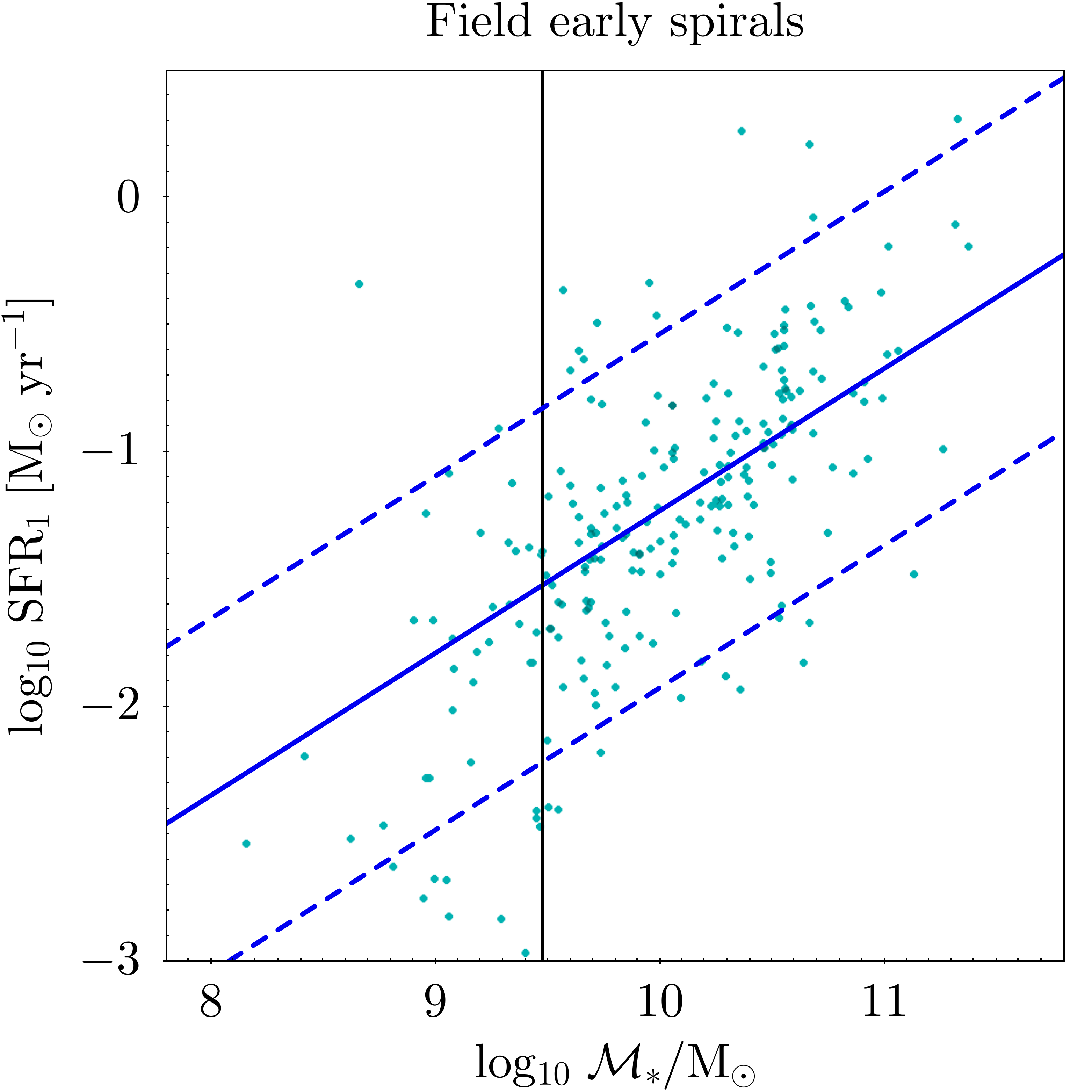}}
\vspace{5px}
     
   \subfloat{\label{fig:SFR1-M_SpL-Cl}
     \includegraphics[width=0.23\textwidth]{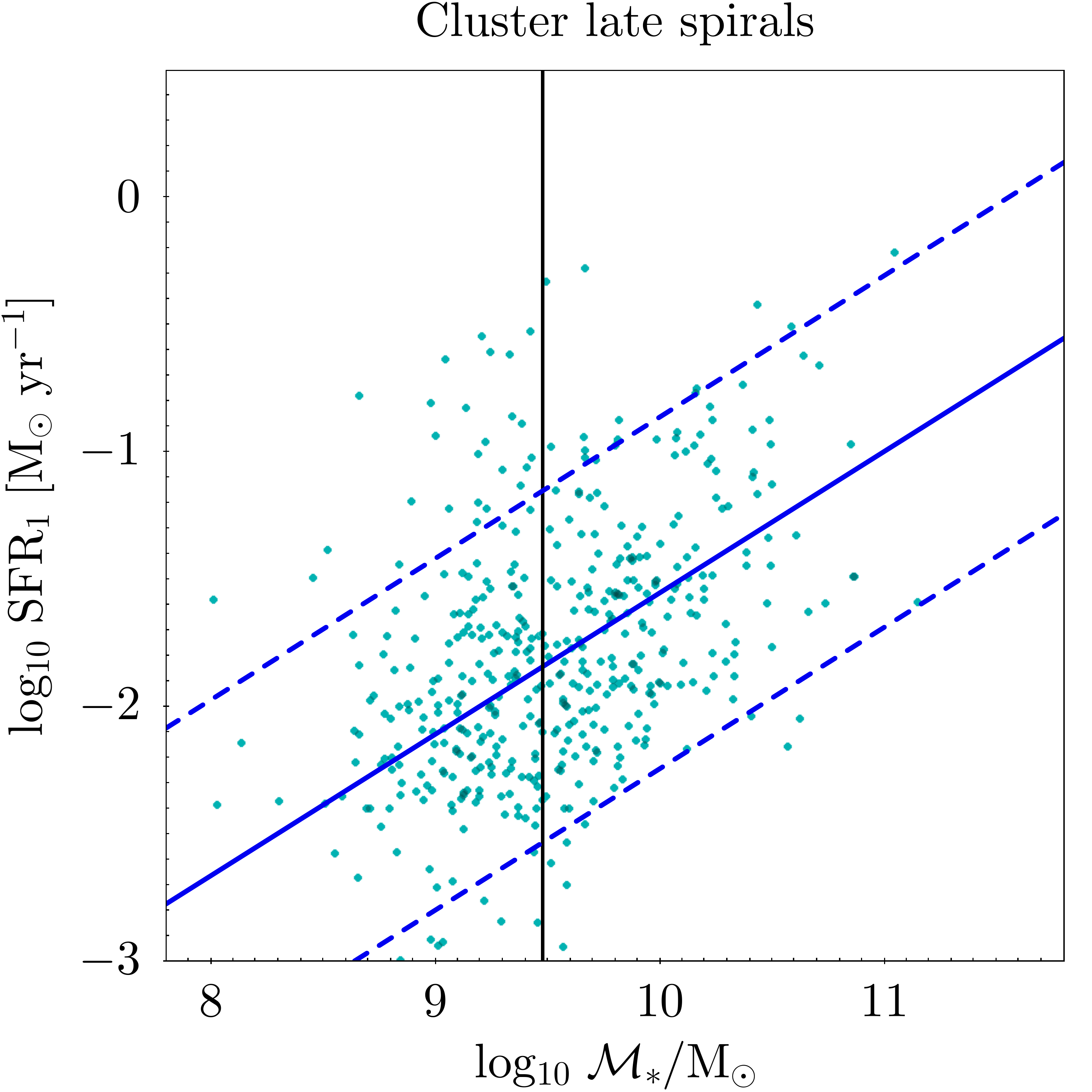}}
   \subfloat{\label{fig:SFR1-M_SpL-Field}
     \includegraphics[width=0.23\textwidth]{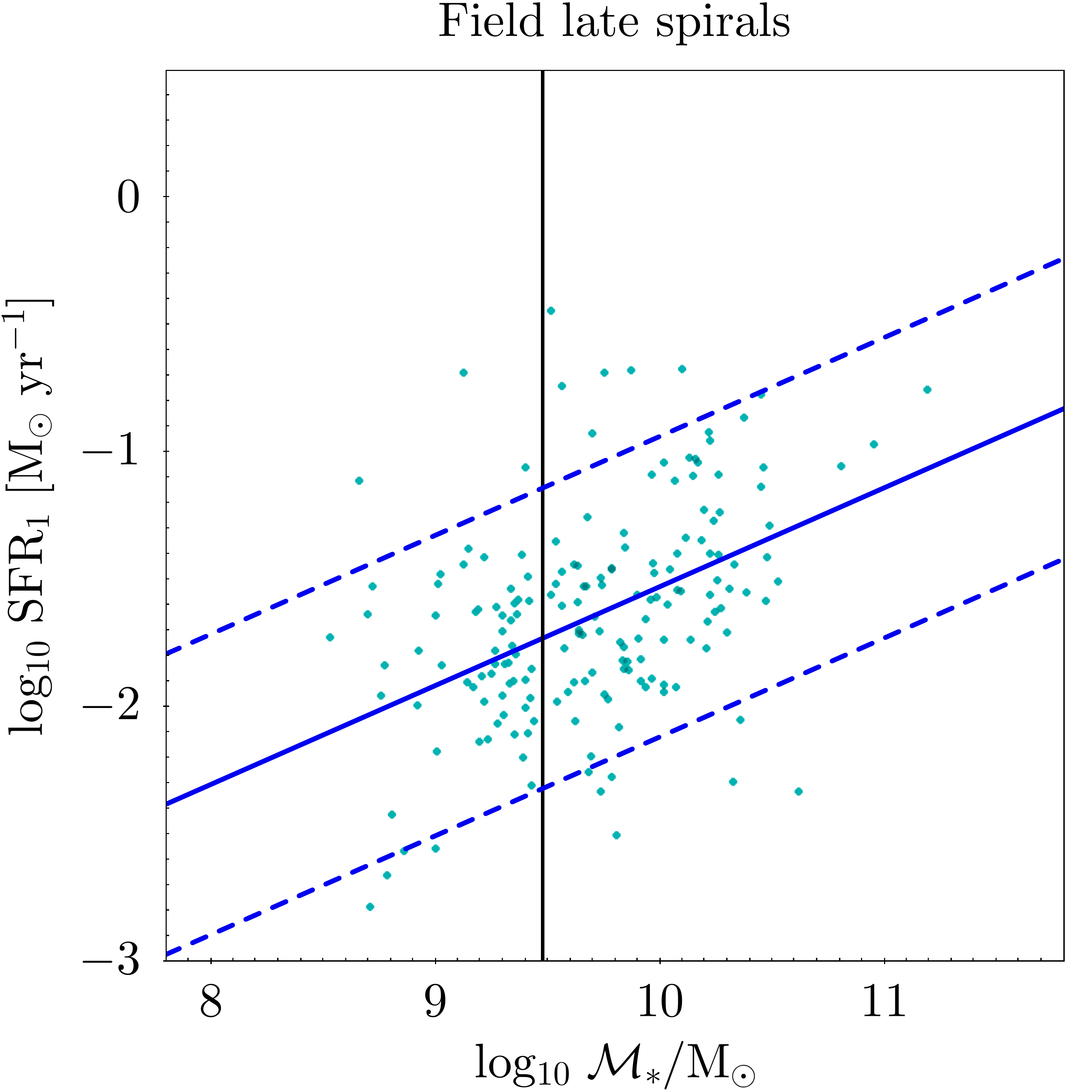}}
     
\caption{SFR-stellar mass relation for early ({\it top}) and late ({\it bottom}) spirals, in clusters ({\it left}) and in the field ({\it right}). {\it Solid blue lines:}  least-square best fits; {\it dashed blue lines:}  $\pm 1.5\times \sigma$ dispersion. {\it Vertical black lines:}  mass lower limit of the sample ($\log_{10} \ \mathcal{M}_*/ {\rm M}_\odot > 9.48$).}
\label{fig:SFR-Mass_relation}
\end{figure}

The SFR-mass relation, also known as the Main Sequence (MS) for galaxies, has been observed in the local universe with SDSS data \citep[e.g.,][]{Kauffmann2003a, Chang2015, Pan2018}, and in high redshift galaxy populations to $z \sim 2$ \citep[e.g.,][]{Lara-Lopez2010, Peng2010, Wuyts2011} and above \citep[e.g.,][]{Speagle2014, Salmon2015, Katsianis2016}. For the sample of WINGS and OmegaWINGS, the SFR-mass relation has been presented and discussed by \cite{Paccagnella2016}. Here, besides using the results from the new stellar population analysis, we make a distinction based on morphology. In particular, we focus on the spiral galaxies, whose MS is presented in \autoref{fig:SFR-Mass_relation}. Only galaxies with SFR$_1 > 1\times 10^{-3} \ \textrm{M}_\odot \ \rm yr^{-1}$ have been used. This limit is obtained from the stellar population synthesis method which, if applied to spectra sampling total stellar masses lower than $10^4 \ \textrm{M}_\odot$ \citep{Weidner2006} might lead to biased results due to the possibly incomplete sampling of the IMF. The threshold value was calculated by taking into account a typical age range of about $10^7$ yr for SFR$_1$.

First of all, we note no significant differences neither in the average values nor in the slopes of the SFR-mass relation of early spirals in the two environments. However, SpE in clusters tend to span a wider range in SFR values at similar stellar masses ($1.5\times \sigma = 0.940$ and $0.695 \ dex$, in clusters and field, respectively). Conversely, cluster late-type spirals have a slope similar to their earlier type counterparts, but a smaller average value (SFR$_1 = 0.1191 $ and $0.0466$ M$_\odot$ yr$^{-1}$, on average, for SpEs and SpLs in clusters, respectively), and also a smaller dispersion ($1.5\times \sigma = 0.691 \ dex$), although the number of galaxies is smaller. On the other hand, SpLs in the field show a less steep relation and a similar dispersion, compared to SpLs in clusters.
To investigate whether the smaller number of spiral galaxies in the field sample introduces biases, we made a statistical test. We extracted 1000 random subsamples of cluster spirals of the same size as the field sample  (the higher the weight of a given galaxy, the probability of being chosen is higher). We confirmed that the main sequence of cluster spirals has a larger dispersion than for field spirals. This result is very evident for SpE, while cluster and field SpL have similar scatters, although the number of objects in this class is also much smaller.

\subsection{The colour-mass relation}
\label{subsec:color-magnitude}
\begin{figure}
 \centering
  \subfloat{
    \includegraphics[trim={0 0 20 0}, clip, height=4.6cm]{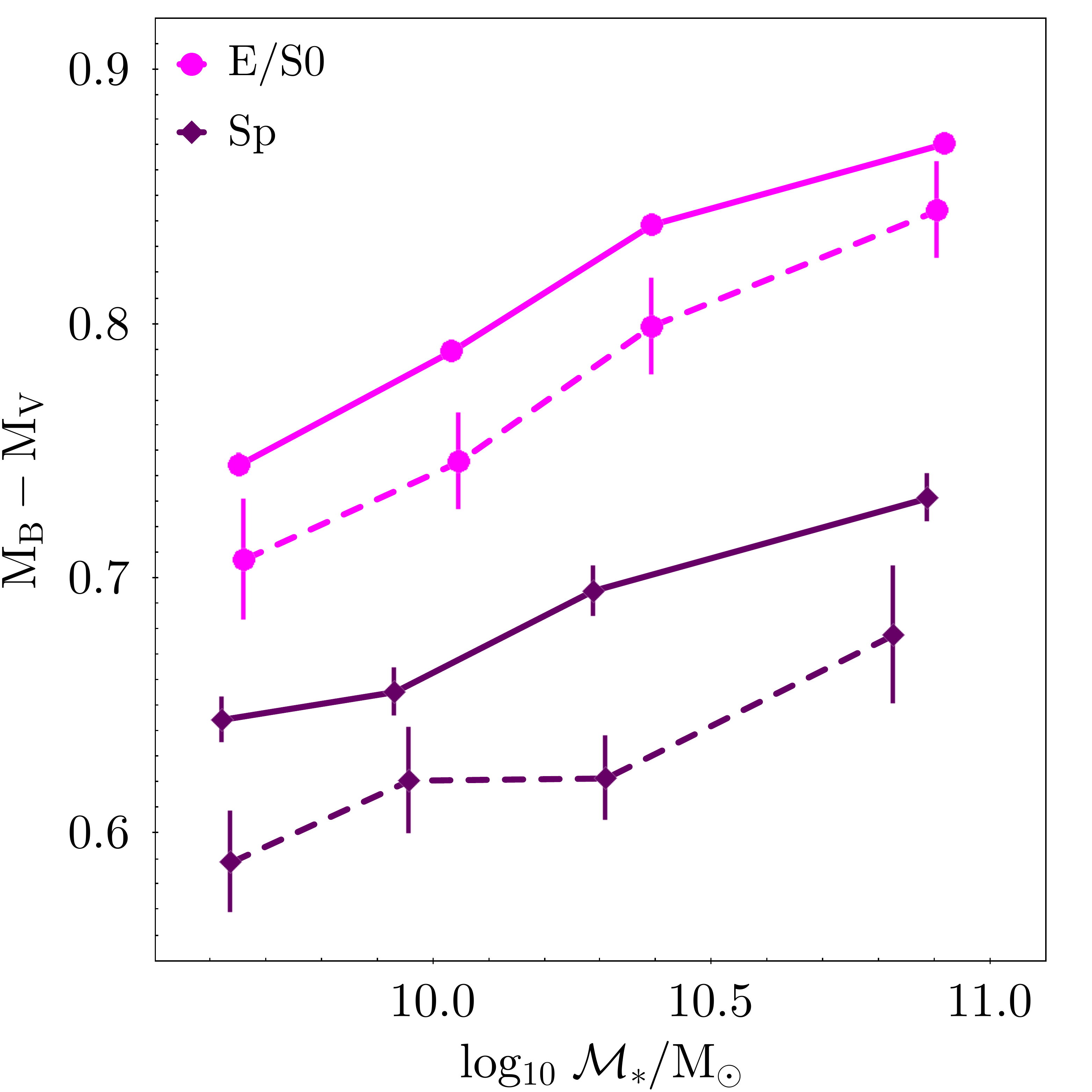}}
  \subfloat{
    \includegraphics[trim={110 0 20 0}, clip, height=4.6cm]{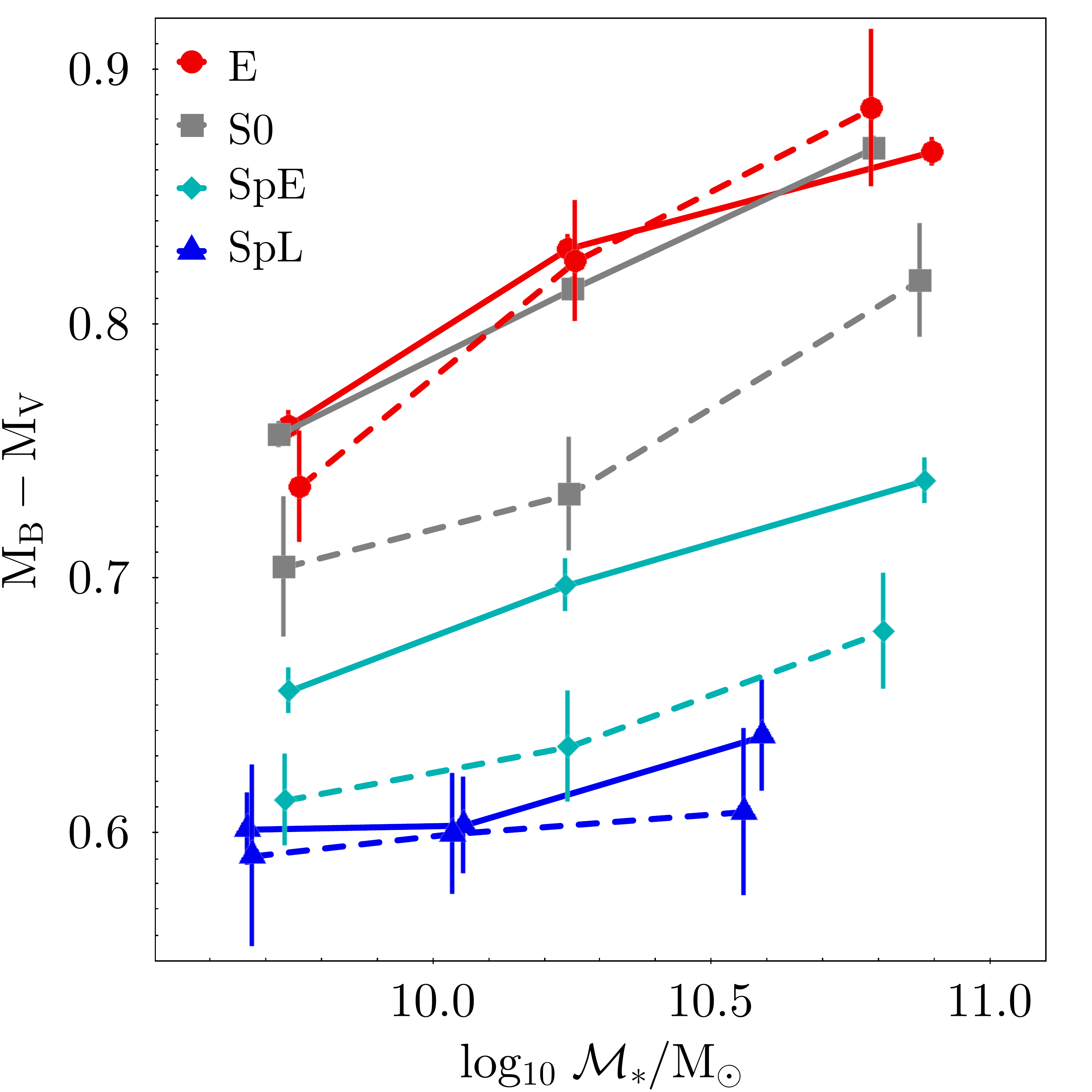}}
  \caption{Mean colours of galaxies in clusters ({\it solid lines}) and in the field ({\it dashed lines}), as a function of stellar mass. {\it Left panel:} early and late-types.  \textit{Right panel:} morphological types E, S0, SpE, and SpL. Error bars were obtained through the bootstrap resampling method.}
 \label{fig:Color-Mass_Cl-Field}
\end{figure}

A broad view of the properties of stellar populations in galaxies can be achieved by looking at the rest-frame colours as a function of stellar mass. To do so, we use the values of these properties as provided by \texttt{SINOPSIS}, based on the spectral model, which include K-corrections automatically (we have assumed that K-corrections are the same for aperture and total magnitudes). The advantage of this approach is that we do not need to rely on pre-calculated K-correction values that depend on morphological types, and  which might be problematic for cluster spirals, since they are normally redder than their field counterparts. Furthermore, this formulation has already been successfully tested and used in \cite{Valentinuzzi2010}.

When analysing the relation between mass and colour for spiral galaxies in clusters and the field, we witness the well-known Butcher-Oemler effect \citep[e.g.,][]{Butcher1978a}, i.e., the scarcity of blue, star-forming, galaxies in local clusters or, more in general, the evolution of the blue galaxy fraction with redshift. 

In \autoref{fig:Color-Mass_Cl-Field}, we show this relation for cluster and field galaxies, by morphological type, in stellar mass bins. Besides displaying the clear, well-known, dependence of colour on mass \citep[e.g.,][]{McGaugh2014, Wel2016}, cluster galaxies are redder, at fixed mass, than their field counterparts (e.g., \citealp{Martinez2010}; see the right panels in their \autoref{fig:Color-Mass_Cl-Field}). When this relation is analysed as a function of morphology, we observe that late spirals have practically the same colours in clusters and the field, at all masses.  Similarly, ellipticals in both environments show negligible differences, at fixed mass. On the other hand, cluster early spirals are redder than those in the field by 0.04 $-$ 0.06 mag. The strongest difference is found for S0 galaxies, which in clusters are as red as ellipticals, but in the field are bluer by 0.05 $-$ 0.08 mag.

\cite{Tanaka2004} found a correlation between colour and LD, but without separating by morphology, with redder and more luminous galaxies located in environments with higher LD. We also explore the possibility of a correlation between colour and LD, and find that, while cluster spirals are redder, their colours do not seem to depend significantly on local density: the colour turns out to be mostly flat with environment, except for a possible small increase close to the cluster centre, as also found by \cite{Baldry2004}, for the highest local densities.

\section{SFH\lowercase{s} in the cluster and field environment}
\label{sec:SFH_environment}

One of the properties that are most influenced by the cluster environment is, surely, galaxy morphology. Several works, starting with the one by \citet[][but see also more recent works, such as \citealt{Vulcani2012, Fasano2015}]{Dressler1980}, have shown that galaxy morphology is strongly affected by local environment, with early-types being more common at high local density values and, more specifically, in the innermost regions of clusters. For this reason, if a cluster galaxy is still observed with a spiral or, more generally, a late-type morphology, this likely means that it entered the cluster environment relatively recently, i.e., a few Gyr ago, at most \citep[e.g.,][]{Sellwood2011}. On the other hand, there are physical mechanisms that affect also the stellar content and the ability of a galaxy to form stars, that act on shorter timescales, such as ram-pressure stripping and starvation \citep{Boselli&Gavazzi2006}. 

In this section, we study if and how the star formation history of a galaxy is affected by its local environment. To do so in the most unbiased way possible, we first analyse the dependencies of the SFH on the galaxies' main properties: their morphology (\autoref{subsec:SFH_morph-type}), and their stellar mass (\autoref{subsec:SFH_mass}). We focus especially on spiral galaxies, i.e., the objects most likely to be recently acquired by the cluster, hence those where evolutionary processes should be more evident.

When considering the overall lifespan of galaxies, to better visualise differences in the SFH, it is useful to normalise the SFR in each bin to the oldest one (SFR$_4$). Given that the bulk of the total stellar mass of galaxies is usually formed at early epochs, this representation is similar to the specific SFR (sSFR, i.e., the SFR divided by the corresponding stellar mass). This procedure allows us to explore how star-forming activity has changed over time, relative to its initial value, as a function of morphological type, galaxy mass, and cluster properties, among other parameters. 

Furthermore, to better detect and quantify variations in the SFH, we define a ``{\it quenching index}'', i.e., the ratio between the SFRs in different, usually contiguous, age bins:
\begin{equation}
    S_{i,j}=\frac{SFR_i}{SFR_j},    
\end{equation}
with $i>j\;\in[1,4]$ (see \autoref{tab:stellar_ages} for the age bins definitions). With quenching indices thus defined, $S_{i,j}>1$ indicates a diminishing SFR in time. Different values of this index quantify diverse quenching (or enhancement, if $S_{i,j}<1$) intensities, and can be used to quantitatively compare the SFH of galaxies with varied properties and in disparate environments. Hence, $S_{4,3}$ is a proxy of the initial build-up efficiency of the stellar mass. $S_{3,2}$ can be used to quantify changes involving earlier epochs, when the galaxy was likely in the first stages of interaction with the cluster environment, while $S_{2,1}$ is related to the current epoch quenching timescale. Hereafter, the quoted uncertainties on the mean SFRs and quenching indices were evaluated with the boostrap resampling method, where the error bars denote the 68\% confidence interval drawn from a 1,000 random realisations of the original sample, and the minimum weighted number of galaxies used in each bin is 30 (around 20 objects), unless otherwise specified. SFHs are plotted as a function of the age of the galaxy stellar populations or lookback time, i.e., 10 Myr corresponds to the most recent lookback time, and 10 Gyr to the oldest. 

\subsection{The role of morphology in shaping the SFH}
\label{subsec:SFH_morph-type}
A connection is known to exist between morphology and the SFH \citep[e.g.,][]{Kennicutt89}. This is observationally suggested by the colours of galaxies of different Hubble types (see e.g. \autoref{fig:Color-Mass_Cl-Field}) and by spectral analysis \citep[e.g.][]{Garcia-Benito17}, but is found in cosmological simulations as well \citep[e.g.][]{Tacchella19}. Here, we analyse how the SFHs depend on morphology for cluster galaxies; we also compare those SFHs to the ones of field galaxies.

\begin{figure}
 \centering
  \subfloat{
    \includegraphics[width=0.23\textwidth]{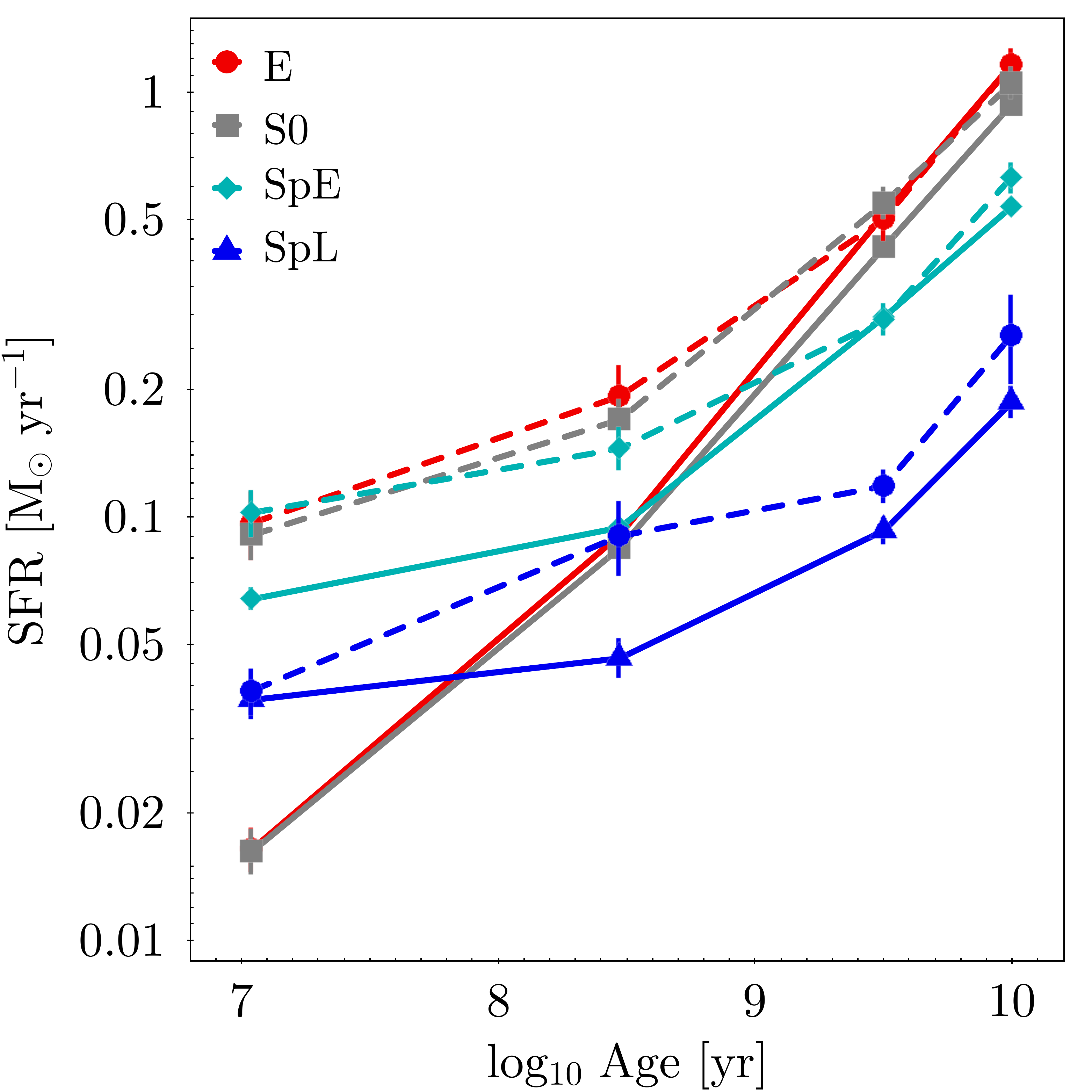}}
  \subfloat{
    \includegraphics[width=0.23\textwidth]{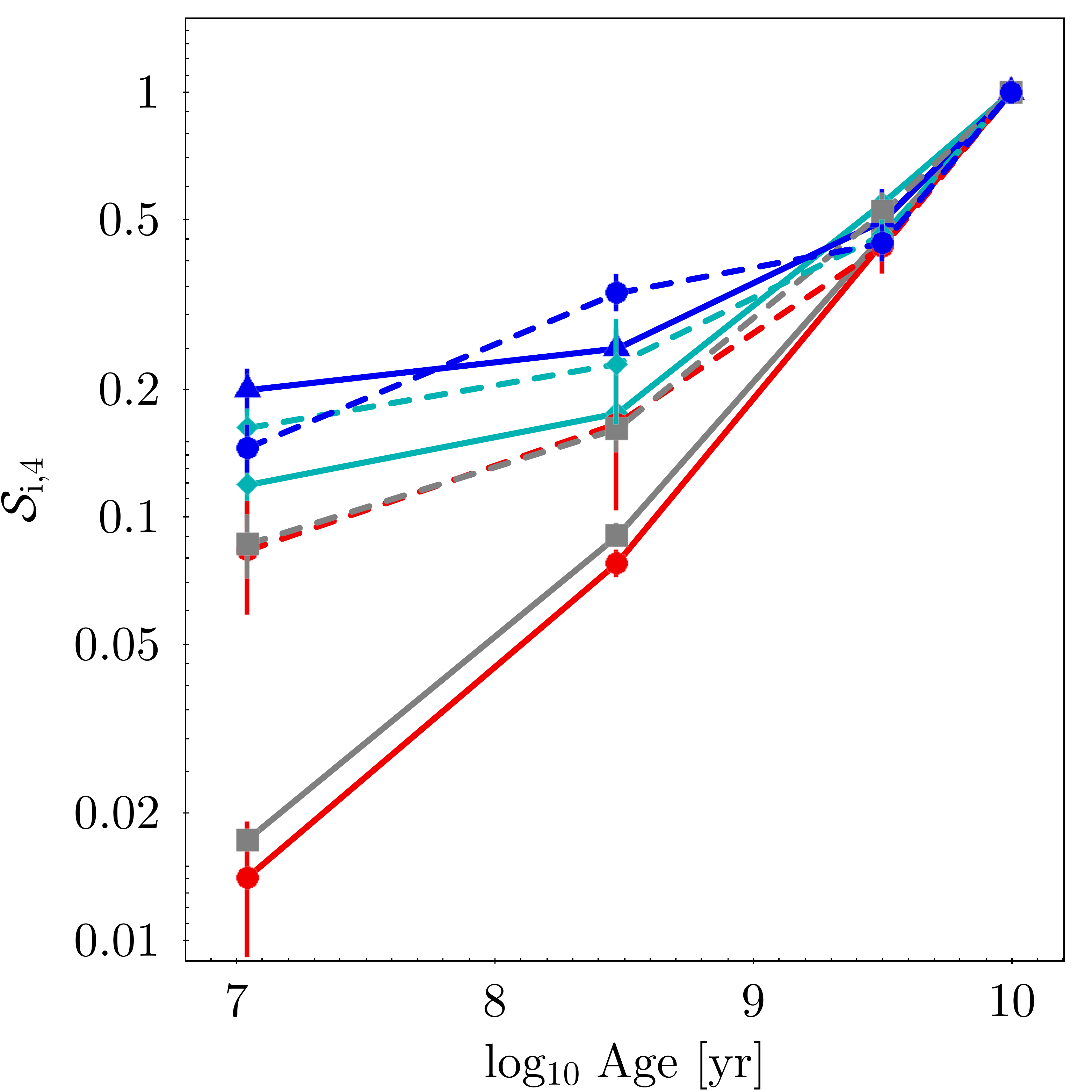}}
    \vspace{1px}
 \caption{SFHs for the final sample of cluster ({\it solid lines}) and field ({\it dashed lines}) galaxies. \emph{Left:} for the main galaxy morphological types. \emph{Right:} same SFHs, normalised to the oldest age bin (SFR$_4$). See \autoref{tab:stellar_ages} for the definition of the SFR bins.}
\label{fig:SFH_morphology}
\end{figure}

\autoref{fig:SFH_morphology} presents the SFHs of cluster and field galaxies, divided into the four main morphological classes. In the left panel, it is possible to discern a sequence in both the average values of the SFR and the slopes, as a function of morphological type. The average value of SFR$_4$, which is a quite good proxy for the total stellar mass, diminishes monotonically from ellipticals to SpL, hence confirming the already known morphology-mass relation \citep[e.g.,][but see also the mass distribution as a function of morphology for the sample in this work, presented in the Appendix, \autoref{fig:Mass_distr_types_sample}]{Vulcani2011,Calvi2012,Wilman2012}. This is even more evident in the right panel of \autoref{fig:SFH_morphology}, where the SFR at each epoch is normalised by the oldest one. What we observe here is a consequence of both downsizing (more massive galaxies are quenched at earlier epochs) and morphology. 

At fixed morphology, the average normalised values of both SFR$_1$ and SFR$_2$ are systematically higher for field galaxies, reflecting a recent quenching effect (or tendency) of the cluster environment. The SFR at earlier cosmic epochs is similar in the two environments at all morphologies. 

Differences in the SFH for the various morphological types are more clearly quantified with the quenching indices, presented in \autoref{tab:quench-index-Morph}. The build-up of stellar mass at the earliest epochs, probed by the $S_{4,3}$ index, is remarkably similar in the four morphological classes and two environments here considered. 

Differences are instead found when inspecting the values of $S_{3,2}$: we first notice that this index is much higher for Es and S0s than for spirals. Secondly, both Es and S0s show important disparities between environments: the highest quenching indices are observed for both cluster ellipticals and S0s, and they are significantly higher than in the field. Spiral galaxies display in general lower $S_{3,2}$ values, indicative of a more constant SFH. Again, they are about 50\% higher for clusters than in the field. 
\begin{table}
\centering
\caption{Quenching indices for cluster and field galaxy samples, according to morphological type. Uncertainties were calculated with the bootstrapping method.}
\label{tab:quench-index-Morph}
\resizebox{\columnwidth}{!}{
\begin{tabular}{cccccc} \hline
                         & Index     & E             & S0            & SpE           & SpL           \\ \hline\hline
\multirow{3}{*}{Cluster} & $S_{4,3}$ & $2.25\pm0.08$ & $2.16\pm0.18$ & $1.82\pm0.11$ & $2.00\pm0.19$ \\
                         & $S_{3,2}$ & $5.71\pm0.42$ & $5.12\pm0.39$ & $3.14\pm0.23$ & $2.00\pm0.35$ \\
                         & $S_{2,1}$ & $5.51\pm0.70$ & $5.00\pm1.09$ & $1.47\pm0.09$ & $1.25\pm0.14$ \\ \hline
\multirow{3}{*}{Field}   & $S_{4,3}$ & $2.30\pm0.36$ & $1.91\pm0.21$ & $2.16\pm0.18$ & $2.20\pm0.29$ \\
                         & $S_{3,2}$ & $2.62\pm0.42$ & $3.24\pm0.47$ & $2.02\pm0.25$ & $1.35\pm0.28$ \\
                         & $S_{2,1}$ & $2.02\pm0.66$ & $1.87\pm0.28$ & $1.42\pm0.34$ & $2.27\pm0.25$ \\ \hline
\end{tabular}}
\end{table}

The quenching indices for the most recent epochs ($S_{2,1}$) are generally much lower than those for earlier epochs (in line with what is presented later, in \autoref{subsec:SFH_mass}). The highest values are found in cluster Es and S0s; they can be, strikingly, up to 3 times larger than for their field counterparts. Conversely, this index is very similar for cluster and field SpE, and 80\% larger for SpL in the field than in clusters. This might be related to the shorter timescale probed by the recent SFR and will be discussed later on. Note that the confidence intervals in the values for late spirals are wider, probably due to the small number of galaxies in this population, particularly in the field. 

Finally, we note that SFR$_1$ is significant for all types (\autoref{fig:SFH_morphology}), with a measurable value even for early-type galaxies. Possible differences in the SFH of cluster and field ellipticals may also come from slightly discrepancies in the mass distribution of the two samples, being weighted towards slightly more massive values for cluster ellipticals.

There is a clear sequence of  sSFR$_1$ that follows the morphological classification, and partially reflects a sequence in stellar mass as well. This trend is most likely a lower limit to the real one, given the fact that the central parts of the galaxies have a higher weight in the fibre observations. Had we had spectra covering the whole galaxies, this trend would have likely been more pronounced, since star formation typically proceeds in discs \citep{Munoz2011, Hirschmann2015}. 

\subsection{SFH and stellar mass}
\label{subsec:SFH_mass}
There is a known correlation between morphology and stellar mass, with higher mass galaxies having an increased probability of being ellipticals or S0s, rather than spirals \citep[see, e.g.][]{Vulcani2011}. With the purpose of isolating the effect of stellar mass on the SFHs, we group galaxies in the four morphological classes into three mass bins.

Low-mass galaxies of all morphologies present a much flatter SFH than massive ones. In all mass bins, however, Es and S0s have the steepest decline, and the difference in SFH between early and late-types (i.e., spheroidals and spirals) is much more pronounced in the massive bin (although we should recall that the most massive galaxies in our sample are found among Es and S0s).

\begin{figure}
 \centering
  \subfloat{
    \includegraphics[trim={0 92 0 0}, clip, height=4cm]{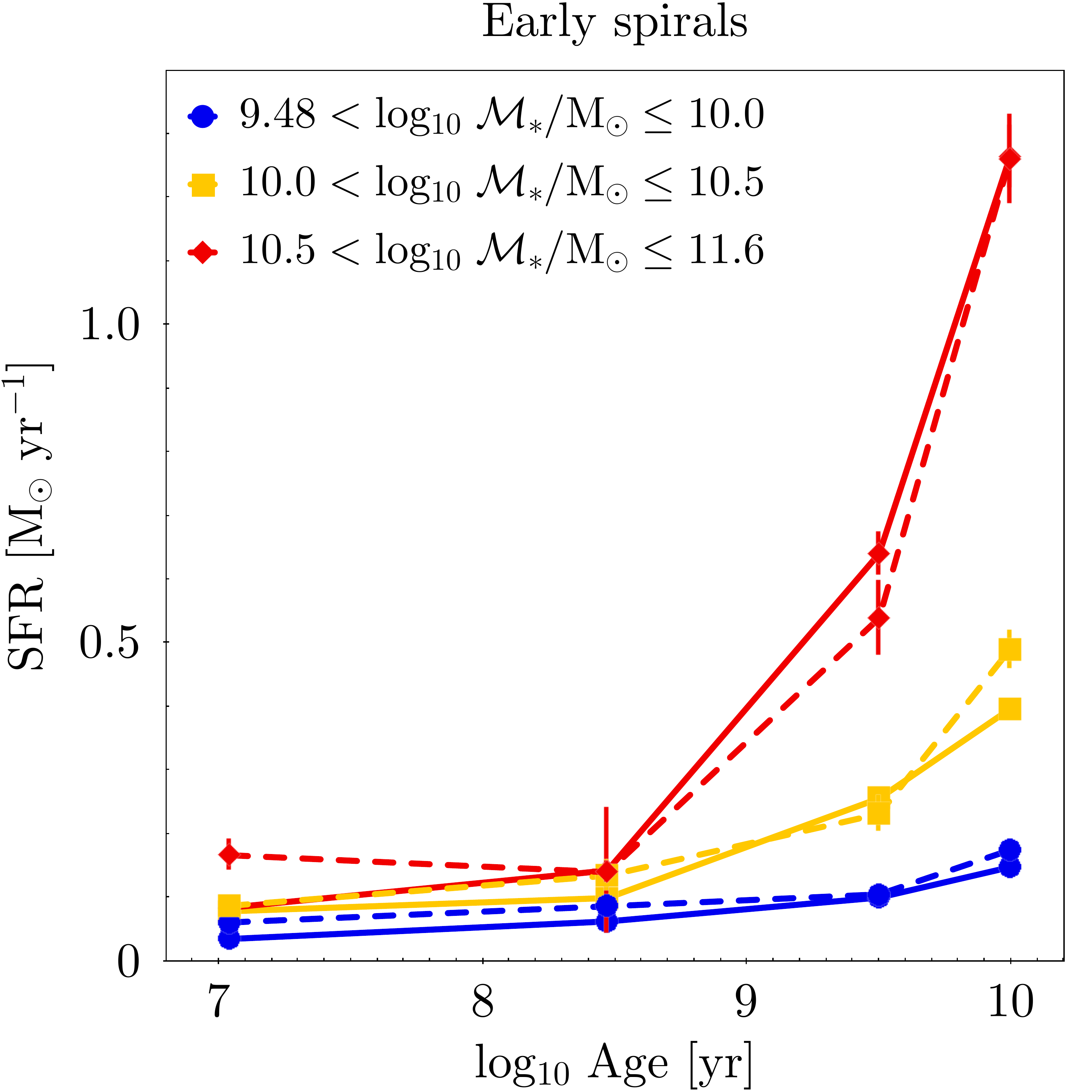}}
  \subfloat{
    \includegraphics[trim={135 92 0 0}, clip, height=4cm]{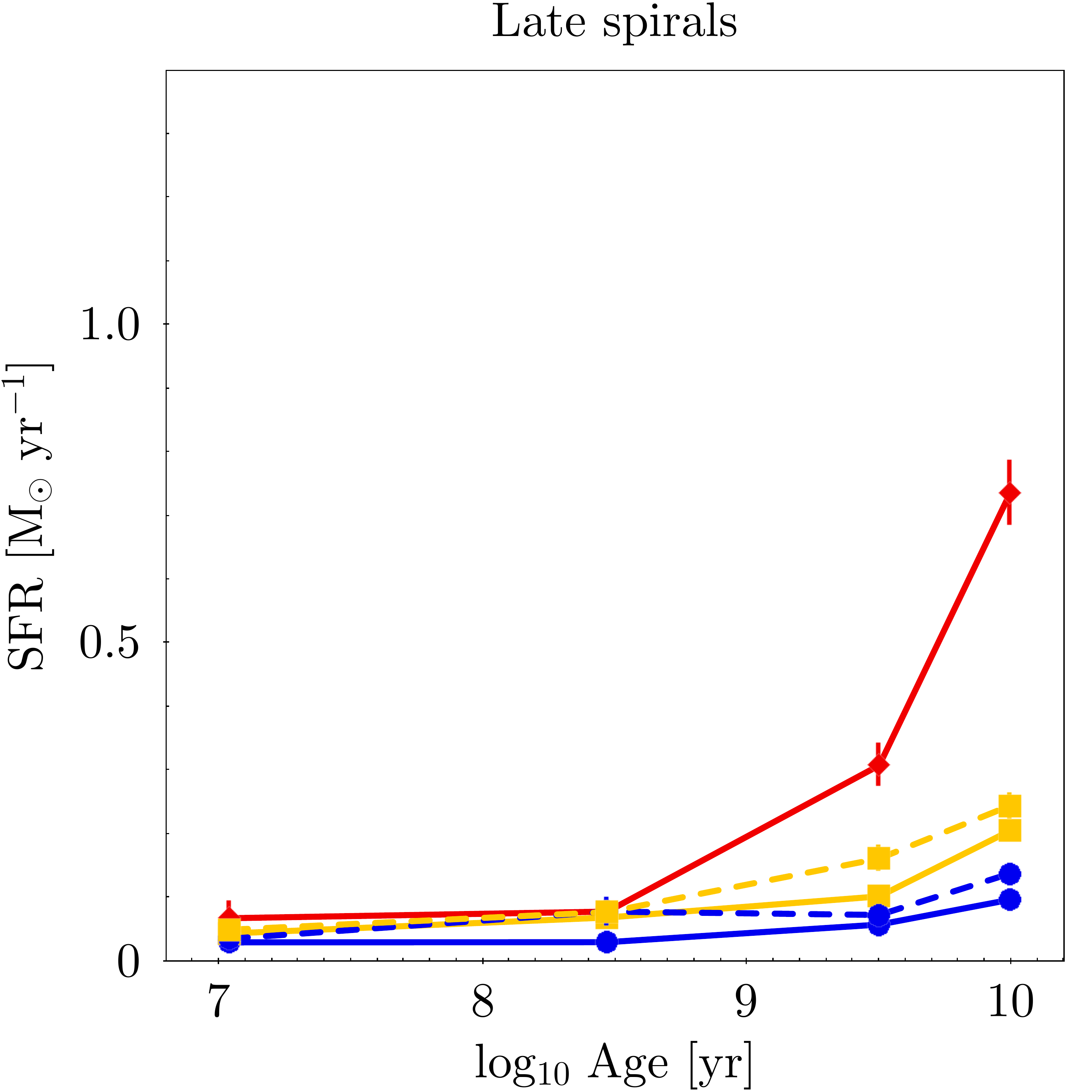}}
\vspace{-8px}
    
  \subfloat{
    \includegraphics[trim={0 0 0 45}, clip, height=4.2cm]{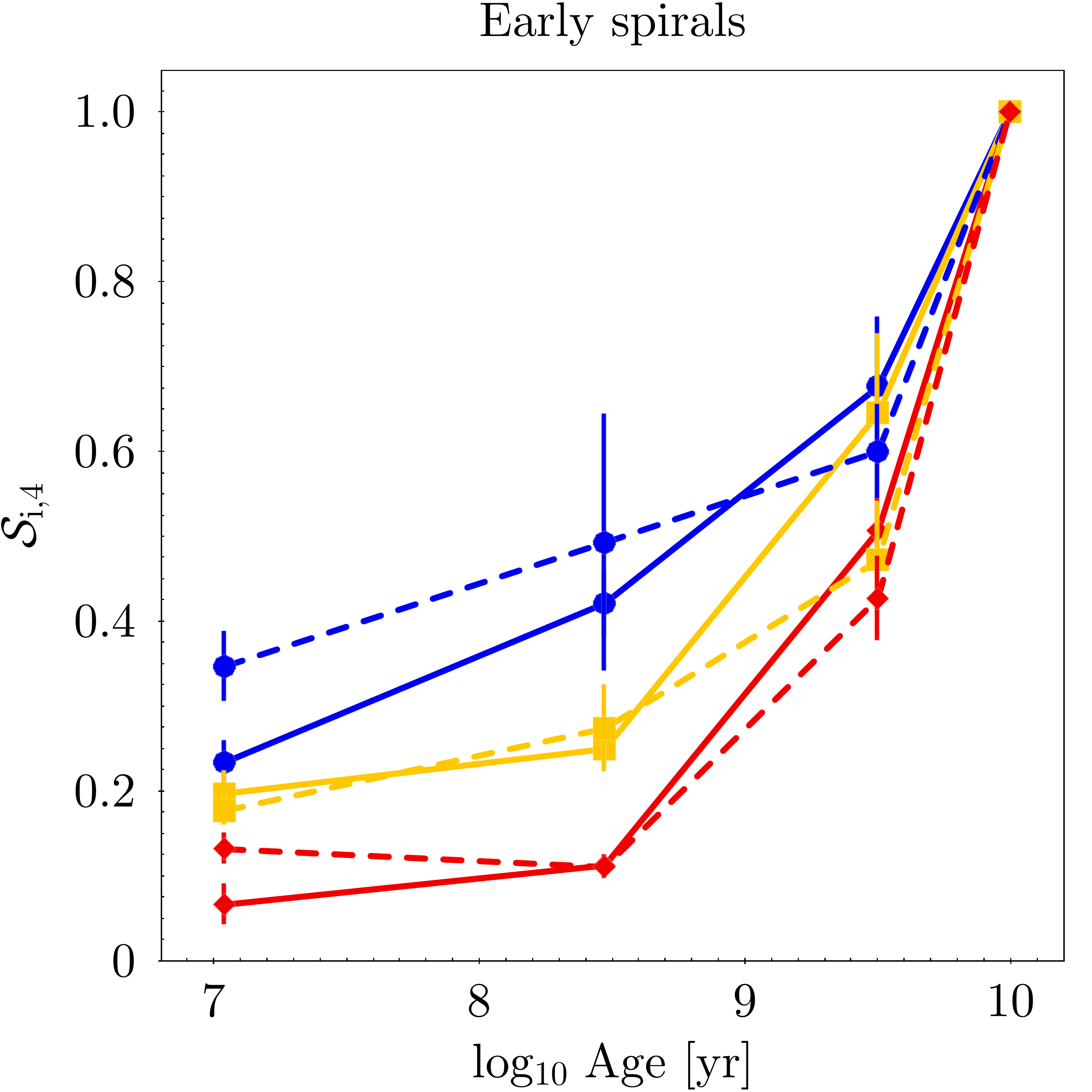}}
  \subfloat{
    \includegraphics[trim={130 0 0 45}, clip, height=4.2cm]{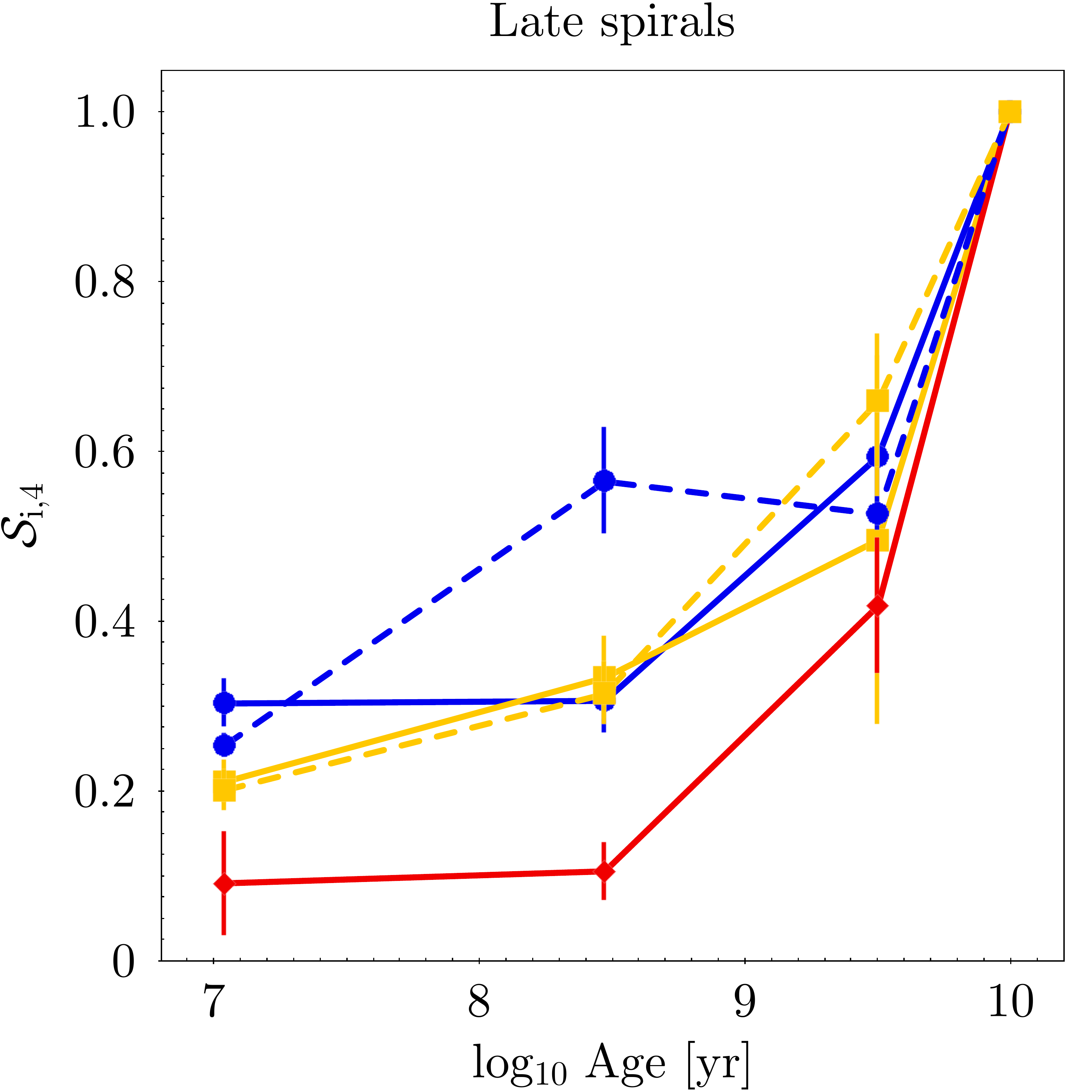}}
        
  \caption{SFHs ({\it  top row}) and SFH normalised by the oldest age bin ({\it  bottom row}) for spiral galaxies in the final sample, separated into bins of stellar mass. {\it Solid lines:} cluster galaxies; {\it dashed lines:} field galaxies. {\it Left column:} early spirals; {\it right column:} late spirals.} 
 \label{fig:SFH_Cl-Field_mass-bins_Sp}
\end{figure}

To compare cluster and field, we now focus on spirals; any possible effect should be more easily spotted in these galaxies, given their recent entry into the cluster. The result is shown in \autoref{fig:SFH_Cl-Field_mass-bins_Sp} and summarised, in terms of the quenching indices, in \autoref{tab:quench-indices-Sp-mass}.

The SFHs patterns agree very well, on average and from several points of view, with a downsizing scenario \citep{Cowie1996}. Consistently with previous well-known results \citep[e.g.,][]{Brinchmann2004, Chen2009, Guglielmo2015}, there is a correlation between galaxy mass and the slopes of the SFH: less massive galaxies have a flatter SFH, reflecting a more continuous star formation process, while massive galaxies are dominated by old stellar populations, indicating a much quicker build-up of their stellar mass.

This result holds both for the whole sample (see \autoref{subsec:SFH_morph-type}), and for spirals considered separately, which means that the SFHs of cluster SpEs and SpLs are dictated by their stellar mass. In particular, the $S_{4,3}$ index assumes, on average, higher values for larger stellar masses, and monotonically decreases from more to less massive galaxies. For field galaxies, the same index takes slightly different values, but still within the confidence interval (substantially wider for the small number of SpLs). No significant differences are hence found when comparing cluster and field galaxies, hinting at a common formation scenario. 

Following with the analysis of the quenching indices, we find that $S_{3,2}$ also decreases monotonically from massive to low-mass galaxies and, again, is larger in clusters than in the field. At fixed mass and morphological type, differences in the quenching indices can be interpreted as a pure, direct, consequence of environmental mechanisms.

Quenching effects on the most recent SFR, probed by $S_{2,1}$, are much less evident, at least as far as spiral galaxies are concerned. Possible explanations to this will be proposed later on, and they involve mechanisms typically found in clusters that can promote star formation activity on short timescales.

\cite{Guglielmo2015} analysed the SFHs of cluster and field galaxies using WINGS and PM2GC data, and found a surprising similarity, at fixed mass and environment, in galaxies of different morphological types. Contrariwise, our investigation through the quenching indices points to non-negligible  differences as a function of morphology. Several factors might explain these divergent results. First of all, quenching indices, defined as ratios, are well suited to highlight differences in the SFR at different epochs. Second, via our use of OmegaWINGS data, we include outer parts of the clusters; these contain larger fractions of late-type galaxies that have a short history of interactions with the cluster itself. Finally, the fact that we do not apply aperture corrections affects late and early-types differently, since the latter generally display flatter radial colour gradients.

\begin{table}
\centering
\caption{Quenching indices of early and late spirals in the cluster and field samples, divided into the same stellar mass bins used in \autoref{fig:SFH_Cl-Field_mass-bins_Sp}. Uncertainties are calculated with the bootstrapping method.}
\label{tab:quench-indices-Sp-mass}
\resizebox{\columnwidth}{!}{
\begin{tabular}{cccccc} \hline
                        & Index    & low-mass      & intermediate-mass           & high-mass  & Type           \\ \hline\hline
\multirow{6}{*}{Cluster}& $S_{4,3}$ & $1.48\pm0.17$ & $1.55\pm0.15$ & $1.95\pm0.18$ & \multirow{3}{*}{SpE} \\
                        & $S_{3,2}$ & $1.60\pm0.25$ & $2.58\pm0.47$ & $4.51\pm0.37$ &  \\
                        & $S_{2,1}$ & $1.80\pm0.25$ & $1.27\pm0.12$ & $1.71\pm0.18$ &  \\ \cline{2-6}

                        & $S_{4,3}$ & $1.68\pm0.21$ & $2.02\pm0.23$ & $2.31\pm0.43$ & \multirow{3}{*}{SpL}\\
                        & $S_{3,2}$ & $1.94\pm0.37$ & $1.49\pm0.50$ & $3.27\pm1.22$ &  \\ 
                        & $S_{2,1}$ & $1.01\pm0.13$ & $1.58\pm0.31$ & $0.87\pm0.46$ &  \\ \hline
                        
\multirow{6}{*}{Field}  & $S_{4,3}$ & $1.67\pm0.20$ & $2.08\pm0.29$ & $2.08\pm0.22$ & \multirow{3}{*}{SpE}\\
                        & $S_{3,2}$ & $1.22\pm0.34$ & $1.72\pm0.36$ & $4.06\pm0.54$ &  \\
                        & $S_{2,1}$ & $1.42\pm0.40$ & $1.97\pm0.37$ & $0.89\pm0.15$ &  \\ \cline{2-6}
                        
                        & $S_{4,3}$ & $1.90\pm0.10$ & $1.52\pm0.42$ & -- & \multirow{3}{*}{SpL}\\
                        & $S_{3,2}$ & $0.93\pm0.47$ & $2.10\pm0.28$ & -- &  \\
                        & $S_{2,1}$ & $2.23\pm0.23$ & $1.56\pm0.29$ & -- & \\ \hline
\end{tabular}}
\end{table}

\subsection{SFH and projected radial distance}
\label{subsec:SFH_proj-dist} 

\begin{figure*}
 \centering
   \subfloat{
   \label{fig:MS_All}
    \includegraphics[trim={0 94 0 0}, clip,  height=4.4cm]{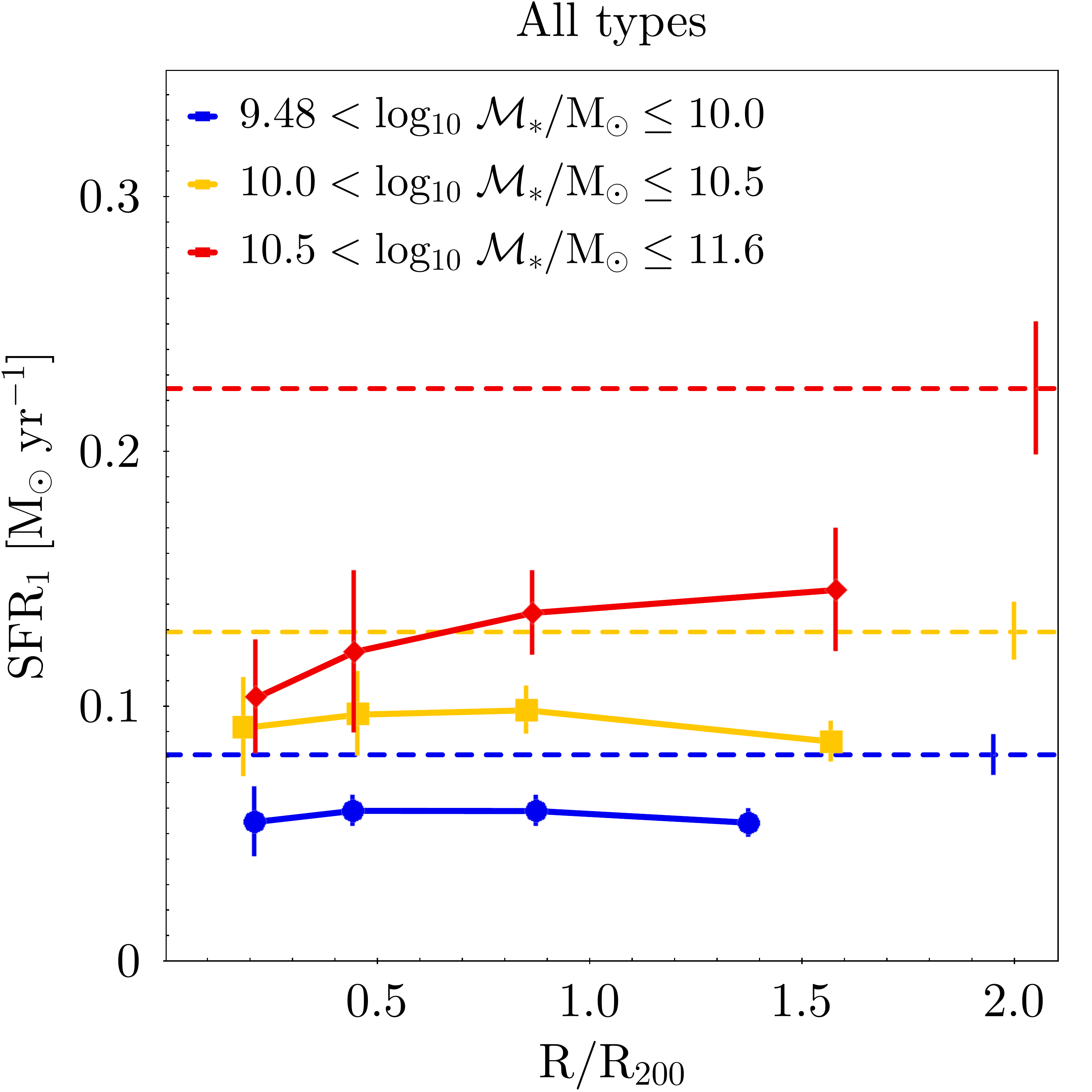}}
  \subfloat{
   \label{fig:MS_E}
    \includegraphics[trim={135 94 0 0}, clip,  height=4.4cm]{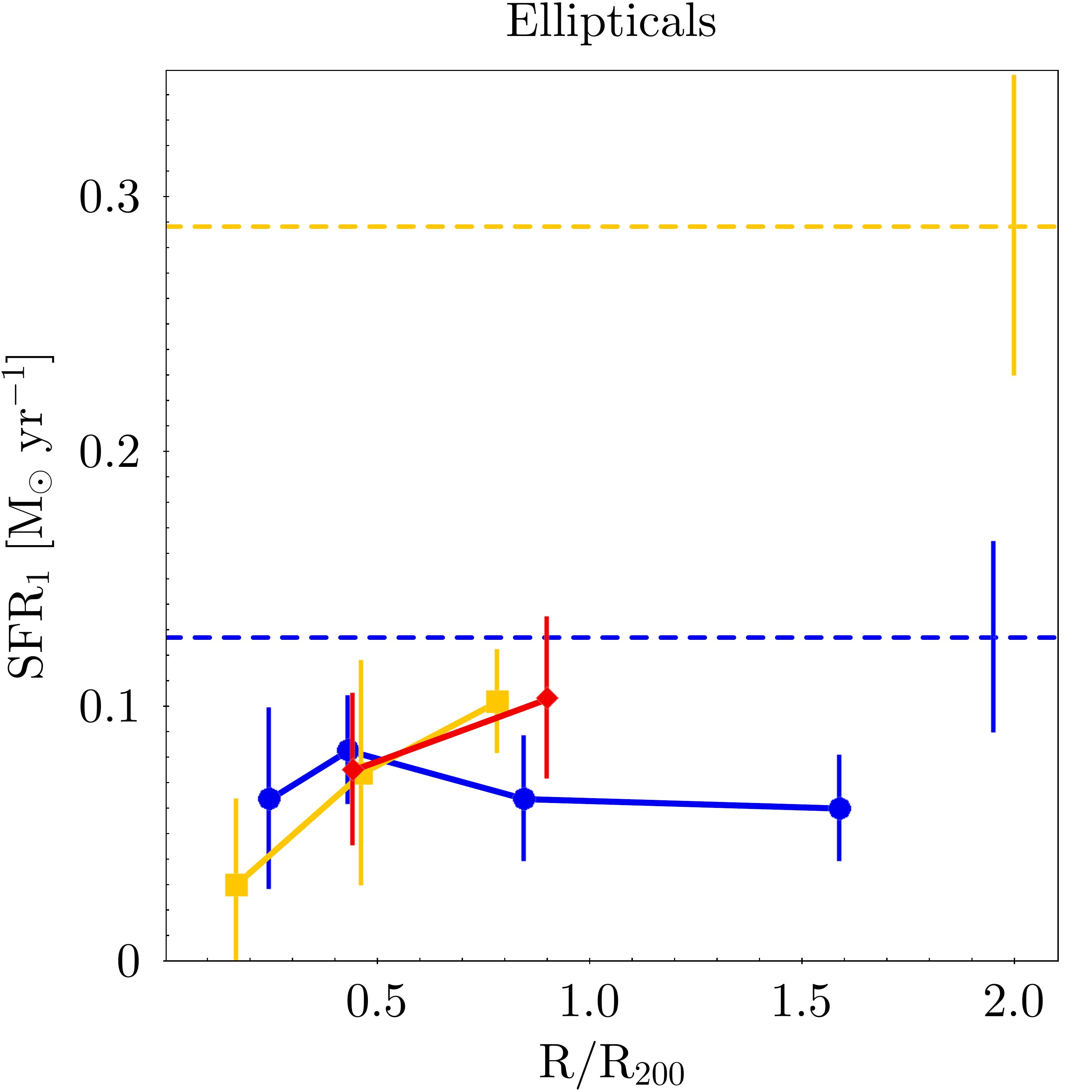}}
  \subfloat{
   \label{fig:MS_S0}
    \includegraphics[trim={135 94 0 0}, clip,  height=4.35cm]{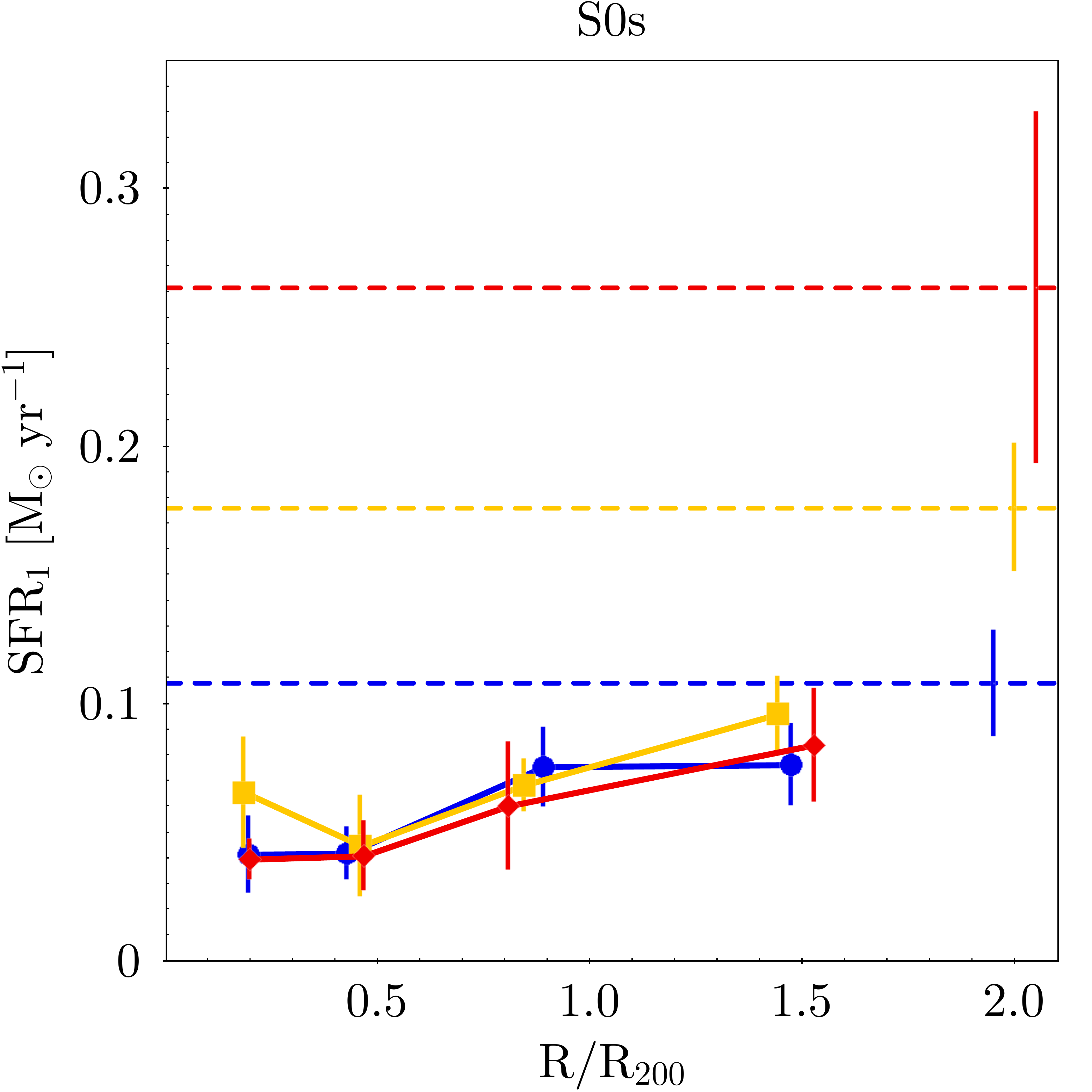}}
  \subfloat{
   \label{fig:MS_Sp}
   \includegraphics[trim={135 94 0 0}, clip,  height=4.4cm]{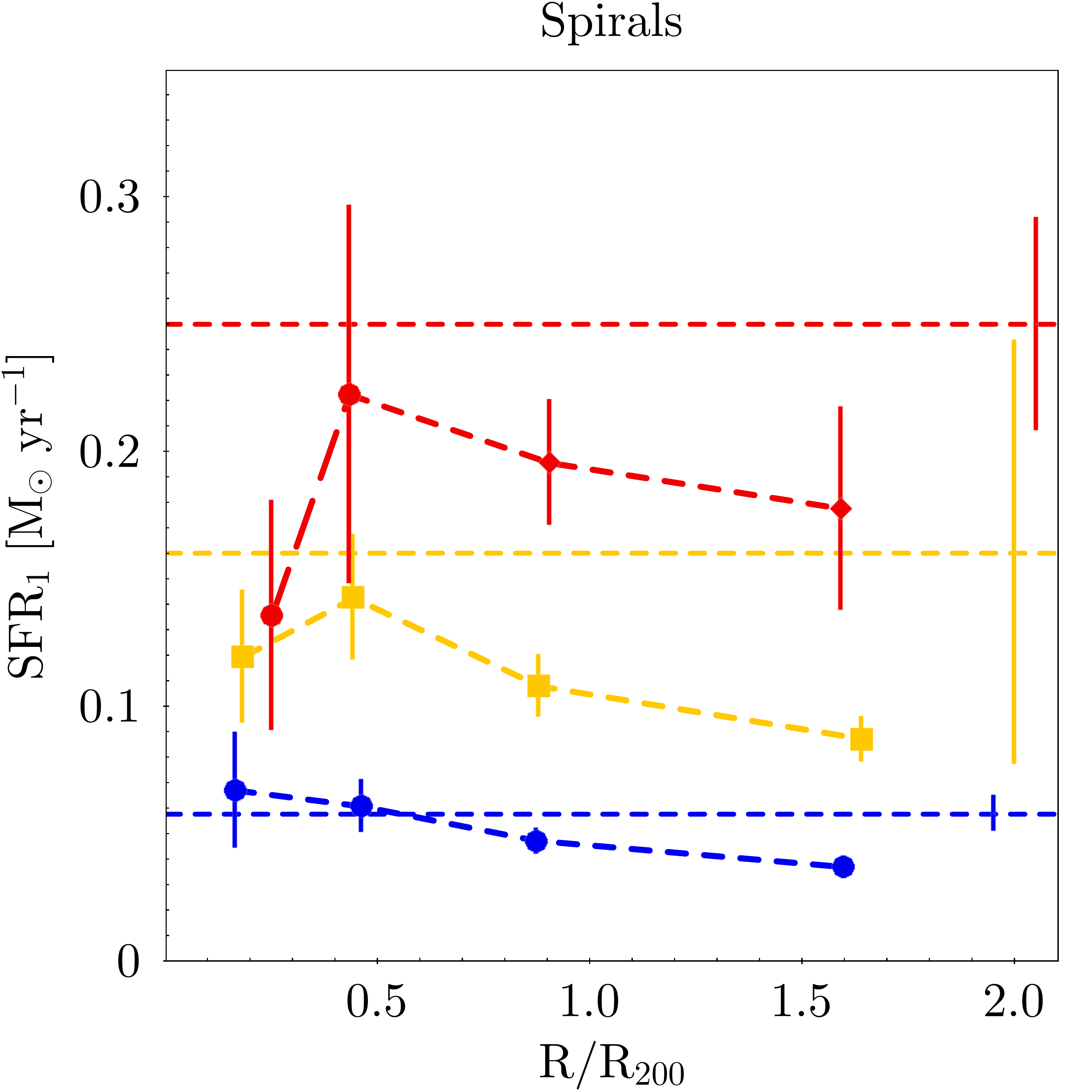}}
     \vspace{-5px}
    
   \subfloat{
   \label{fig:MS2_All}
    \includegraphics[trim={0 0 0 43}, clip,  height=4.7cm]{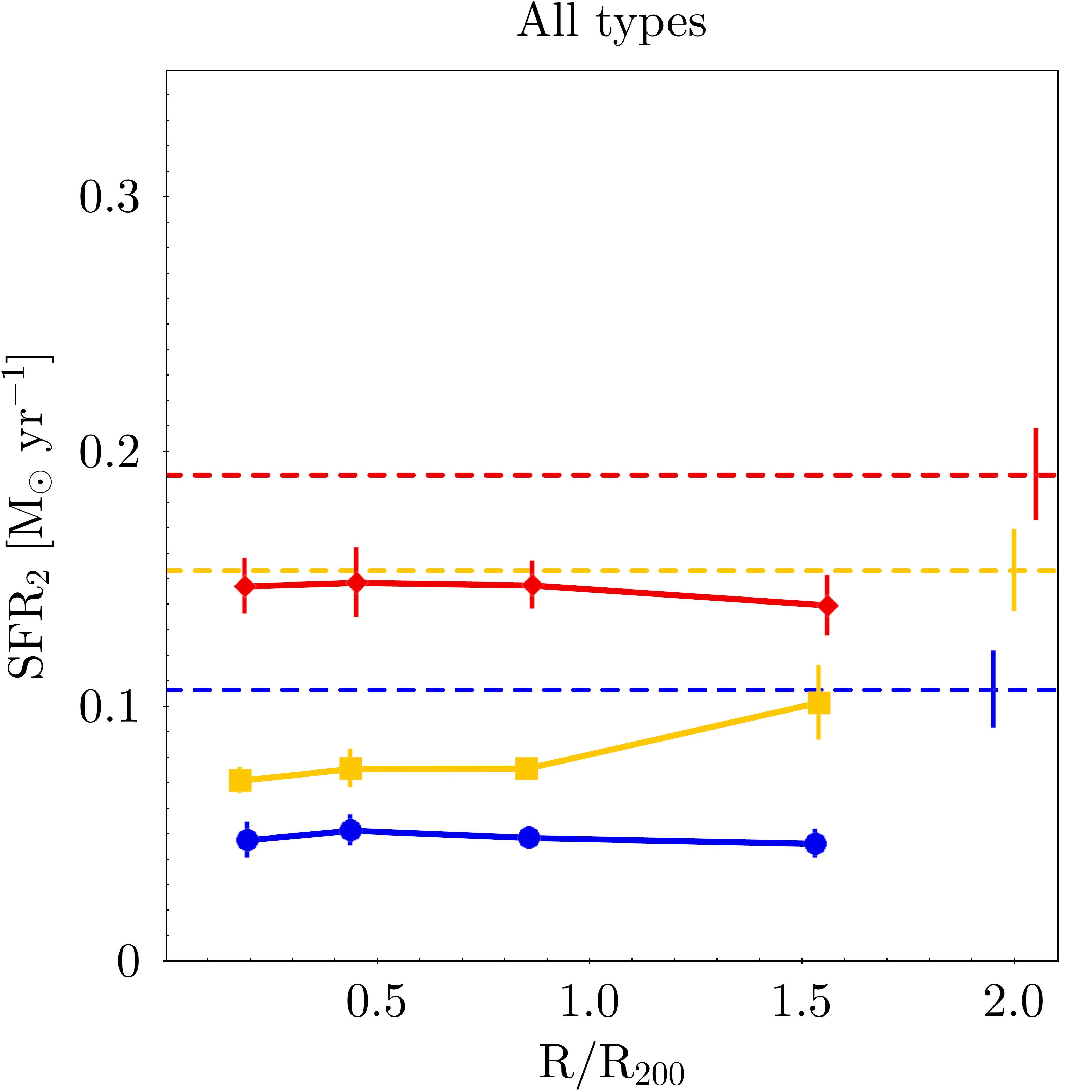}}
  \subfloat{
   \label{fig:MS2_E}
    \includegraphics[trim={135 0 0 43}, clip,  height=4.7cm]{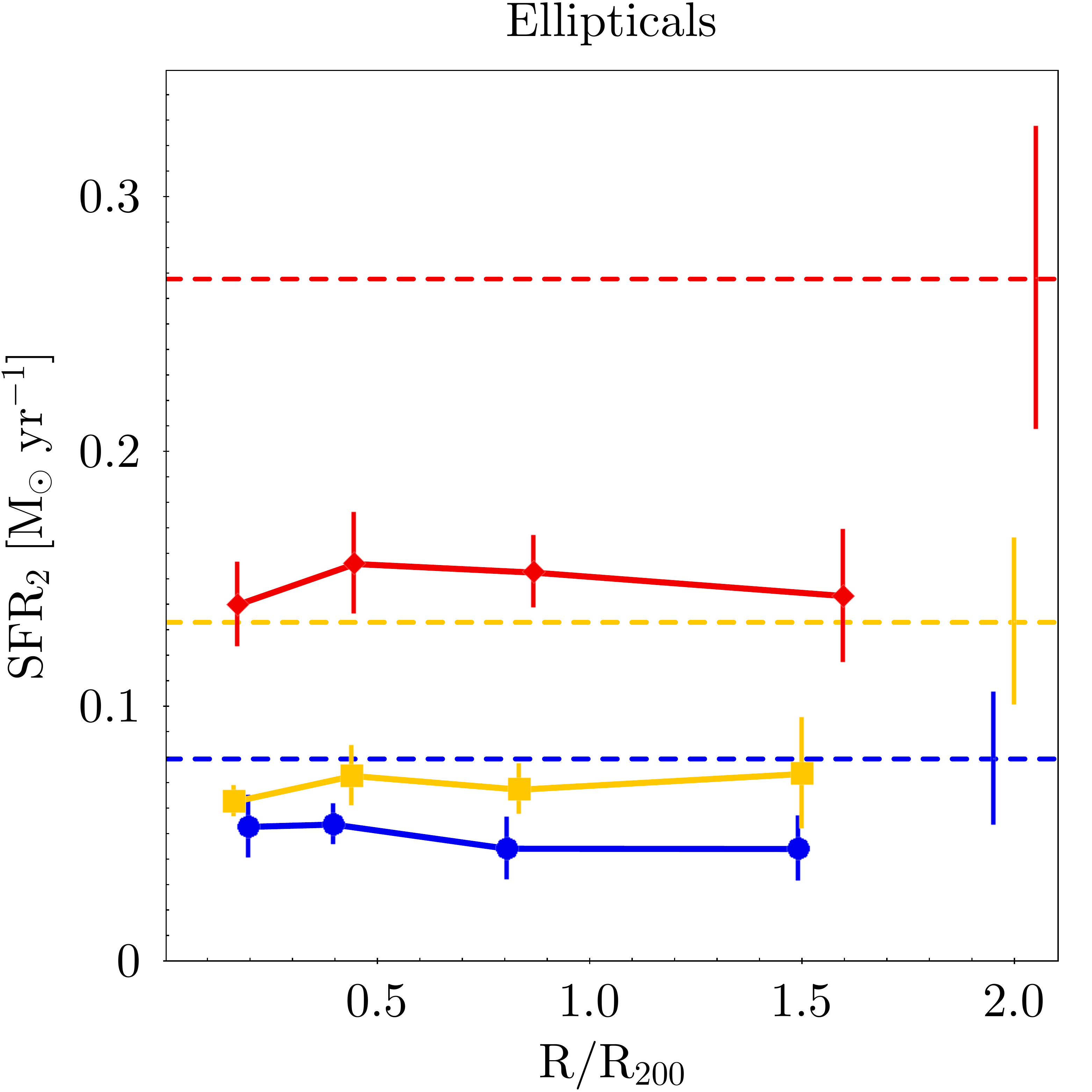}}
  \subfloat{
   \label{fig:MS2_S0}
    \includegraphics[trim={135 0 0 43}, clip,  height=4.65cm]{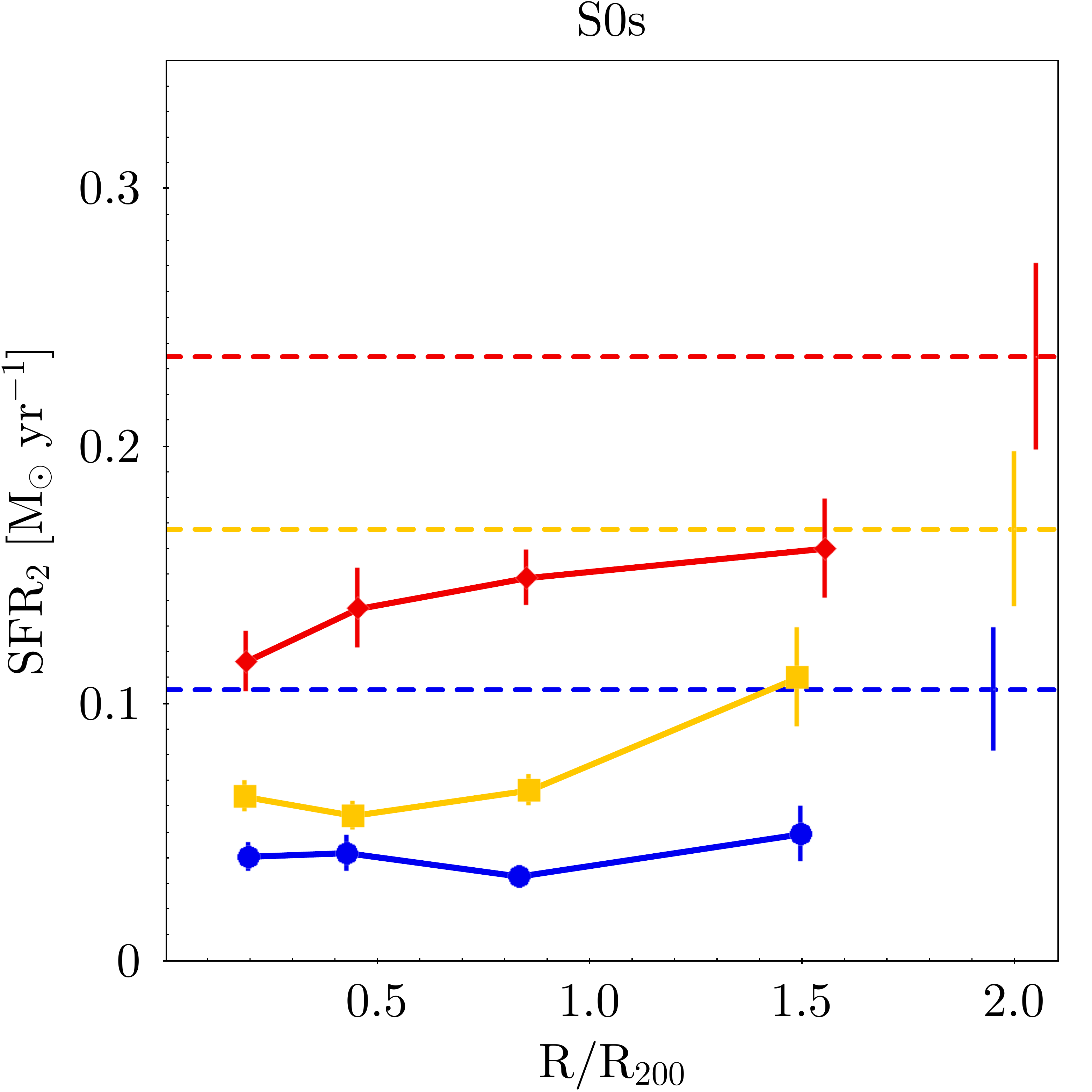}}
  \subfloat{
   \label{fig:MS2_Sp}
    \includegraphics[trim={135 0 0 43}, clip,  height=4.7cm]{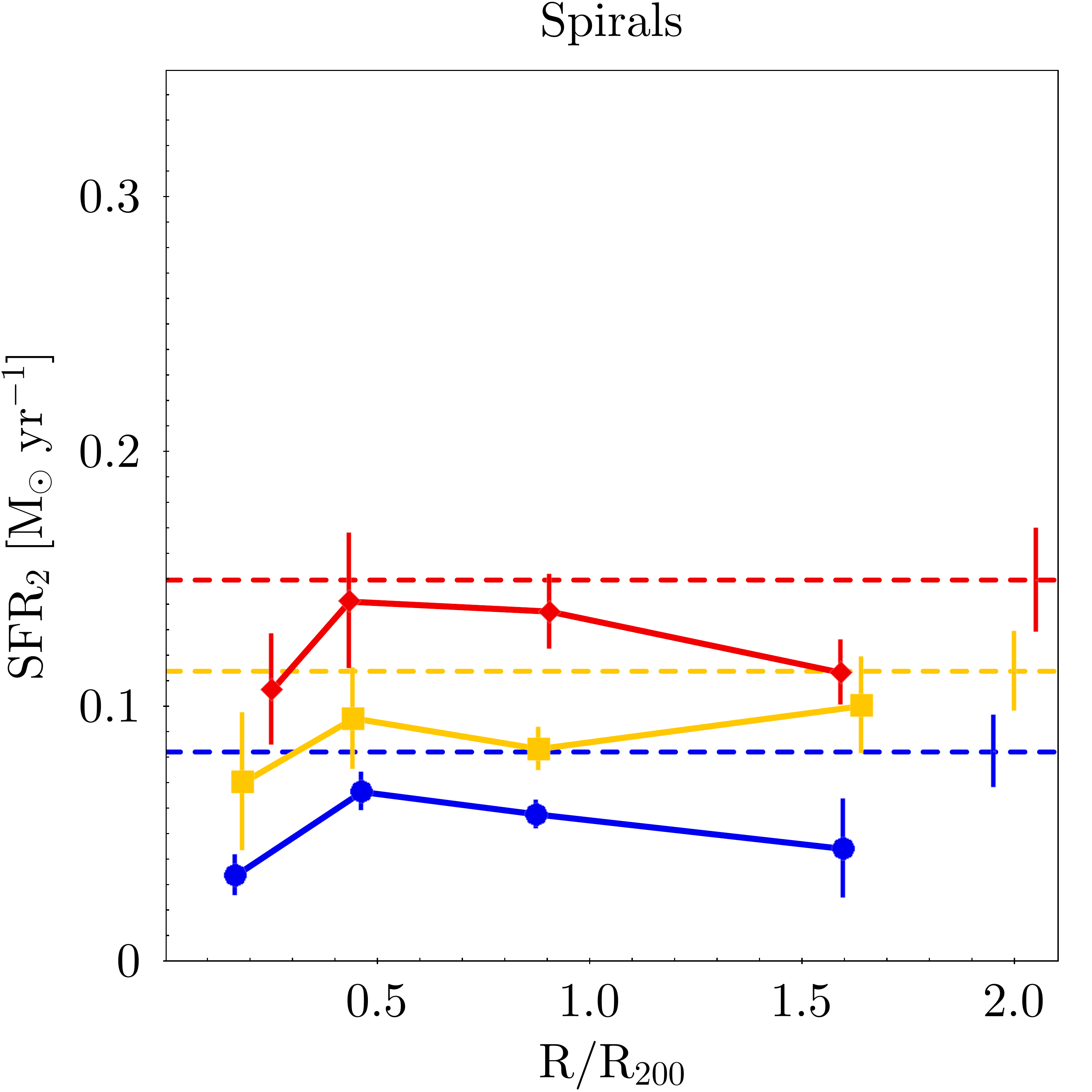}}
    
 \caption{SFR versus mean projected distance for the final sample, in bins of stellar mass, denoted by different colours (see legend in the {\it top left} panel), and separated by morphological class: all types {(\it first column)}, ellipticals {(\it second column)}, S0s {(\it third column)}, and spirals {(\it fourth column)}. {\it Top row:} SFR$_1$ (only actively star-forming galaxies in the sample). {\it Bottom row:} SFR$_2$ (full galaxy sample). {\it Solid lines:} cluster galaxies; {\it dashed lines:} field galaxies. The {\it long-dashed red line} in the top right panel means that the number of star-forming, massive, cluster spirals in the bin closest to the center is very small, and this trend should be taken with caution.}
 \label{fig:Main-sequence}
\end{figure*}

The comparison between the SFH of field and cluster galaxies shows that the environment can affect the ability of galaxies to form stars. In what follows, we look into the mechanisms that are responsible for this in more detail, through the analysis of the recent SFR.

\cite{Paccagnella2016} have studied the changes in the SFR--mass relation for WINGS/OmegaWINGS galaxies, as a function of the clustercentric projected distances. Here, we will extend this analysis, investigate possible effects due to morphology, and explore differences in the older stellar populations, with a particular focus on SFR$_2$.  This parameter samples star formation processes that most likely happened during the first galaxy-cluster interaction for the recently accreted population. To this end, we will consider spiral galaxies as one unique population, making no distinction between early and late-types, to avoid poor statistics.

In \autoref{fig:Main-sequence}, top row, we show the mean SFR$_1$ of star-forming galaxies (i.e., those displaying SFR$_1 > 1\times 10^{-3} \ \rm M_\odot \ yr^{-1}$; see \autoref{subsec:SFR-Mass_relation}), as a function of projected distance to the host cluster centre, normalised by $R_{200}$ and divided into four bins. In the bottom row, we display the mean SFR$_2$ of the whole sample, regardless of the SFR$_1$ value, also as a function of projected clustercentric distance. From left to right, the columns exhibit all morphological types, ellipticals, S0s, and spirals, respectively. In each panel, objects have been separated into three mass bins, denoted with different colours. Cluster galaxies are represented by solid lines and field galaxies, by dashed lines.

In the top left panel (average SFR$_1$ for all morphological types), we observe roughly flat trends as a function of projected distance, except for the most massive bin, where SFR$_1$ decreases towards the centre. When separating into different morphologies, SFR$_1$ declines with diminishing distance for S0s, and for ellipticals of intermediate masses (note that the statistics are poor for Es in the cluster outskirts). Interestingly, the SFR-mass relation --on average-- seems to break down in this representation. SFR$_1$ appears only weakly correlated with stellar mass for early-types, while a dependency still holds for late-types. Regarding the trend with distance, in the case of spirals, SFR$_1$ grows for all masses towards the centre, up to a clustercentric distance of $0.5 R_{200}$. At smaller distances, there is a possible hint of a decreasing SFR$_1$ with smaller distance for high- and intermediate-mass galaxies. Although this observed tendency is driven by a small number of galaxies, it is confirmed by bootstrap resampling analysis. These late-type galaxies, having preserved their spiral morphology, may have also kept a substantial fraction of their gas until close to the centre, and the high-density medium there may have stimulated vigorous star formation, through hydrodynamical or gravitational interactions. 

We analyse the SFR$_2$ in a similar way (\autoref{fig:Main-sequence}, bottom row), to search for possible consequences of the environment on stellar populations at older ages. With this aim, we consider all galaxies in the sample (and not only actively star-forming ones). First, we observe once again that field galaxies have on average a higher SFR$_2$ than cluster members in the same mass bins, most evidently in the case of spheroidal galaxies. The differences between field and clusters, though, are smaller than for SFR$_1$. Taking all the morphological types together (leftmost panel), galaxies within $1 R_{200}$ show a roughly flat SFR$_2$ at any mass. Unlike what is observed for SFR$_1$, when separating by morphology, the SFR-mass relation is again in place, with a clear trend between mass and SFR. 

\begin{figure}
 \centering
  \subfloat{
   \label{fig:Quench_frac_R-proj}
    \includegraphics[height=4.5cm]{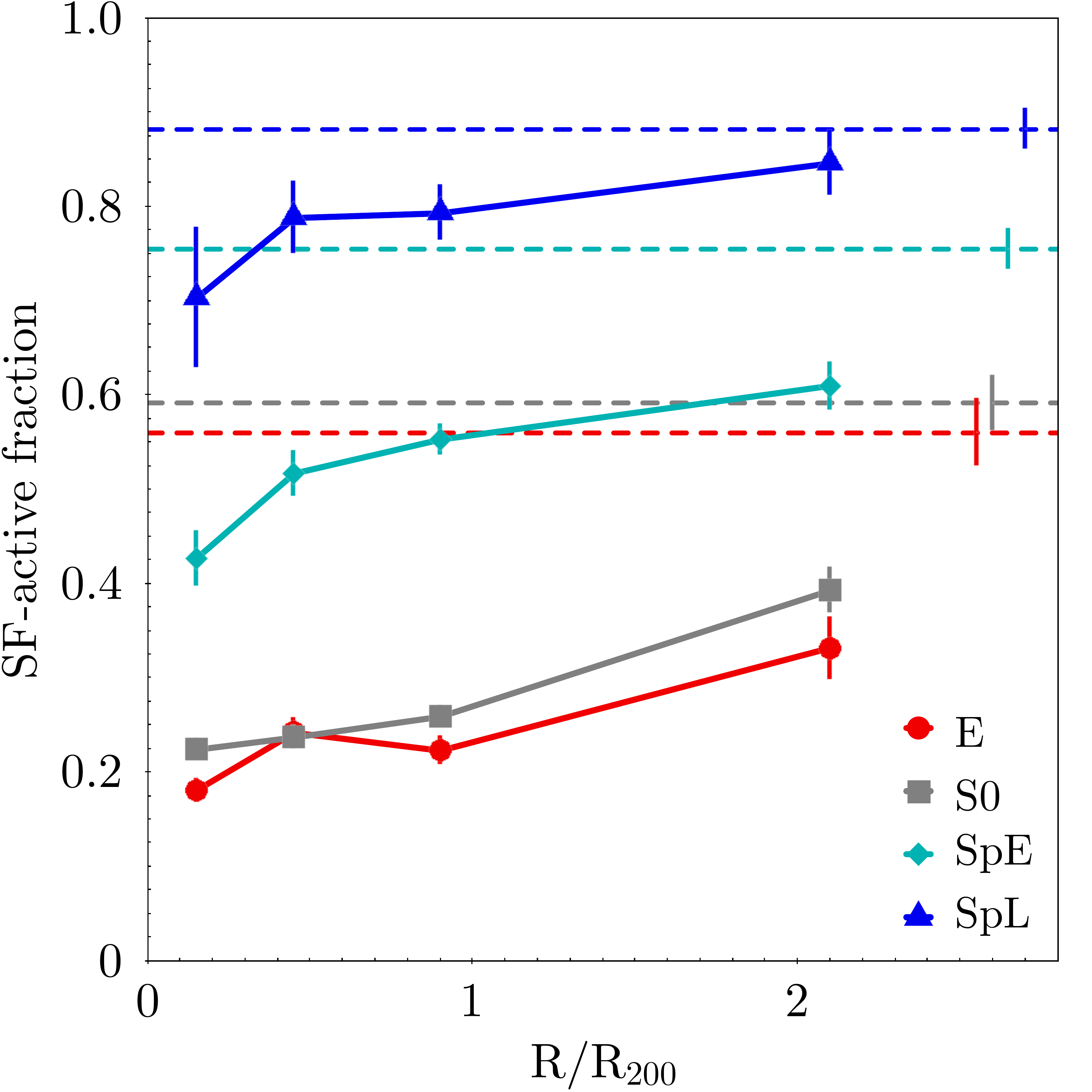}}
  \subfloat{
   \label{fig:Quench_frac_LD}
    \includegraphics[trim={110 0 0 0}, clip,  height=4.5cm]{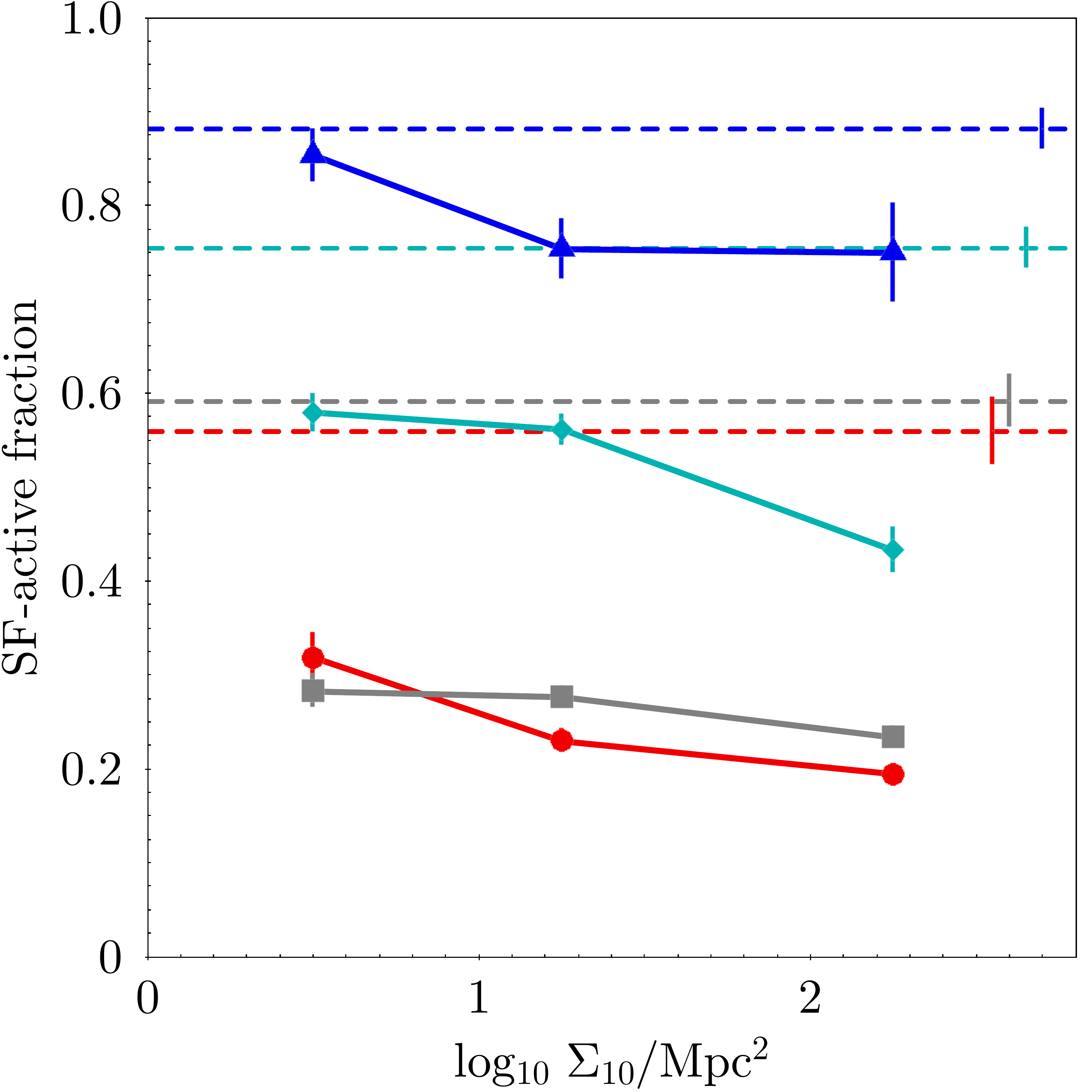}}
    
 \caption{Fractions of star-forming galaxies as a function of clustercentric projected distance ({\it left}) and local density ({\it right}), for the four morphological classes in the sample. The {\it horizontal dashed lines} represent the fractions for field galaxies. Error bars indicate binomial errors.}
 \label{fig:SFing_fraction}
\end{figure}

The run of SFR$_2$ with distance is again flat for Es (second panel) of all masses. As for S0 galaxies (third panel), SFR$_2$ is higher for larger projected distances, and high- and intermediate-masses. In spiral galaxies, SFR$_2$ increases slightly between the cluster outskirts and projected distances of $0.5 R_{200}$, but it drops closer to the cluster centre.

\begin{figure*}
 \centering
  \subfloat{
   \label{fig:MS-LD_All}
    \includegraphics[trim={0 94 0 0}, clip,  height=4.4cm]{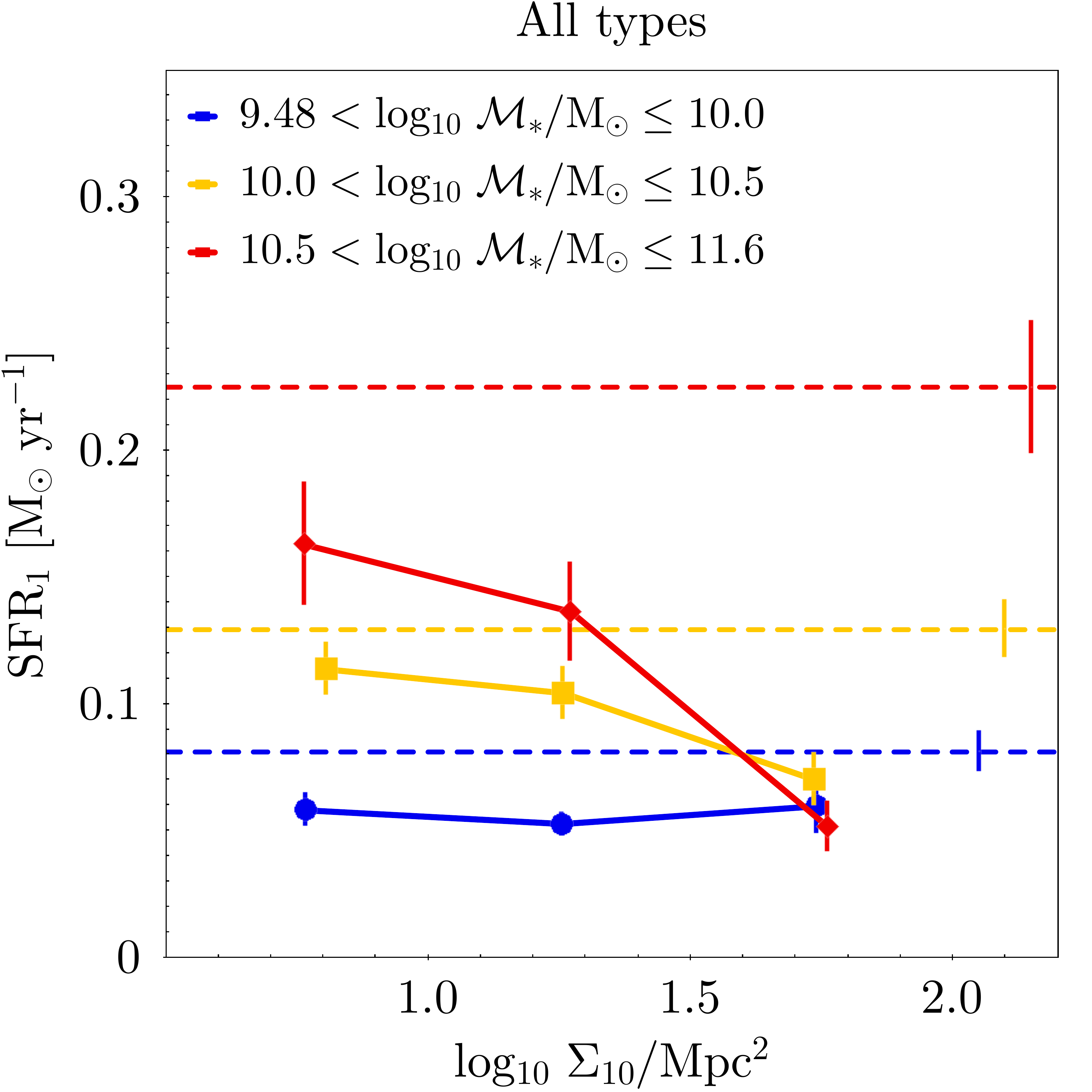}} 
  \subfloat{
   \label{fig:MS-LD_E}
    \includegraphics[trim={135 94 0 0}, clip,  height=4.4cm]{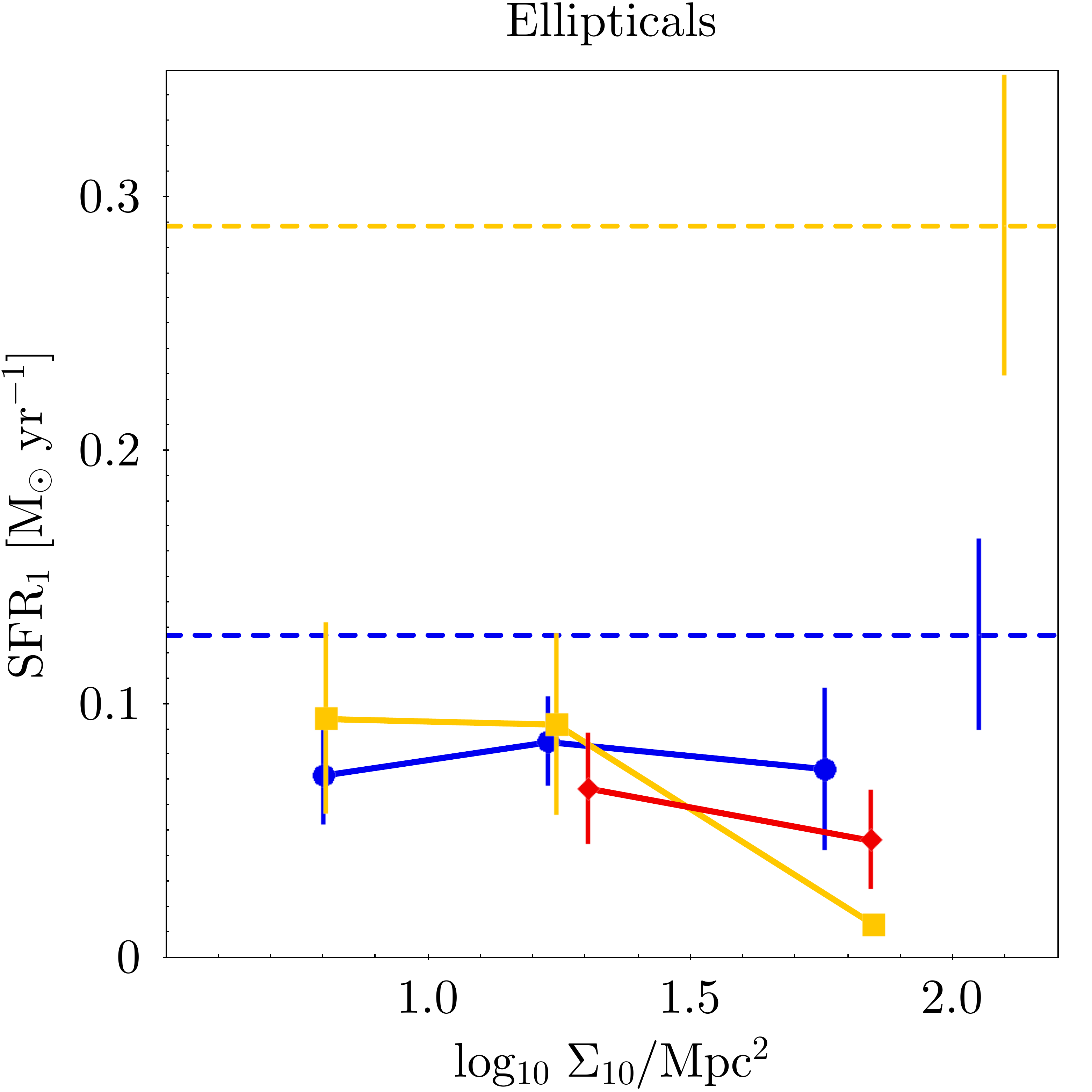}}
  \subfloat{
   \label{fig:MS-LD_S0}
    \includegraphics[trim={135 94 0 0}, clip,  height=4.35cm]{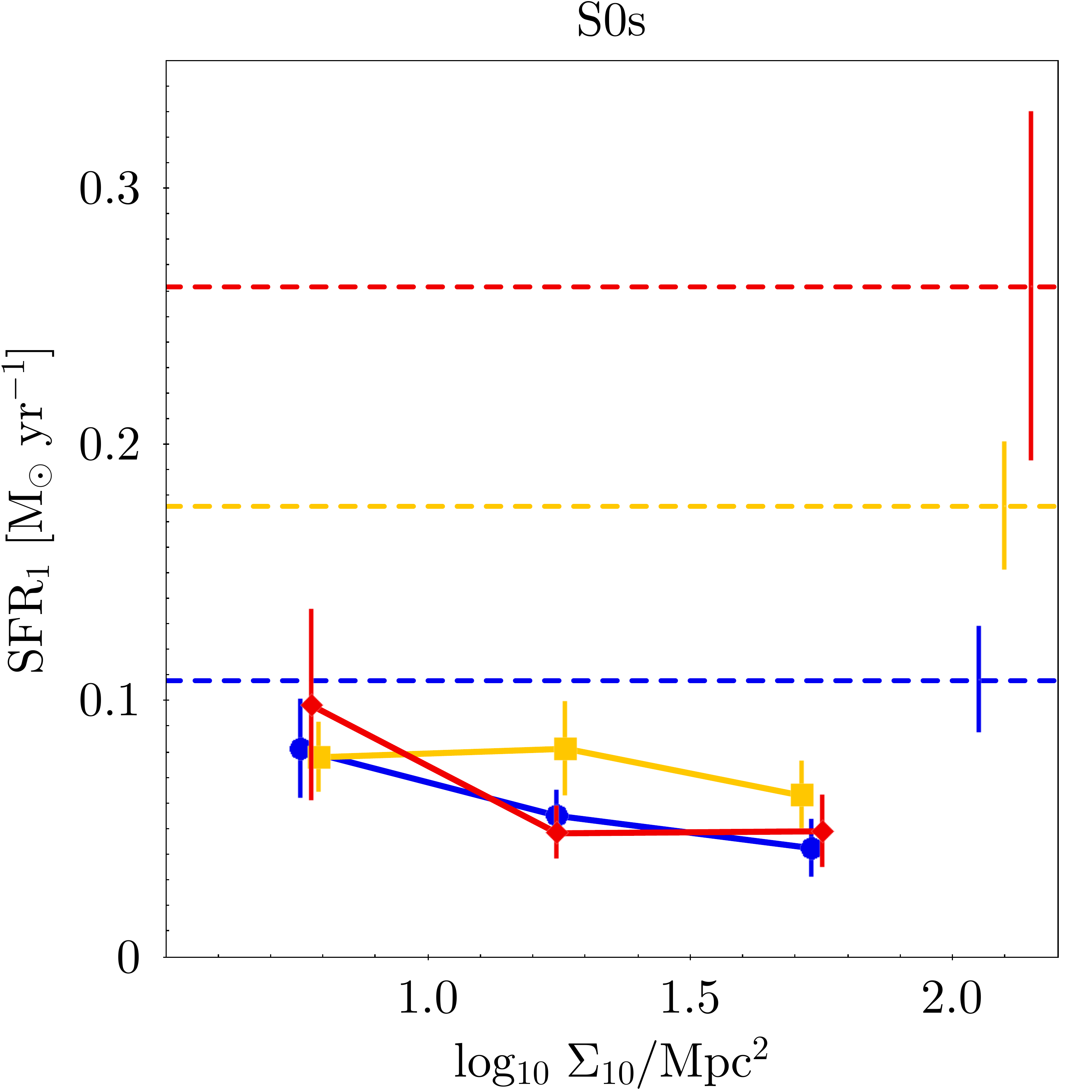}}
  \subfloat{
   \label{fig:fig:MS-LD_Sp}
    \includegraphics[trim={135 94 0 0}, clip,  height=4.4cm]{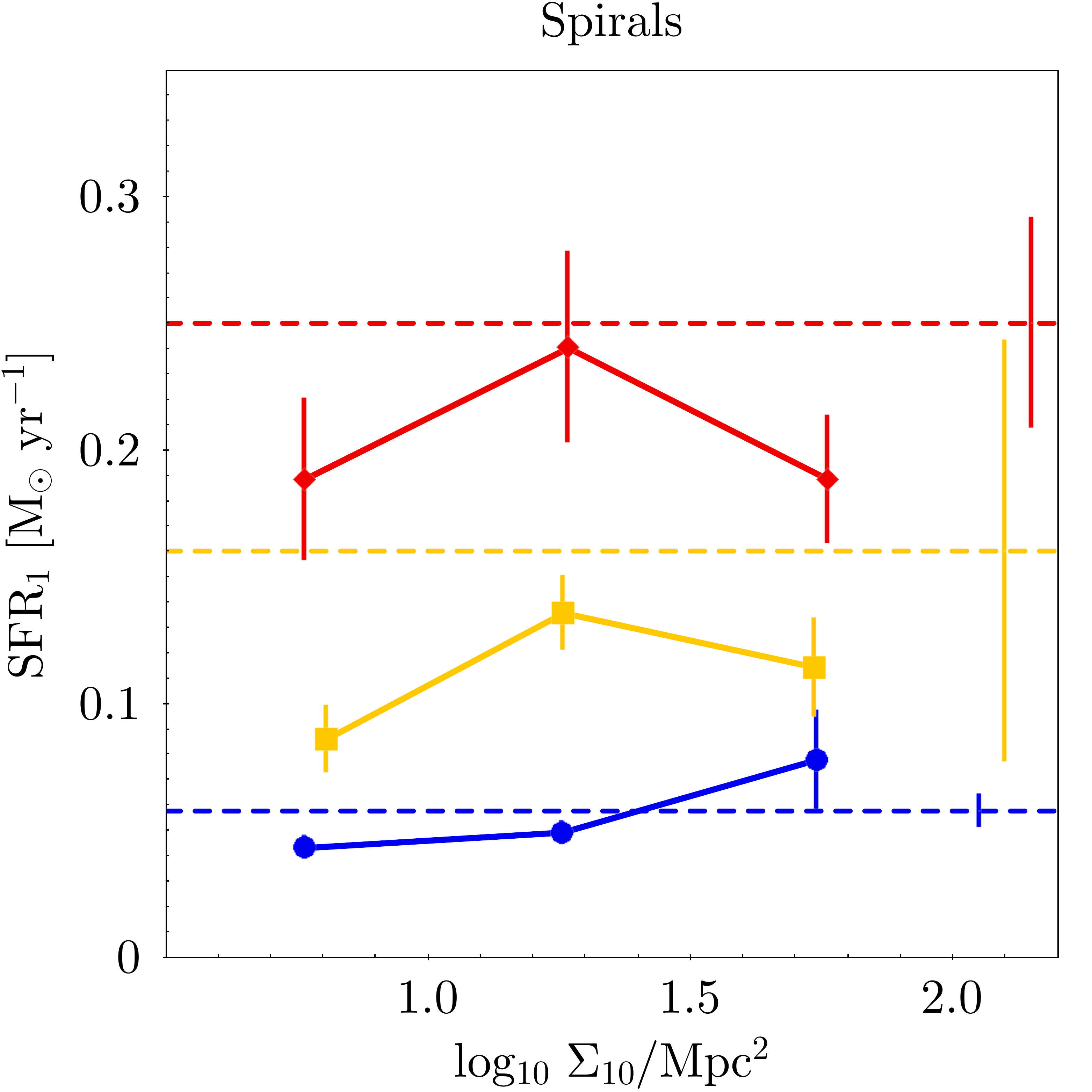}}
    \vspace{-5px}
    
  \subfloat{
   \label{fig:MS2-LD_All}
    \includegraphics[trim={0 0 0 43}, clip,  height=4.7cm]{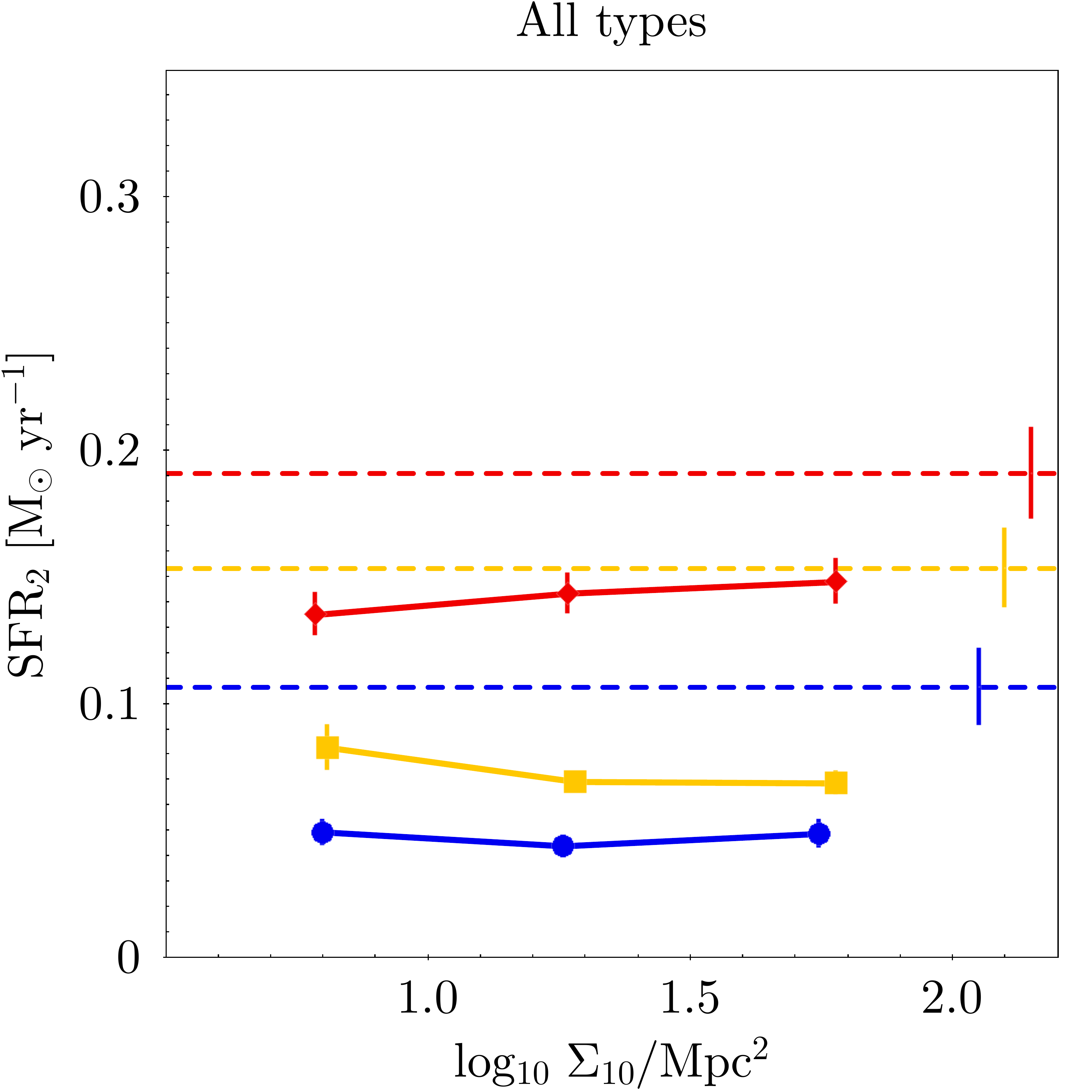}}    
  \subfloat{
   \label{fig:MS2-LD_E}
    \includegraphics[trim={135 0 0 43}, clip,  height=4.7cm]{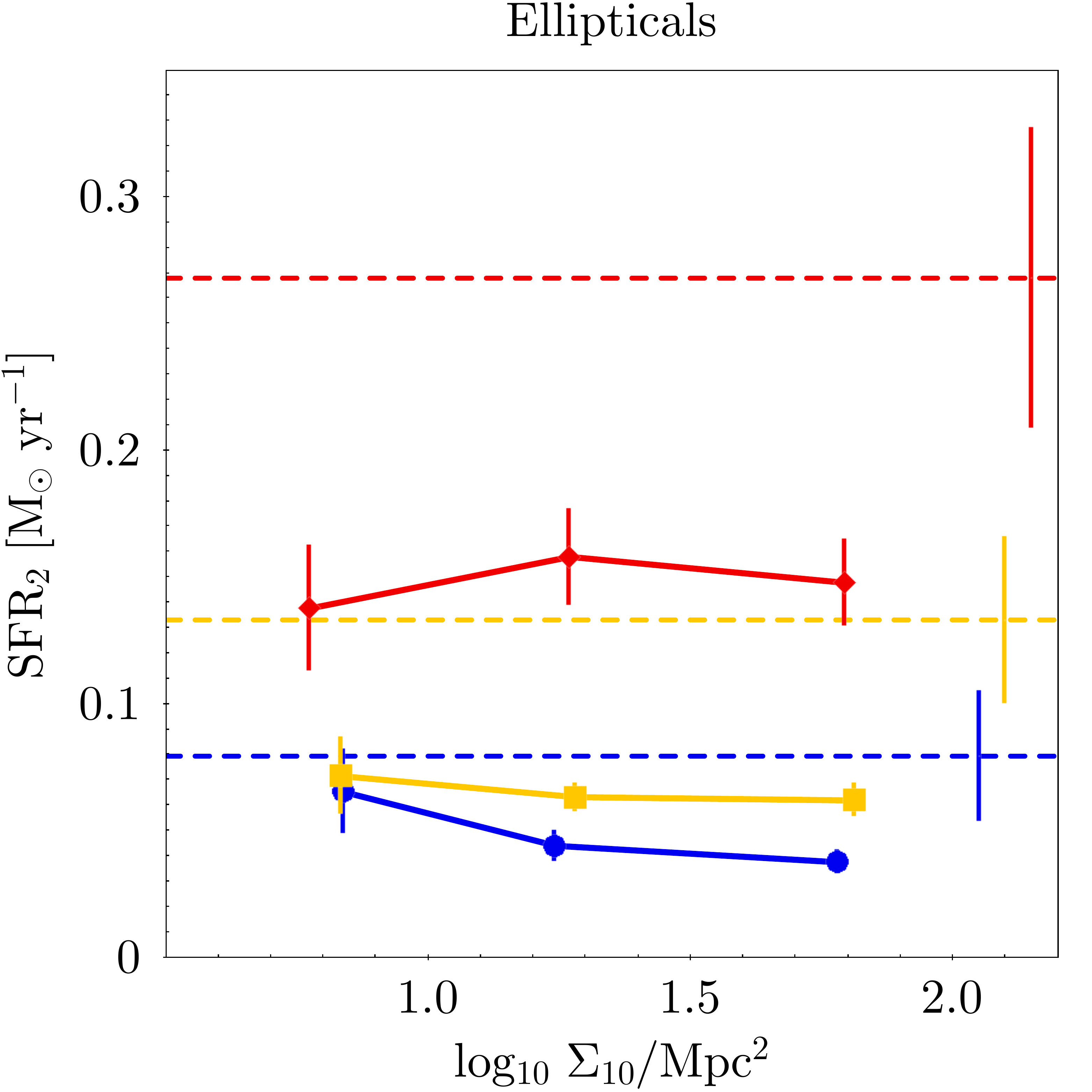}}
  \subfloat{
   \label{fig:MS2-LD_S0}
    \includegraphics[trim={135 0 0 43}, clip,  height=4.65cm]{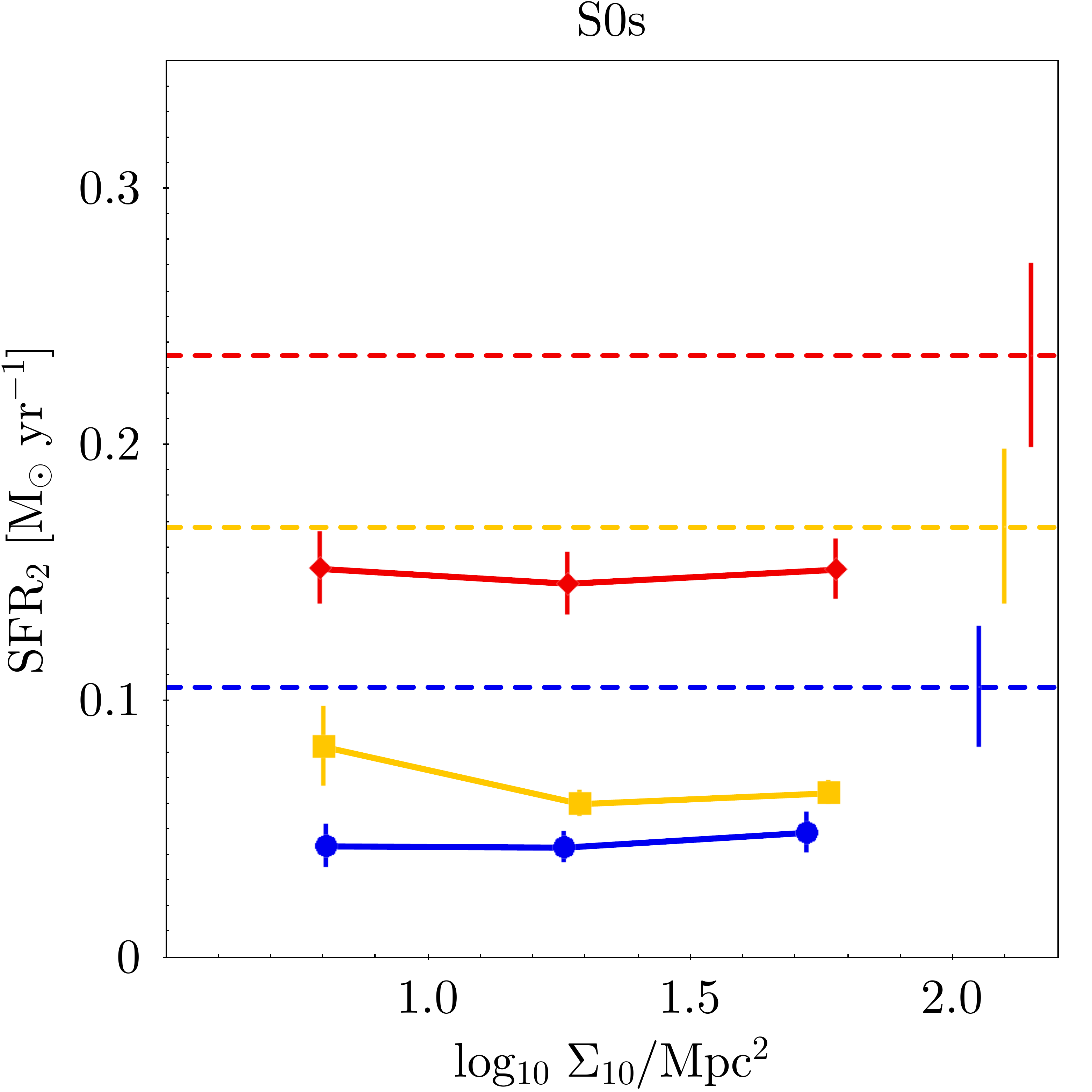}}
  \subfloat{
   \label{fig:MS2-LD_Sp}
    \includegraphics[trim={135 0 0 43}, clip,  height=4.7cm]{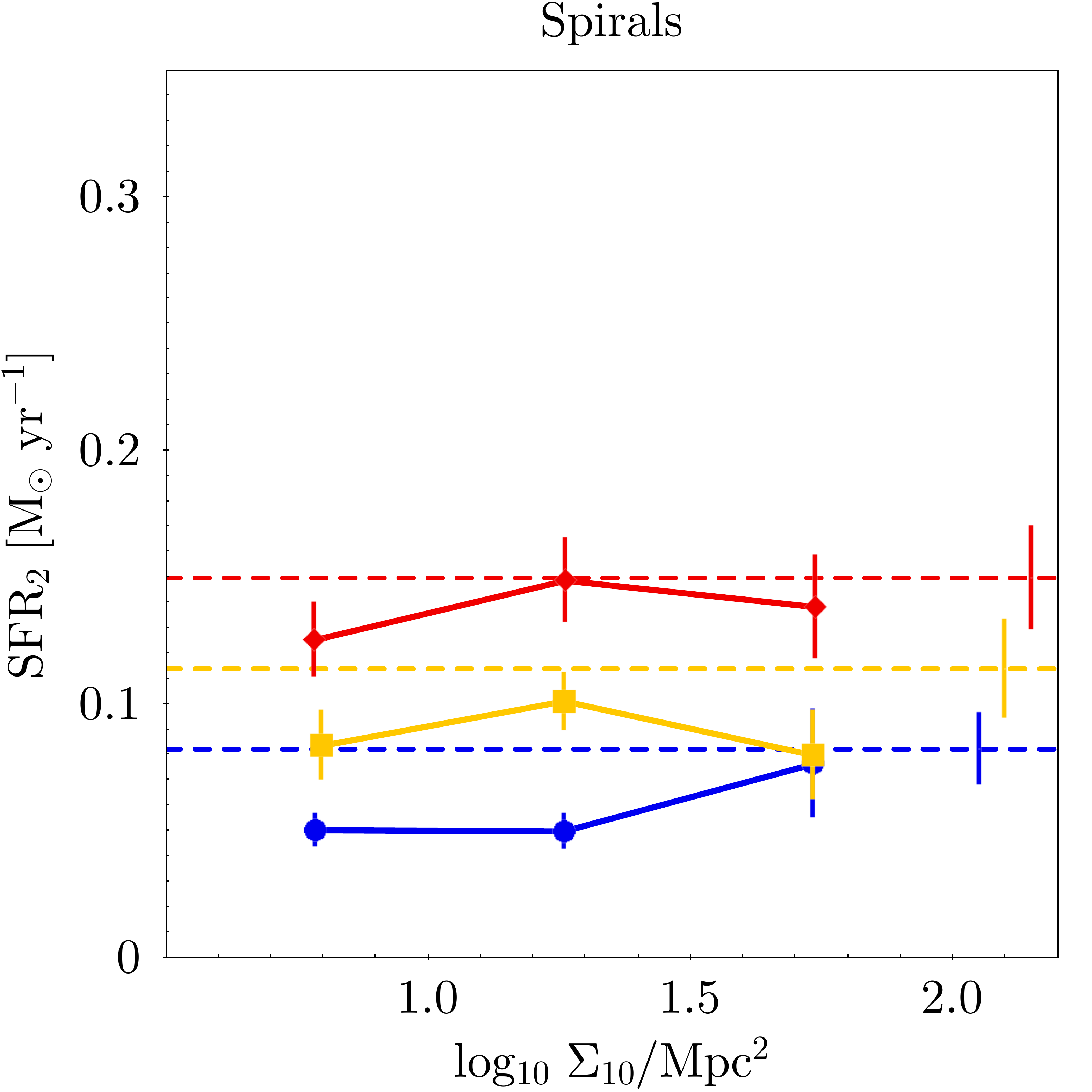}}
    
 \caption{SFR$_1$ ({\it top row}) of star-forming galaxies and SFR$_2$ ({\it bottom row}) for the galaxy sample, versus local density, separated in bins of stellar mass, as indicated in \autoref{fig:Main-sequence}.}
 \label{fig:Main-sequence-LD}
\end{figure*}

On the other hand, when we consider the percentage of actively star-forming galaxies (\autoref{fig:SFing_fraction}) as a function of clustercentric projected distance (left panel) and local density (right panel), we note a decline towards the centre. Between $\sim 2$ and 0.2 cluster virial radii, the star-forming fractions go from $61.0\pm2.6\%$ to $42.7\pm3.0\%$ for SpE, from $84.6\pm3.5\%$ to $70.3\pm7.5\%$ for SpL, from $33.2\pm3.4\%$ to $18.1\pm1.3\%$ for Es, and from $39.3\pm2.5\%$ to $22.4\pm1.2\%$ for S0s. The star-forming fractions of both spiral types exhibit a sudden drop of 9\% between 0.5 and 0.2 virial radii, while for early-types they taper slowly within 0.9 virial radii and display fluctuations that may be due to projection.

For all morphological classes, star-forming fractions in the field are higher than in clusters, at any distance from the centre or LD value. The fractions as a function of LD (\autoref{fig:SFing_fraction}, right panel) are comparable to those obtained with respect to projected distances. When we separate by morphological classes, early-types decrease slightly with increasing LD. The quenching effect of the environment, parameterized by both projected clustercentric distance and LD, is somewhat stronger for spirals. In \autoref{subsec:dissentangle}, we try to disentangle both environment tracers.

\subsection{SFH and galaxy local density}
\label{subsec:SFH_LD}
Controversial results have been found when analysing the relation between SFR and LD \citep[see, e.g.][]{Pintos-Castro2019}. \cite{Tyler2013} studied cluster and field galaxies, and found no difference in the SFR distribution of both sets in relation to the LD while, e.g., \cite{Calvi2018} did. 

The existence of a relation between LD and SFH might provide insights into the mechanisms that affect stellar populations, whether due to the influence of the cluster in general or caused by the higher density of galaxies. Of course, the local density of galaxies is a function of clustercentric distance: the closer to the cluster centre,  the probability of finding a high number of galaxies per unit area is higher \citep[][and also our \autoref{fig:LD-R_R-neigh}]{Fasano2015}. Thus, disentangling the possible effects of galaxy position within the cluster from those of the higher galaxy density is definitely not straightforward.

To explore the possible influences of the LD, we look at the mean SFR$_1$ of star-forming galaxies, and SFR$_2$ of all the galaxy sample, separated in stellar mass bins, as a function of local density, divided in three ranges: $\log_{10} \ \Sigma_{10}/$Mpc$^2$ = $\text{-}0.05 - 1.0$, $1.0 - 1.5$, and $1.5-3.0$. We show the result in \autoref{fig:Main-sequence-LD}, for cluster members (solid lines) and field galaxies (dashed lines). For the latter, the average SFR$_1$ and SFR$_2$ do not change significantly with LD, and the span of LD values is significantly smaller (the mean log$_{10} \ \Sigma_{10}/\rm{Mpc}^2$ = 1.4 for cluster members, while it is 1.0 for field galaxies). For this reason, for field galaxies we take the average SFR in each stellar mass bin. 

When not separated by morphology, the SFR$_1$ (\autoref{fig:Main-sequence-LD}, top row) of field galaxies is always higher, at fixed LD and mass, than for their cluster counterparts. This is also true when ellipticals and S0s are considered separately, but not if only spirals are accounted for (even though the dispersion around the average value can be large). Taking all morphological types together, SFR$_1$ is insensitive to LD for galaxies in the lowest mass bin, while a decreasing trend with increasing LD is observed for intermediate- and high-mass ($\log_{10}\mathcal{M}_*/ \rm M_\odot \gtrsim 10.0$). This effect is strongest for the most massive bin. As for ellipticals, trends of SFR$_1$ with LD are not clear. For the low and high-mass bins, SFR$_1$ seems flat, while it  decreases for intermediate-mass galaxies in the highest LD bin. The lack of a trend at high-mass may be due to the scarcity of such objects at low LD.

Regarding S0s, the result mirrors the one found for projected distance: SFR$_1$ decreases in denser regions and does not depend on stellar mass. These two morphological types, being the most numerous, drive the large-scale trends. Concerning spirals, SFR$_1$ grows with increasing density in low- and intermediate- mass galaxies. In the case of massive ones, the confidence intervals are larger and we only have a few objects. This result, once again, reflects the trends already spotted for late-type galaxies as a function of clustercentric distance.

When the SFR$_2$ is inspected (\autoref{fig:Main-sequence-LD}, bottom row), no correlation with LD is detected: even when galaxies are divided according to their morphologies, the run of SFR$_2$ with LD is consistent with being flat. What is clear, instead, is the difference in SFR$_2$ between cluster and field galaxies: cluster galaxies have systematically lower SFR$_2$, at all masses. This is especially evident for earlier types. For spirals the difference is more subtle, and it almost vanishes for the highest mass bin. 

\subsection{The influence of the nearest galaxy}
\label{subsec:galaxy-neighbours}
\begin{figure}
 \centering    

   \subfloat{ 
    \includegraphics[trim={0 92 0 0}, clip,  height=4.cm]{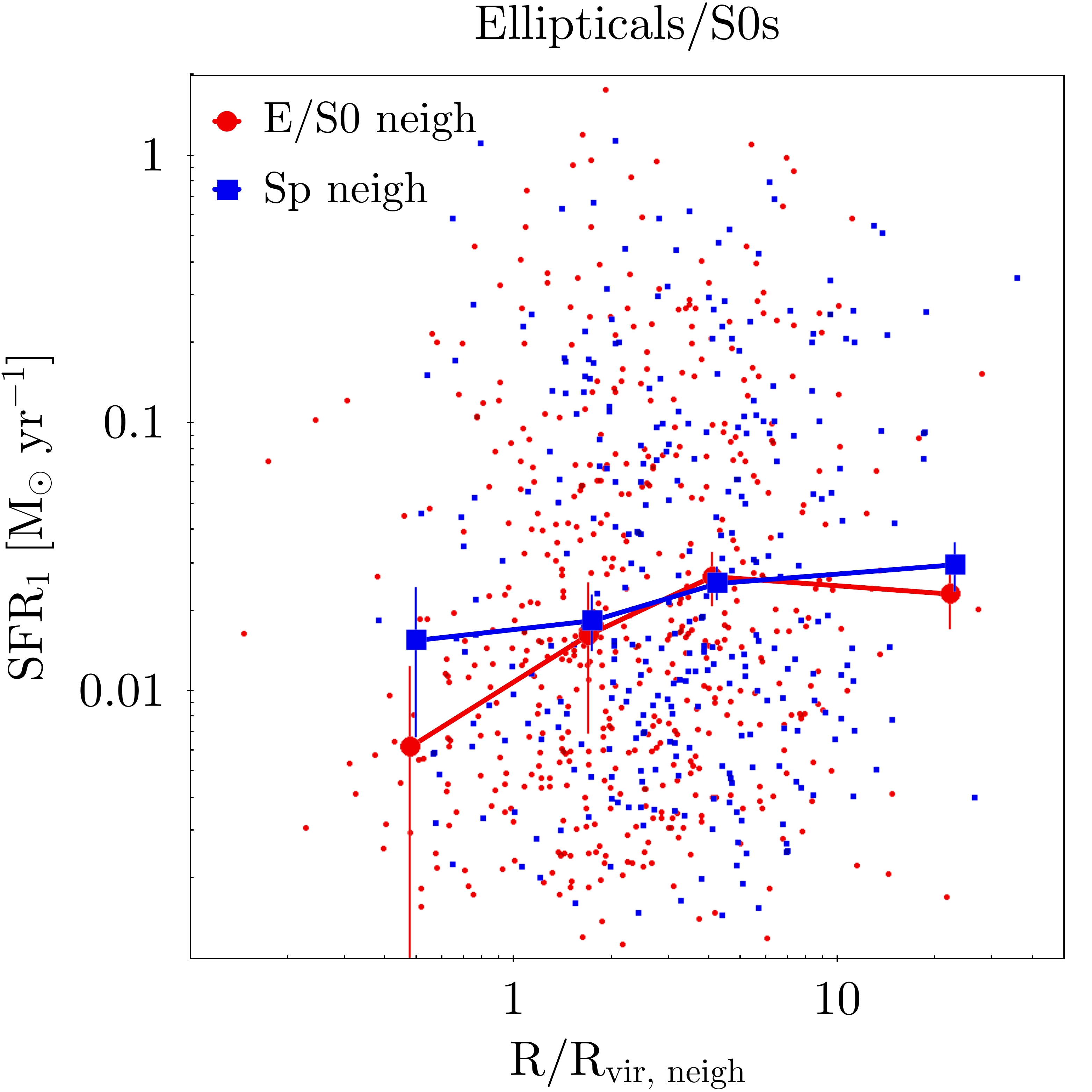}}
  \subfloat{
    \includegraphics[trim={135 92 0 0}, clip,  height=4.cm]{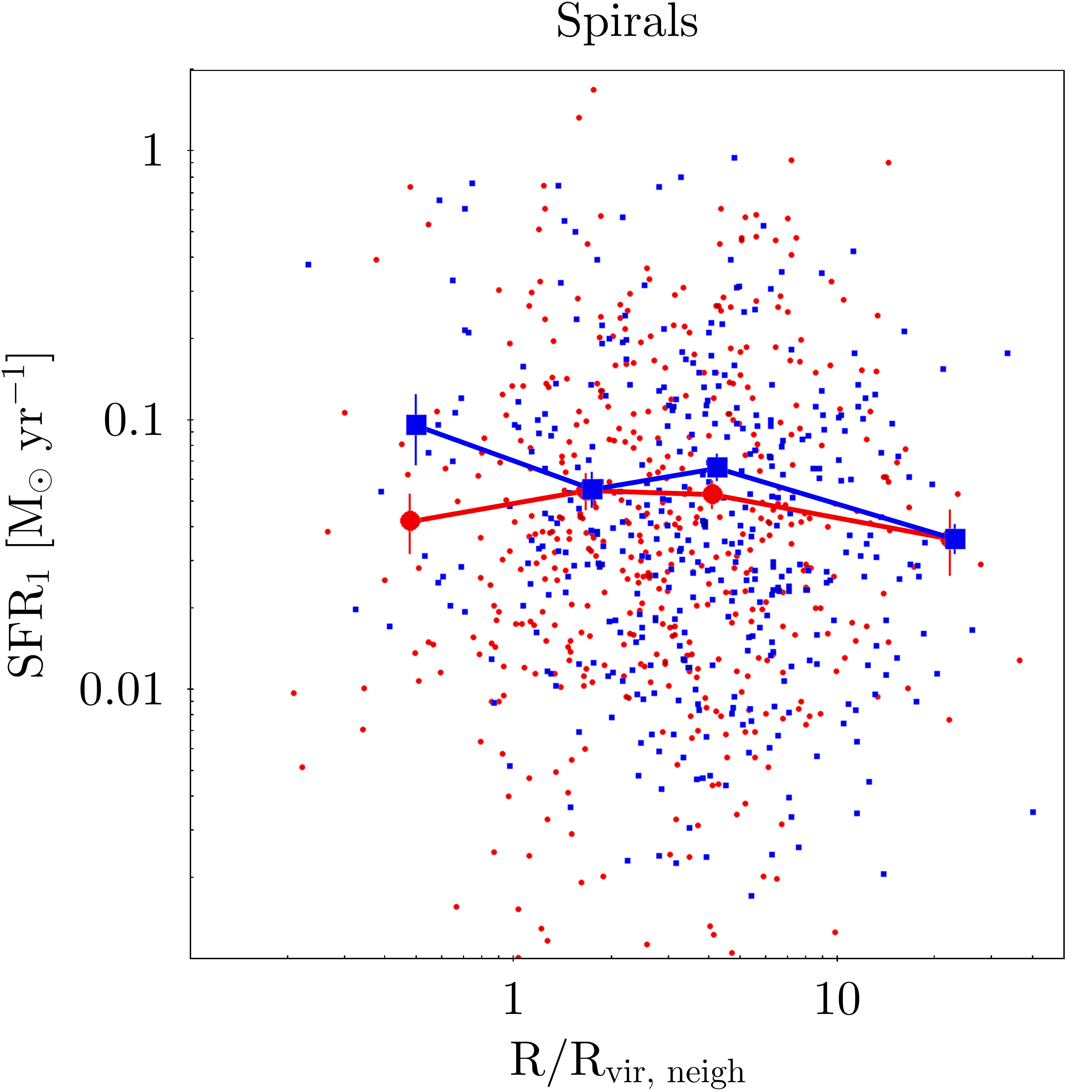}}
    \vspace{-8px}\hspace{2px}
 
    \subfloat{
    \includegraphics[trim={0 0 0 45}, clip,  height=4.15cm]{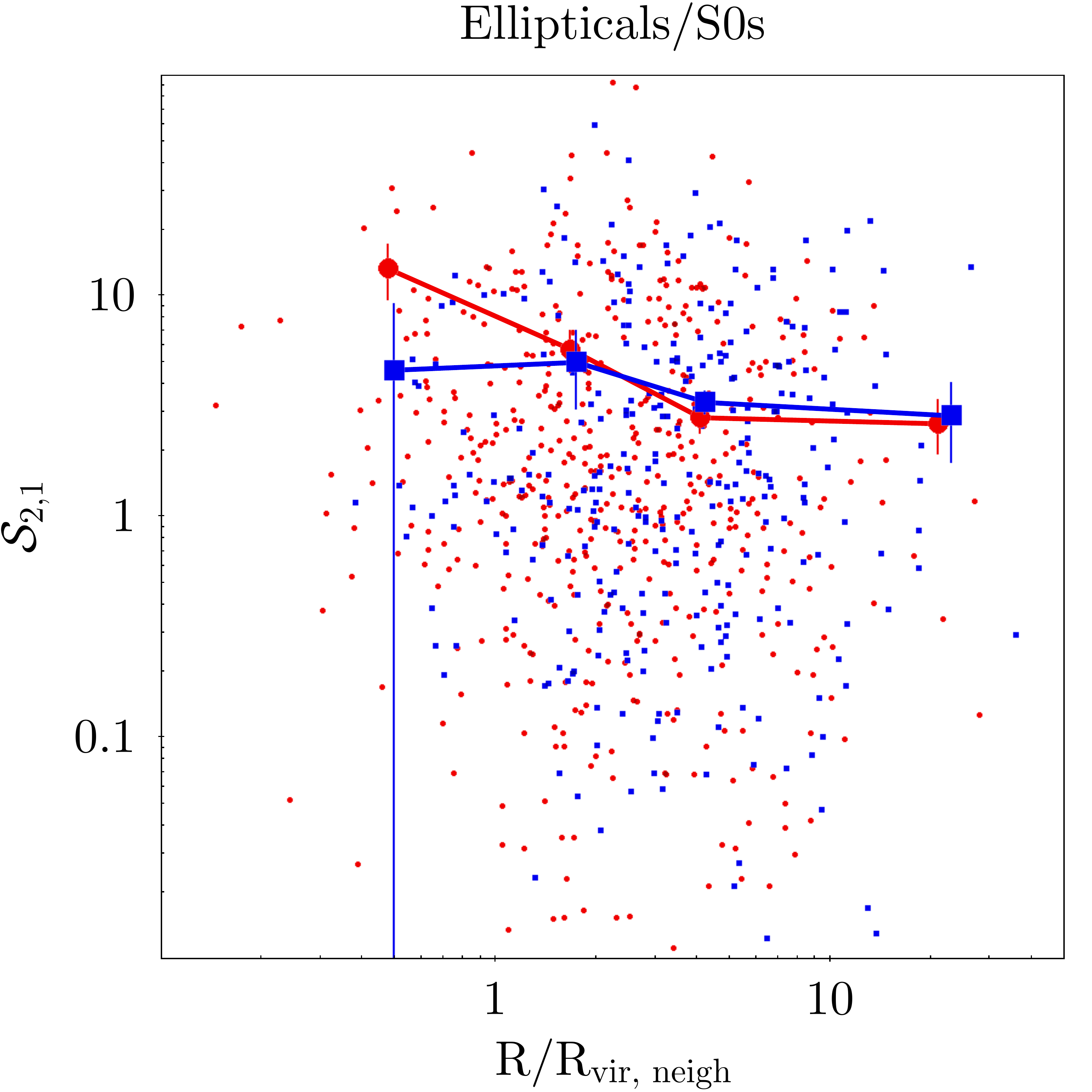}}
  \subfloat{
    \includegraphics[trim={120 0 0 45}, clip,  height=4.15cm]{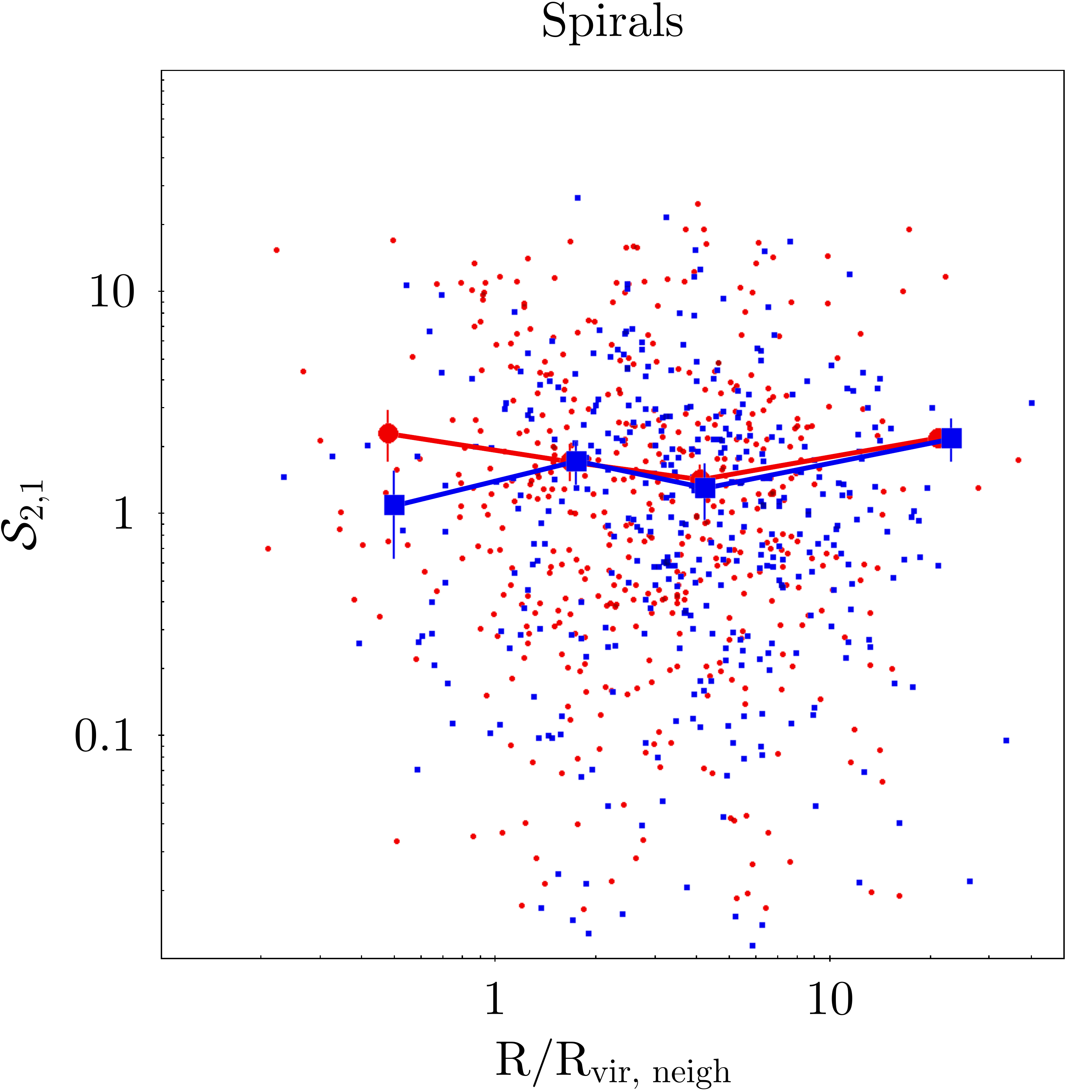}}     
    
 \caption{Weighted means of SFR$_1$ ({\it top row}) and quenching index $S_{2,1}$ ({\it bottom row}), for galaxies with early ({\it left column}) and late ({\it right column}) morphologies, as a function of normalised projected separation. {\it Red/blue} points indicate whether the closest neighbour is an early-/late-type galaxy. {\it Solid red/blue lines:} index weighted means. Error bars are calculated using the bootstrap resampling method.}
 \label{fig:SF_neigh}
\end{figure}

\cite{Park-Hwang2009} have used the distance to the most luminous neighbouring galaxy as a means to study the effects of galaxy-galaxy interactions --hydro-dynamical, gravitational, and of tidal nature--, versus those of the large scale environment, on the properties of galaxies in clusters \citep[but see also][for an analysis of galaxies in lower-density environments]{Park-Choi2009}. Here, we carry out a similar analysis, by testing the occurrence of variations in luminosity, colour, and SFH, as a function of the morphology of the nearest cluster member galaxy and its projected distance, normalised by its virial radius (see \autoref{subsec:environment-tracers}).

It should be highlighted that in the definition of the closest neighbours given in \autoref{subsec:environment-tracers}, we only use galaxies that are spectroscopically confirmed members. Since the spectroscopic completeness is not $100\%$, the results we present in what follows are to be considered as lower limits to the ones we would be able to detect in ideal conditions.

We define four bins for the distance to the nearest neighbour: $R/R_{\textrm{vir}} = 0.1 - 1$, $1 - 2.5$, $2.5-6$, and $>6$. Regarding morphology, we use the broad definition early- or late-type galaxies, for both subject galaxies and neighbours. First, we analyse whether the luminosity of a given galaxy is affected by its closest neighbour, given its distance and morphology. We do not find any remarkable dependence of the weighted mean (or median) luminosity (absolute magnitude in the $V$ and $B$ bands) as a function of either morphology or distance of the closest neighbour.

On the other hand, when rest-frame ($M_B - M_V$) colours are considered, late-type galaxies with an early-type neighbour closer than 1 virial radius are, on average, slightly redder ($\sim 0.085$ mag) than those with a late-type neighbour within the same distance. For closest early-type neighbours located further away than 1 virial radius, the average colour is practically identical. No differences are found for early-type galaxies, regardless of the type and distance of their neighbours.

Repeating the same analysis for the SFH, we do not observe any significant trend or difference in the median SFR$_2$, SFR$_3$ or SFR$_4$, neither for early nor for late-types, as a function of neighbour morphology or distance.  Although the overall distribution of SFR$_1$ values does not show any clear trend, we do find differences in the average weighted values. For early-types (\autoref{fig:SF_neigh}, top left), the average SFR$_1$ decreases with diminishing neighbour distances, both for early- (the number of galaxies is 387, and 884 when weighted) or late- (115 galaxies, 259 weighted) type neighbours; the effect is slightly more pronounced for early-types with an early-type neighbour within 1 virial radius,  albeit the trend is only marginal, as accounted by the large confidence intervals estimated by the bootstrap resampling method.

For late-type galaxies (\autoref{fig:SF_neigh}, top right), the mean SFR$_1$  increases for close pairs, when the target galaxy is within the virial radius of its late-type neighbour, while remaining constant as a function of distance if the neighbour is an early-type. When a late-type has a similar companion within 1 virial radius of the latter (46 galaxies, 78 weighted), the mean SFR$_1$ is twice higher than for an early-type companion (116 objects, 201 weighted). The same behaviour is observed when the median value of the SFR$_1$ is considered. 

If $S_{2,1}$ index is evaluated instead (\autoref{fig:SF_neigh}, bottom row), quenching increases monotonically for early-type galaxies as a function of the distance to the closest neighbour, although this increase is within the confidence interval accounted for by bootstrapping, with identical values independent of the neighbour morphology. However, the average index flattens out when a late-type is found within 1 $R_{\rm vir}$. As for late-types, $S_{2,1}$ is mostly flat with neighbour separation, and independent of companion morphology, with values close to 1. The only difference is observed when a galaxy is located within 1 virial radius of the companion: spirals with early-type neighbours have quenching indices higher than those with another spiral neighbour. 

\begin{figure}
 \centering
    \subfloat{
    \includegraphics[trim={0 0 0 0}, clip,  height=4.4cm]{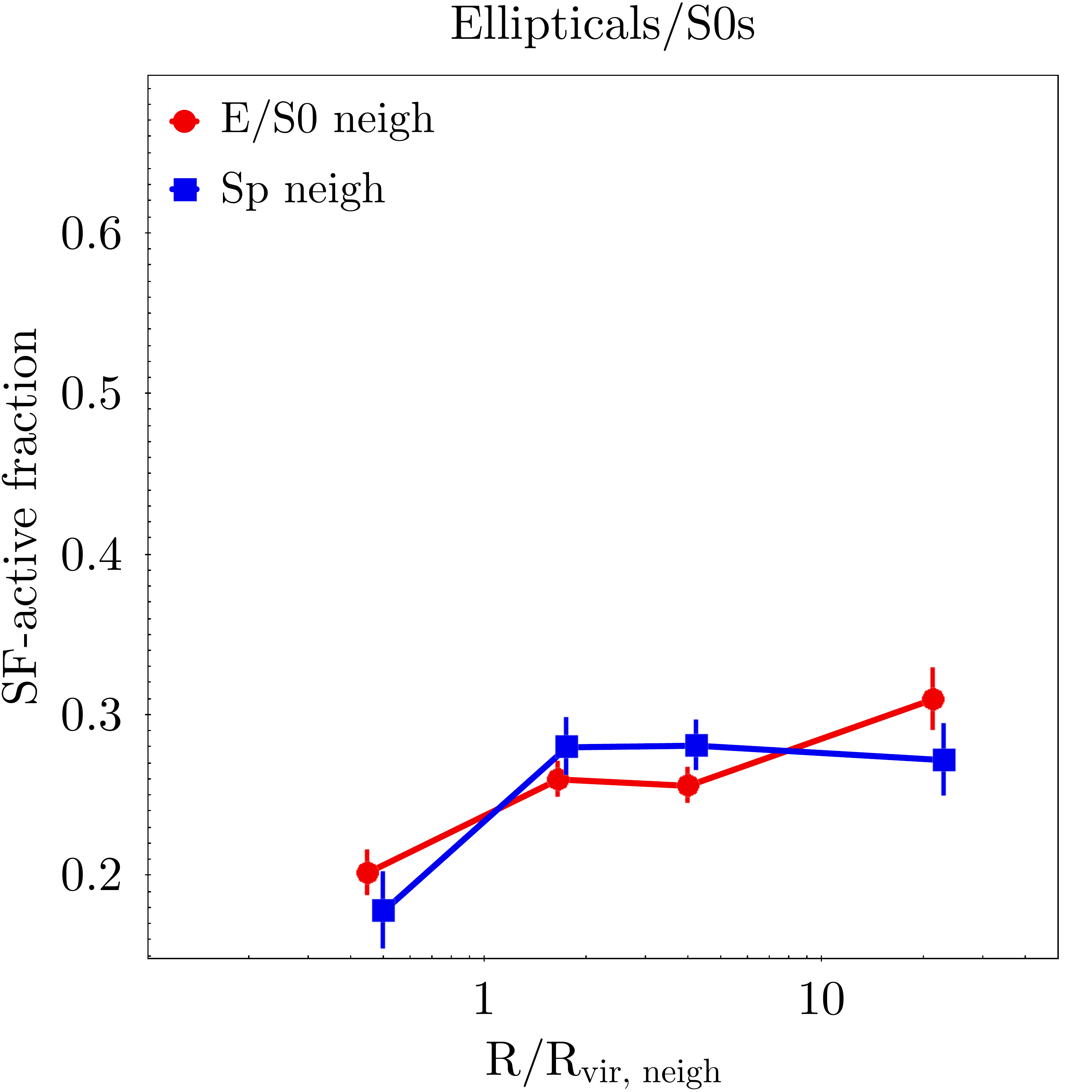}}
 \subfloat{
    \includegraphics[trim={120 0 0 0}, clip,  height=4.4cm]{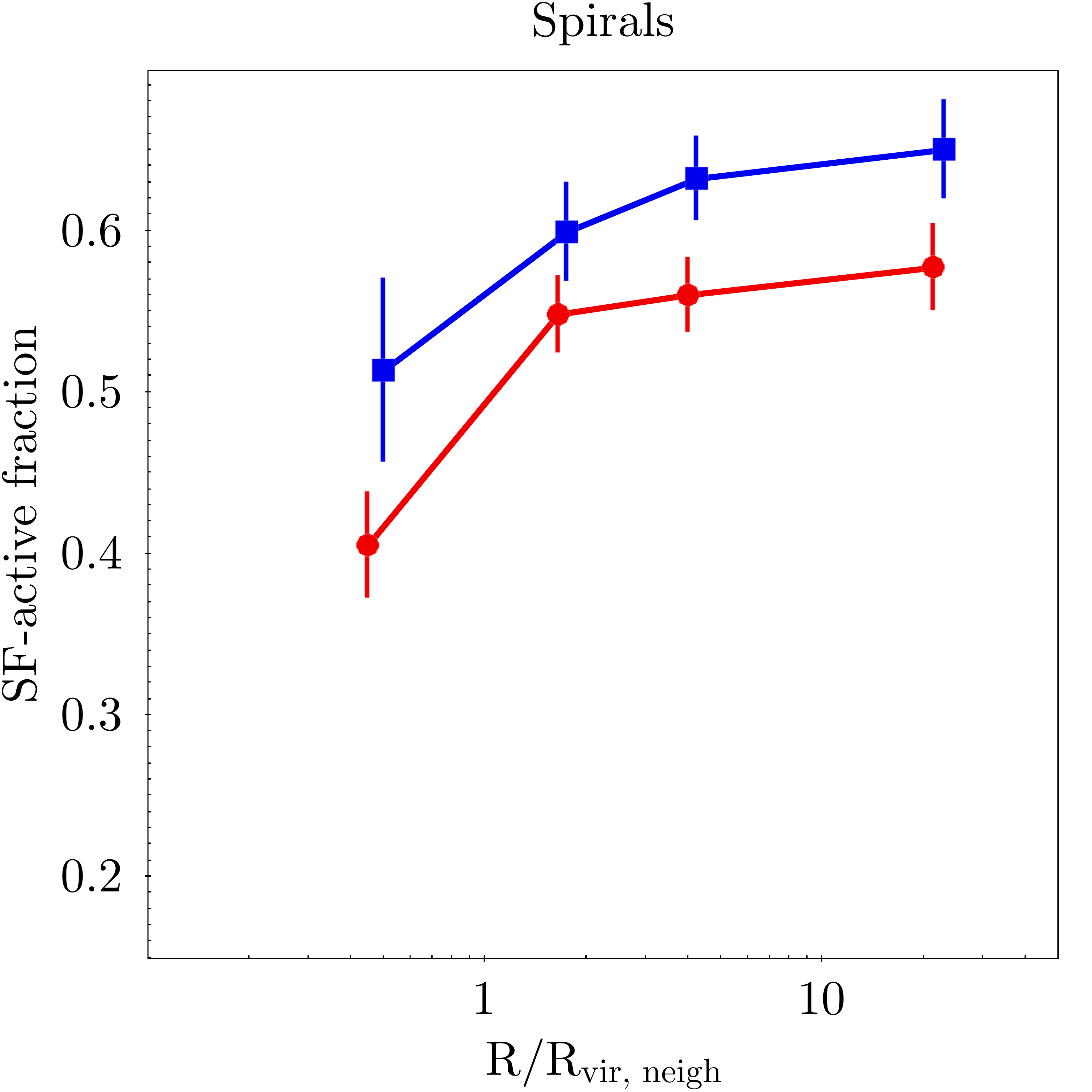}}     

\caption{Fractions of star-forming early- ({\it left}) and late- ({\it right}) type galaxies as a function of distance to the closest neighbour. Colours as in \autoref{fig:SF_neigh}. Error bars represent binomial errors.}
 \label{fig:SF-active-Fract_neigh}
\end{figure}

In \autoref{fig:SF-active-Fract_neigh}, we report the fraction of star-forming galaxies as a function of neighbour distance and type. In the case of early-types (left panel), the fraction of star-forming galaxies drops, from approximately 29\% to 19\%, when a galaxy is closer than 1 neighbour virial radius, regardless of neighbour type. As for late-type galaxies (right panel), the star-forming fraction is higher for a late-late pair than for a late-early pair by a factor of about 10\%, at all neighbour distances. However, regardless of neighbour type, the star-forming fraction is lower when the pair separation is smaller than one virial radius of the companion. Fractions of actively star-forming galaxies decrease with diminishing distance, from 65\% to 51\% for spiral-spiral pairs, and from 58\% to 41\% for spiral-early-type pairs.

This analysis seems to indicate that the distance and morphological type of the closest neighbour have a clear influence on the current star formation rate of a given galaxy. However, we have found no dependence of $S_{4,3}$ and $S_{3,2}$, the oldest quenching indices, on neighbour distance and/or morphology. This would be consistent with the fact that galaxy encounters are very fast events that might leave signatures only on short time scales; some encounters will have a positive effect, and others a negative one, on the SFR; and that close neighbours today would not necessarily be so a few Gyr ago. 

\subsection{Local environment or neighbour proximity?}
\label{subsec:neighbor-LD}
A direct effect of the MD relation is the fact that early-type galaxies are more likely found towards the centres of clusters. Hence, reversing the argument, this gives a higher probability that a galaxy with an early-type closest neighbour is located in the innermost parts of clusters. In an attempt to account for the modulation of environment on the influence of neighbour galaxies, we now analyse the stellar population properties of spirals as done before, but separated into low (log$_{10} \ \Sigma_{10}/\text{Mpc}^2 \leq 1.2$) and high local density (log$_{10} \ \Sigma_{10}/\text{Mpc}^2 > 1.2$) bins. 
\begin{figure}
 \centering
    
   \subfloat{
   \label{fig:SFR1_Sp-neigh_LD1}
    \includegraphics[trim={0 95 0 0}, clip, height=4.cm]{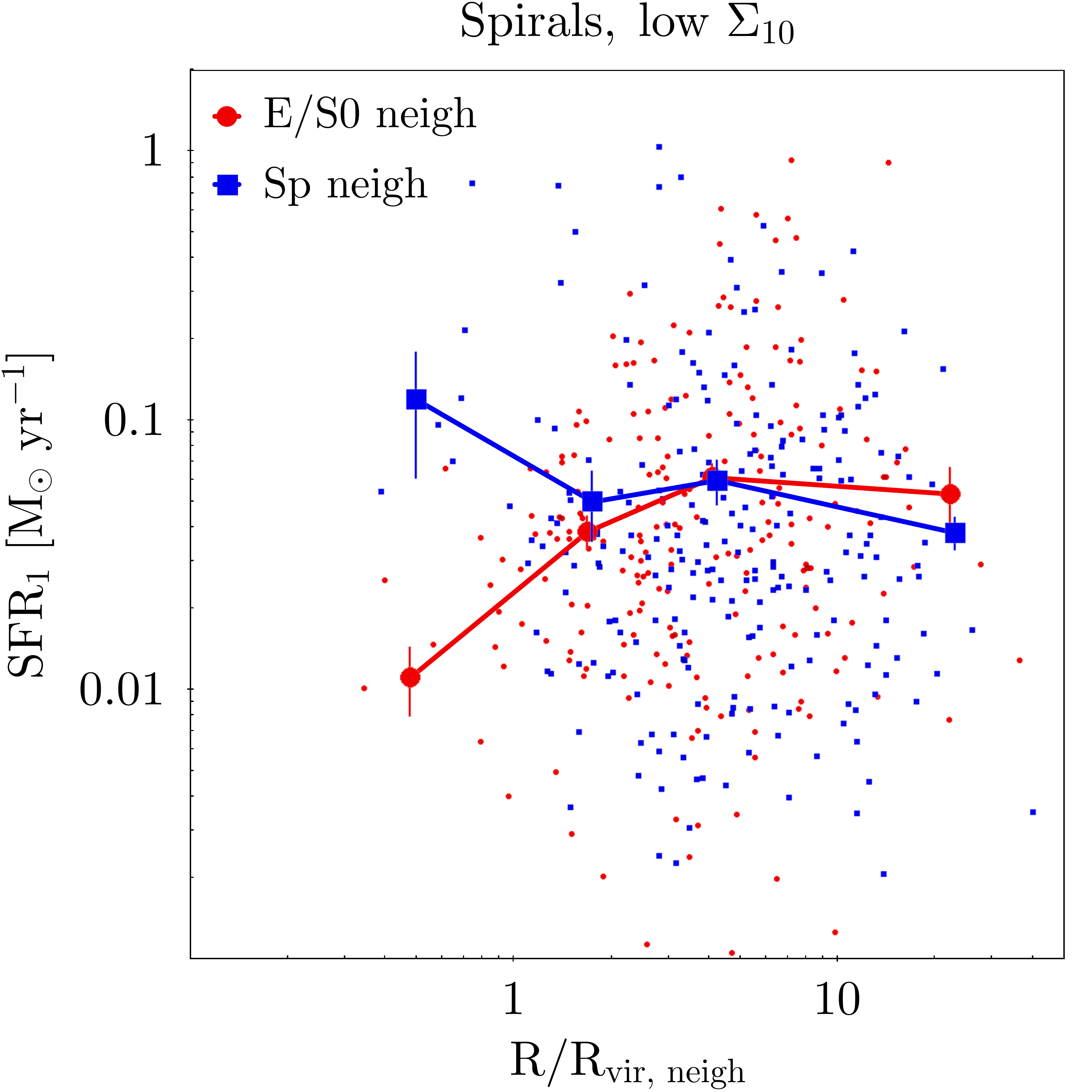}}
   \subfloat{
   \label{fig:SFR1_Sp-neigh_LD2}
    \includegraphics[trim={150 95 20 0}, clip,  height=4.cm]{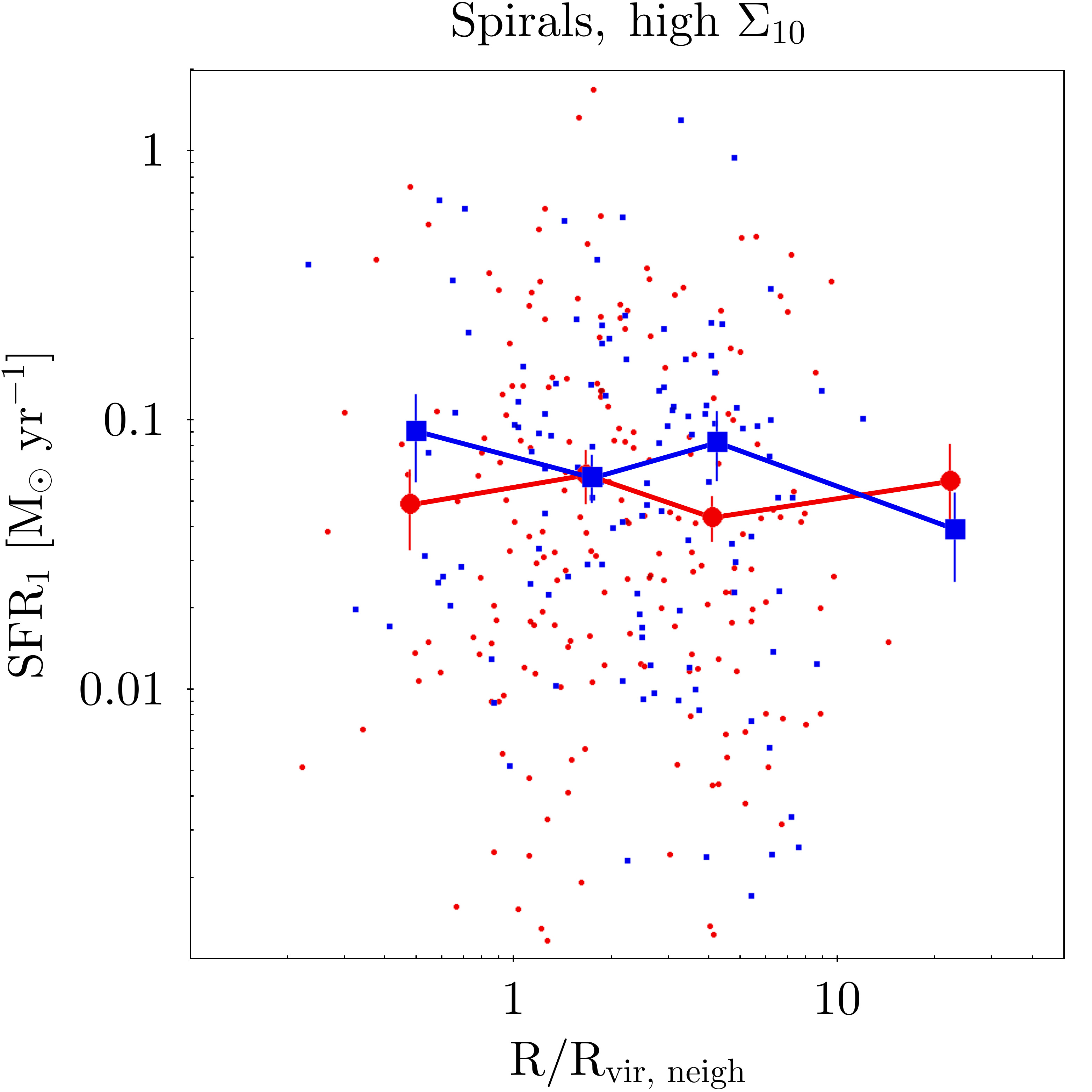}}
    \vspace{-7px} \hspace{5.5px}   

   \subfloat{
   \label{fig:S21_Sp-neigh_LD1}
    \includegraphics[trim={0 0 0 57}, clip, height=4.1cm]{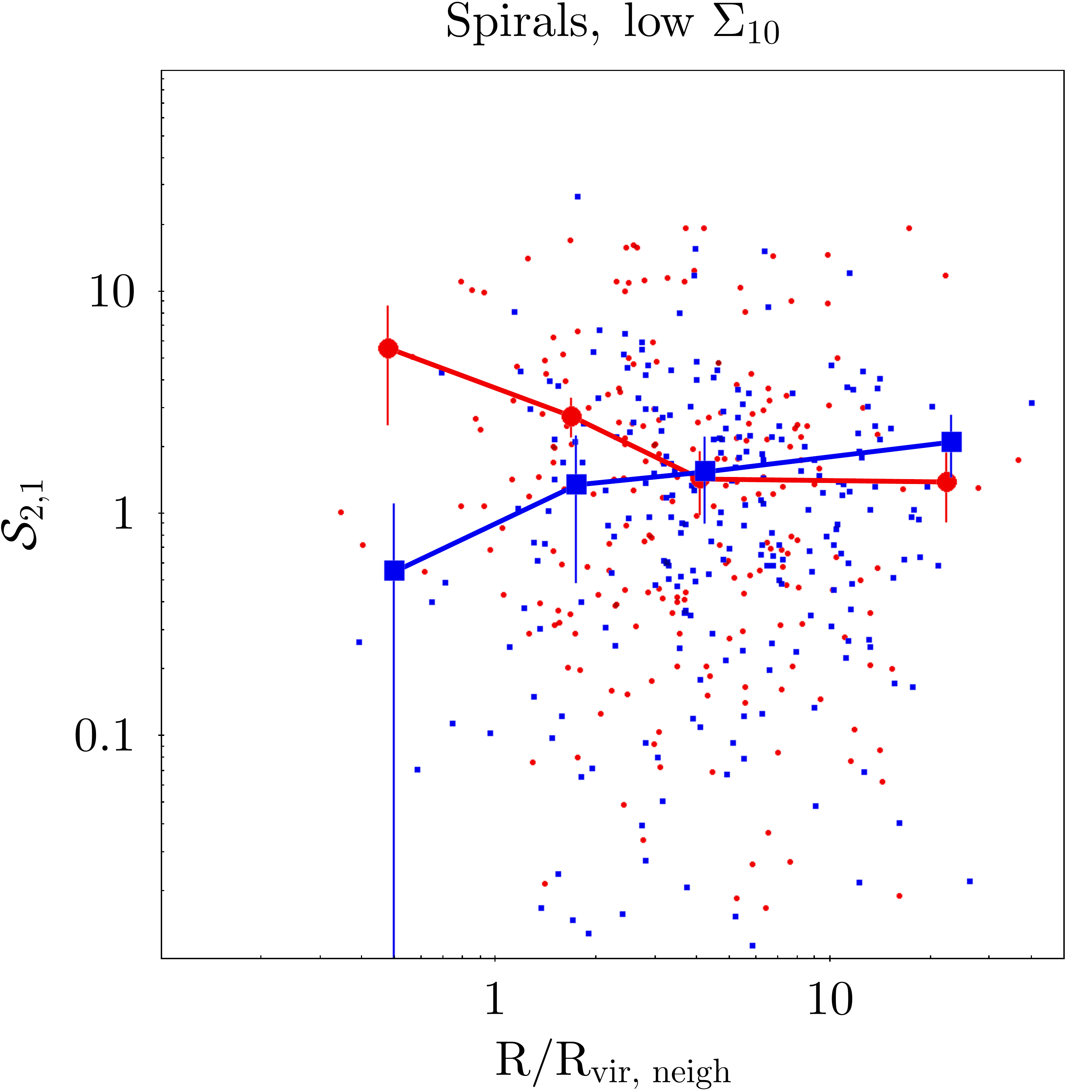}}
   \subfloat{
   \label{fig:S21_Sp-neigh_LD2}
    \includegraphics[trim={130 0 0 57}, clip,  height=4.1cm]{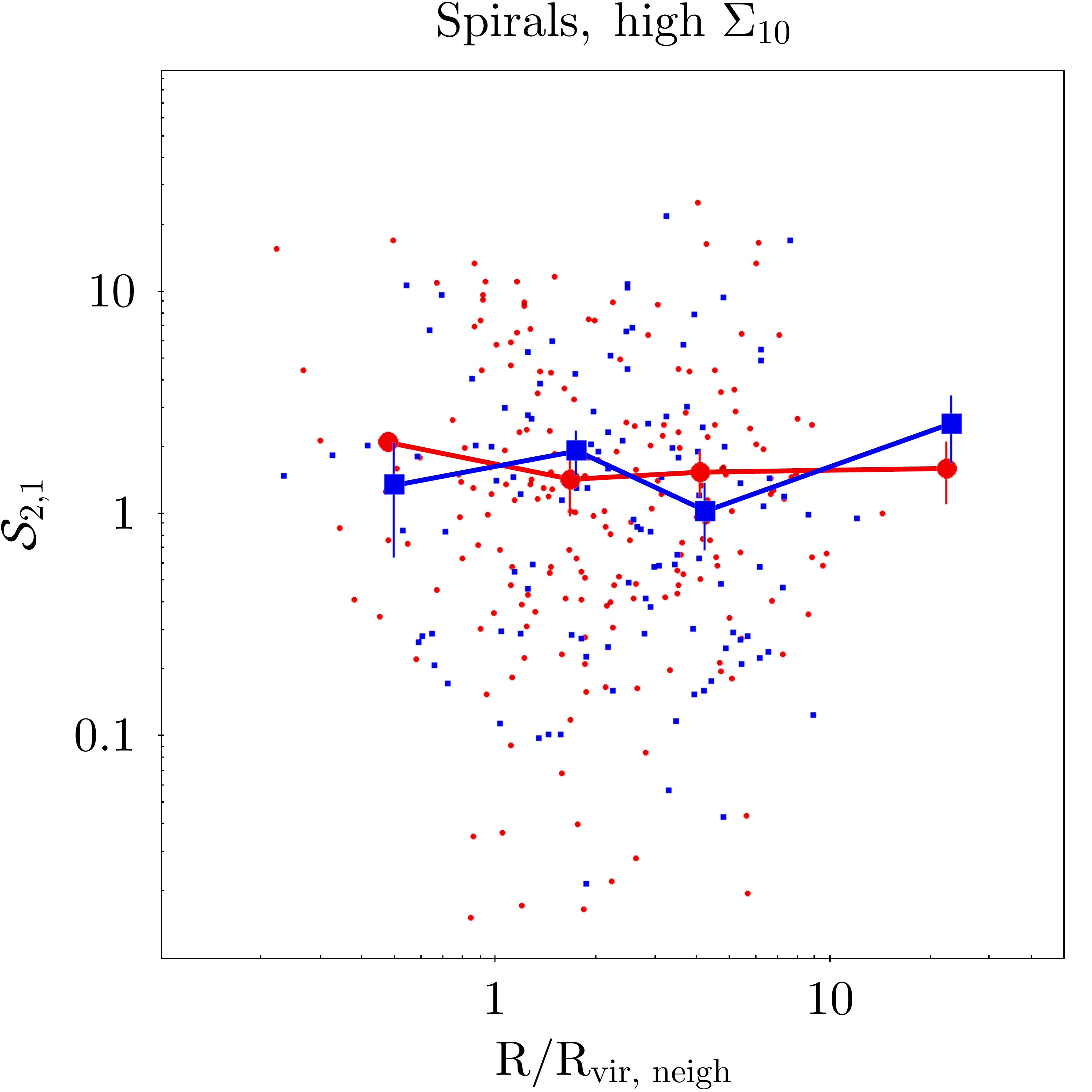}}    

\caption{Weighted means of SFR$_1$ ({\it top row}) and quenching index $S_{2,1}$ ({\it bottom row}), for spiral galaxies in the sample, as a function of normalised projected separation to the closest neighbour, divided into two bins of local density: low LD (log$_{10} \ \Sigma_{10}/\text{Mpc}^2 \leq 1.2$, {\it left column}) and high LD (above this limit, {\it right column}). Colours as in \autoref{fig:SF_neigh}.}
 \label{fig:SF_neigh_Sp_LD}
\end{figure}

In the top row of \autoref{fig:SF_neigh_Sp_LD}, we show the dependence of the average SFR$_1$ on the distance and morphology of the closest neighbour, for spirals located in low (left) and high (right) local density environments. When in low LD, SFR$_1$ of spirals grows by a factor of three when the neighbour galaxy is another spiral (14 galaxies, 20 weighted) with a separation smaller than one virial radius of the companion. If, however, the neighbour is an early-type (24 objects, 31 weighted), the SFR drops by up to a factor of seven (in both cases the comparison is made with the average SFR value when the neighbour is $\sim$ two or more virial radii away). Conversely, in high LD environments (30 galaxies, 55 weighted, with late-type neighbour; and 88 objects, 182 weighted, in the case of an early-type companion), on average, the SFR is flat, regardless of the morphology and distance of the closest neighbour. 

SFR$_2$ and SFR$_3$ (plots not shown) for spirals and early-types, on the other hand, are on average flat, independently of the distance and morphology of the closest neighbour today. SFR$_3$ measures the star formation activity more than 1 Gyr ago; with typical cluster crossing times of $\sim$ 1 Gyr, spirals are likely to be at their first passage within the cluster, and the local density might have been changing throughout this period. Even though SFR$_2$ measures star formation only $\sim$ 0.5 Gyr ago, our time resolution is likely not enough to discern multiple galaxy interactions, each of which may both enhance and quench the SFR during a relatively short time.

Again, quenching indices $S_{4,3}$ and $S_{3,2}$, for both spirals and early-types, do not display any dependence on the closest neighbour (plots not shown). On the other hand, the $S_{2,1}$ index for spirals (\autoref{fig:SF_neigh_Sp_LD}, bottom row) points to a fairly large difference with LD. At low LD, spirals show  an average quenching up to 1 dex higher if the closest galaxy is an early-type. When the companion galaxy is farther away than $4 R_{\rm vir, \ neigh}$, the quenching index is almost flat, with an average value identical to that of spirals at high LD, regardless of neighbour type. The fact that $S_{2,1}$ is on average flat at high LD values might be due to the mix of interactions that are capable of both favouring and suppressing star formation.

\begin{figure}
 \centering
    
    \subfloat{
    \includegraphics[trim={0 0 0 0}, clip,  height=4.4cm]{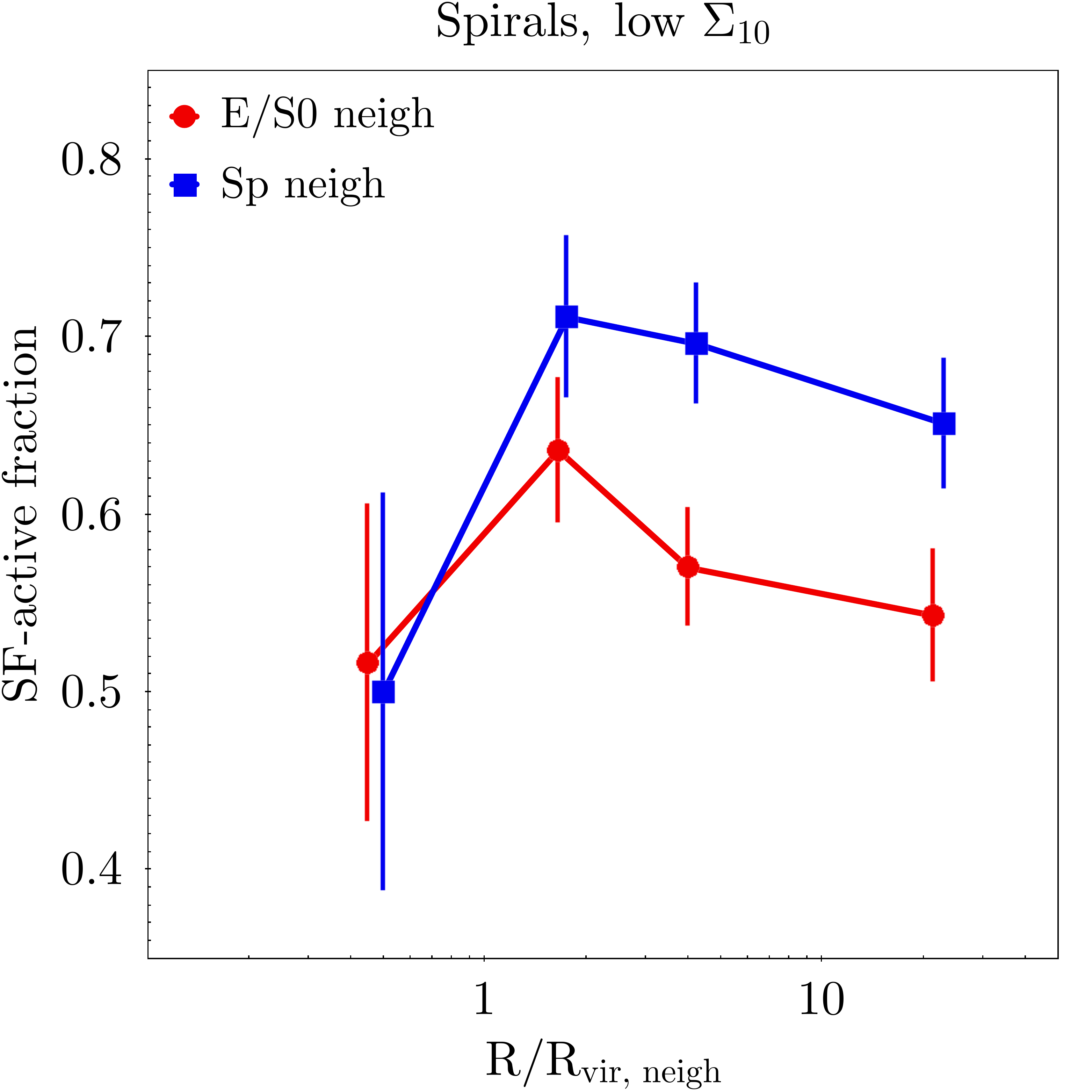}}
 \subfloat{
    \includegraphics[trim={120 0 0 0}, clip,  height=4.4cm]{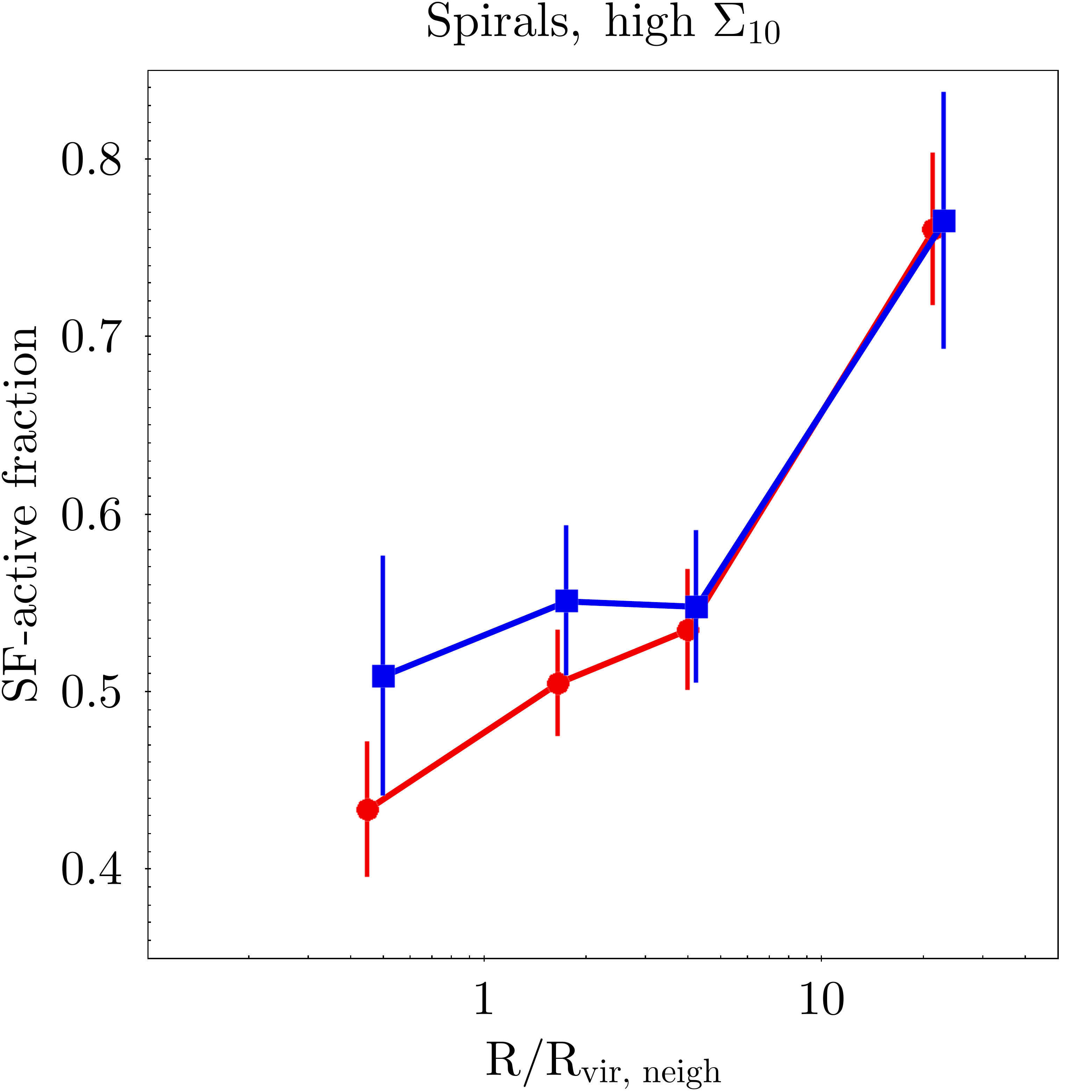}}

\caption{Fractions of star-forming spiral galaxies as a function of distance to the closest neighbour, separated in two bins of local density as in \autoref{fig:SF_neigh_Sp_LD}.
Colours as in \autoref{fig:SF_neigh}. Error bars represent binomial errors.}
 \label{fig:SF-active-Fract_neigh_LD}
\end{figure}

We also analyse the fractions of star-forming spirals in the two bins of high and low LD, as shown in \autoref{fig:SF-active-Fract_neigh_LD}. This fraction is lowest when spirals are located within one virial radius of the closest galaxy, regardless of its morphology and the LD. In the high-LD bin, the SF-active fraction grows with companion distance, independently of its morphology. This suggests that, in this regime, any effect induced by the presence of a close neighbour is convolved with those produced by other mechanisms.

\section{The large-scale environment}
\label{sec:global}
In the previous section, we have analysed the influence of intrinsic and local environmental properties on the stellar content of galaxies. Here, we repeat this analysis, but now taking into account the cluster environment as a whole, parameterized by cluster galaxy velocity dispersion ($\sigma_\textrm{cl}$) and cluster X-ray luminosity (L$_{\rm X}$). Of course, large-scale and local properties of the environment are not clearly separated: massive clusters (large-scale property) will be, in general, richer, both in intra-cluster medium (ICM) and in galaxies (local properties). 

The morphological fraction analysis of WINGS cluster galaxies ($M_V \leq -19.5$) performed by \cite{Poggianti2009} revealed a lack of correlation with $\sigma_{\rm{cl}}$ of the three main morphological classes (E, S0, and spirals), and a weak dependence of the spiral fraction on L$_{\rm X}$, when both $\sigma_{\rm cl}$ and L$_{\rm X}$ are used indistinctly as proxies of cluster mass. Extending that analysis to OmegaWINGS data (see \autoref{sec:Fraction_types_cluster}), we have found low correlations between the fractions of S0s ($r=0.34$) and spirals ($r=-0.26$) with  L$_{\rm X}$, while the correlations with $\sigma_{\rm{cl}}$ are even weaker. For ellipticals, there is no correlation, neither with $\sigma_{\rm{cl}}$ nor with L$_{\rm X}$.

The trends we find of S0 and spiral fractions with (the proxies of) total cluster mass are likely due to the inclusion of galaxies located far from the cluster centre: these have not been yet affected by the cluster environment, or are still undergoing morphological changes. What we can conclude from this simple analysis is that, when we characterise clusters by their total mass, differences can be found in the morphological fraction: this can be either due to the existence of a large-scale effect or to the combination of single local mechanisms. The next step will be to analyse if and which aspects of the stellar populations are affected. Just like for the morphology, we know that a dense environment strongly affects the star formation process; in what follows, we try to ``globally'' quantify how.

\subsection{The star-forming fraction}
\label{subsec:SFfrac_cluster-mass}
We now investigate the influence of large-scale cluster environment on the quenching of star formation. As we already know, the fractions of actively star-forming galaxies are higher in the field than in clusters (\autoref{tab:SF_fraction}), for all morphological types. Strikingly, more than half of ellipticals and S0s in the field are star-forming, versus 25\% or less in clusters. While 80\% or more of the SpLs are active in clusters and in the field, the star-forming fraction of SpEs drops from 75\% in the field to slightly more than 50\% in clusters.

\begin{table}
\centering
\caption{Fraction of star-forming galaxies by morphology, in the cluster and field environments. Uncertainties correspond to binomial errors.}
\label{tab:SF_fraction}
\begin{tabular}{ c c c } 
\hline
Type & Clusters & Field \\ \hline \hline 
E   & $21.3 \pm 0.8 \%$ & $54.1 \pm 3.7 \%$ \\ 
S0  & $26.3 \pm 0.7 \%$ & $58.6 \pm 2.9 \%$ \\ 
SpE & $53.9 \pm 1.1 \%$ & $75.5 \pm 2.2 \%$ \\ 
SpL & $79.4 \pm 1.9 \%$ & $88.1 \pm 2.2 \%$ \\ \hline
\end{tabular}
\end{table}

\begin{figure}
 \centering    
  \subfloat{
   \label{fig:Quench_frac_Sigma-cl}
    \includegraphics[height=4.6cm]{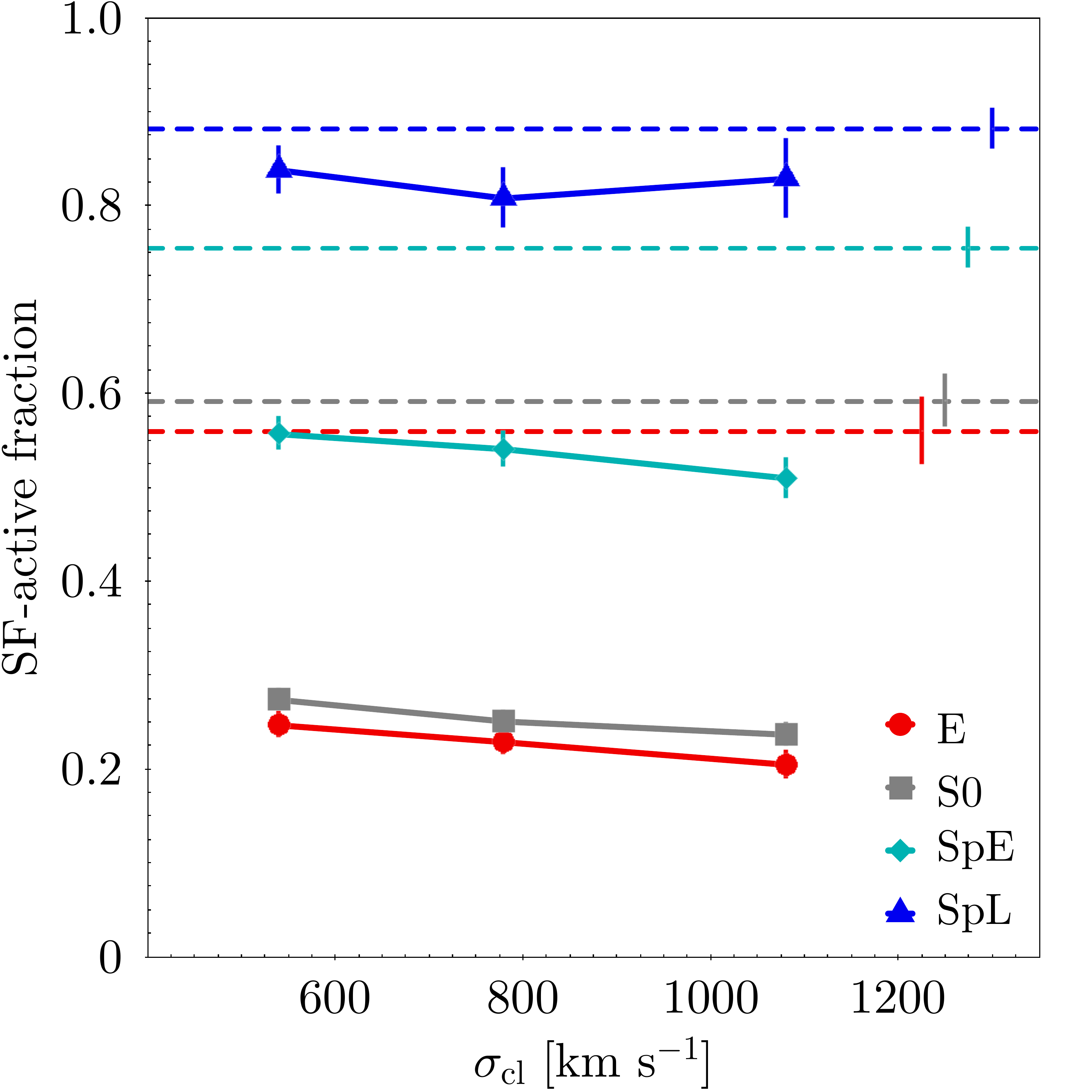}}
  \subfloat{
   \label{fig:Quench_frac_Lx}
    \includegraphics[trim={120 0 0 0}, clip,  height=4.6cm]{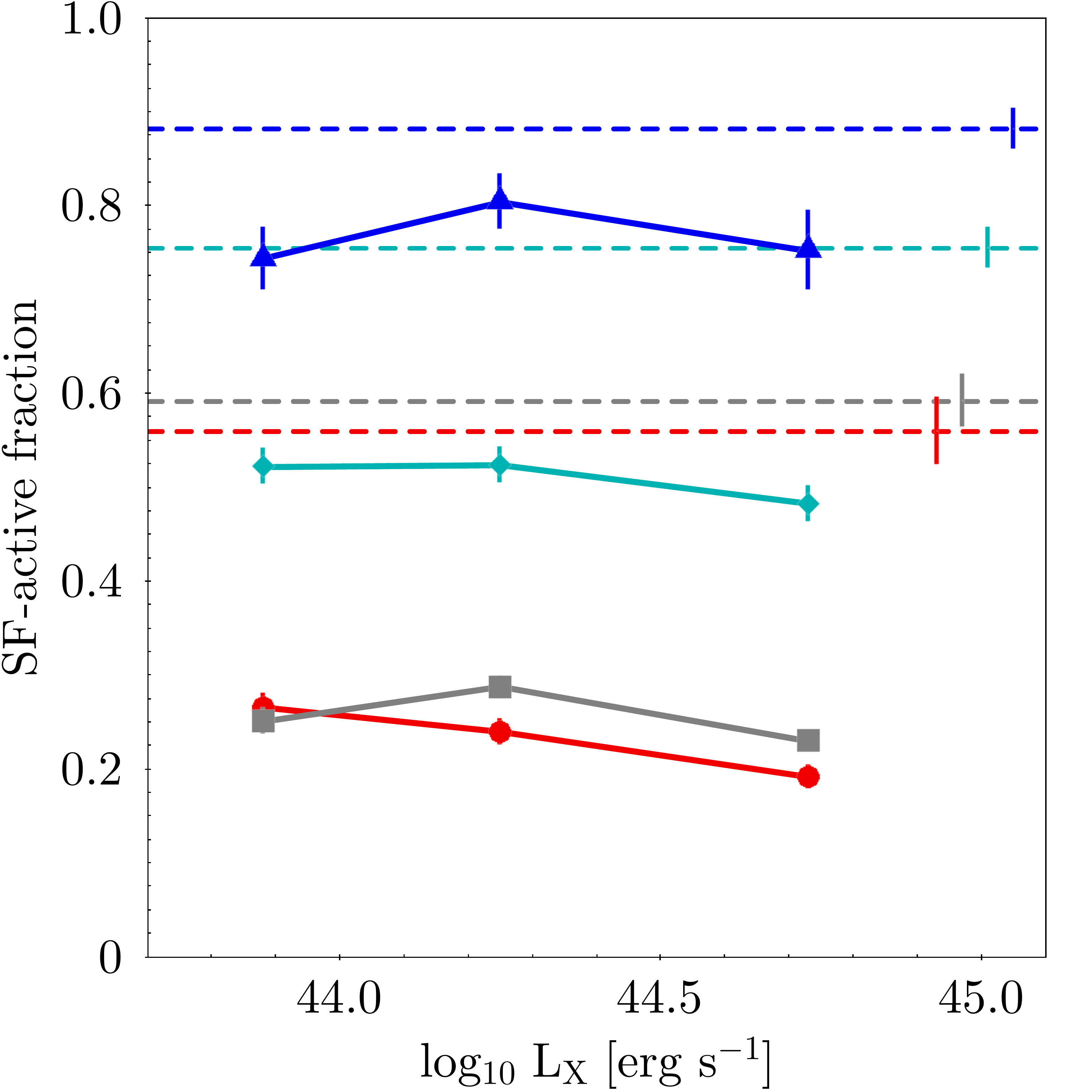}}
    
 \caption{Fractions of star-forming galaxies as a function of cluster galaxy velocity dispersion ({\it left}) and X-ray luminosity ({\it right}), by morphological type. The {\it horizontal dashed lines} show fractions of star-forming galaxies in the field, for the same morphological types. Error bars represent binomial errors.}
 \label{fig:SF_fraction_Cl-mass}
\end{figure}

In \autoref{fig:SF_fraction_Cl-mass}, we show the fractions of star-forming galaxies, by morphological type, as a function of $\sigma_{\rm cl}$ and L$_{\rm X}$, and compare them to the star-forming fractions of the same morphological classes in the field (dashed lines). Only mild trends are observed in clusters, with hints of decreasing fractions with $\sigma_{\rm cl}$ for Es and S0s, and for Es with L$_{\rm X}$.

These results confirm and extend what was previously reported by both \cite{Poggianti2009} and \cite{Fritz2014}, who found that the fraction of emission-line galaxies does not depend on the general properties of the cluster. However, both works only analysed galaxies located within half cluster virial radius. Here, we do not simply rely on the presence/absence of emission lines, but we use a threshold in the H$\alpha$ equivalent width to correct for possible contamination from non-star-forming ionisation processes. 

\subsection{SFH and cluster mass}
\label{subsec:SFH_cluster-mass}
In the following, we investigate whether the large-scale environment affects the ability of a galaxy to form stars not only in the present, but also in earlier cosmic epochs. Once again, we limit the analysis to spirals because these galaxies have most likely spent the shortest amount of time in the cluster, and since star formation dominates in this morphological class. We first calculate the average SFH of SpE and SpL galaxies, separated in three ranges of cluster $\sigma_{\textrm{cl}}$ and L${\rm _X}$ (plots not shown), and find no significant differences as a function of these cluster mass proxies. Even when we divide the spiral sample into low and high mass bins (threshold at $\log_{10} \mathcal{M}_*/ M_\odot = 10.2$), there is no evidence that more massive clusters influence the SFH more effectively.

\begin{figure}
 \centering
 
  \subfloat{
   \label{fig:SFR32_Sp_Cl-sigma}
    \includegraphics[trim={0 95 0 0}, clip, height=4.cm]{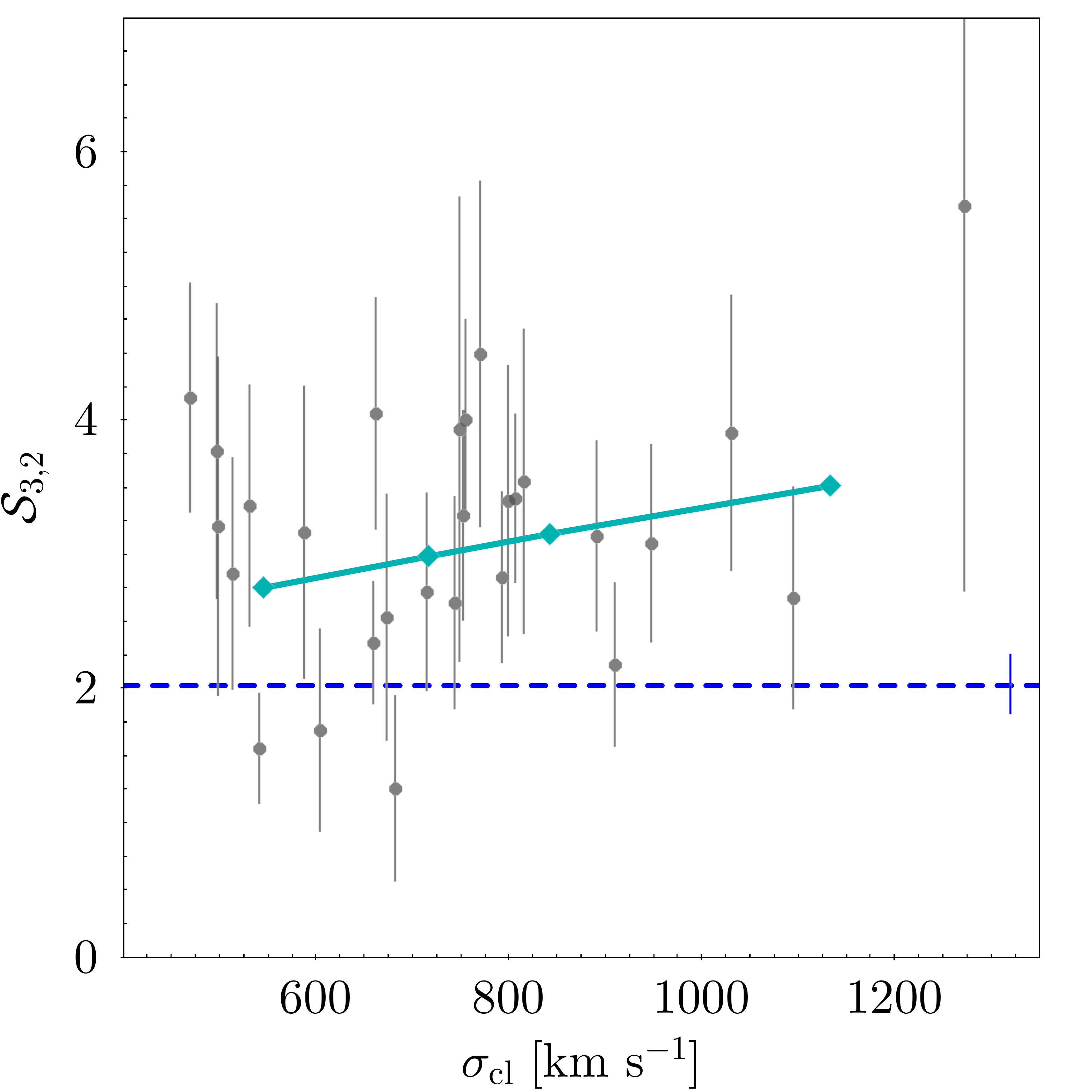}}
  \subfloat{
   \label{fig:SFR32_Sp_Cl-Lx}
     \includegraphics[trim={100 95 0 0}, clip, height=4.cm]{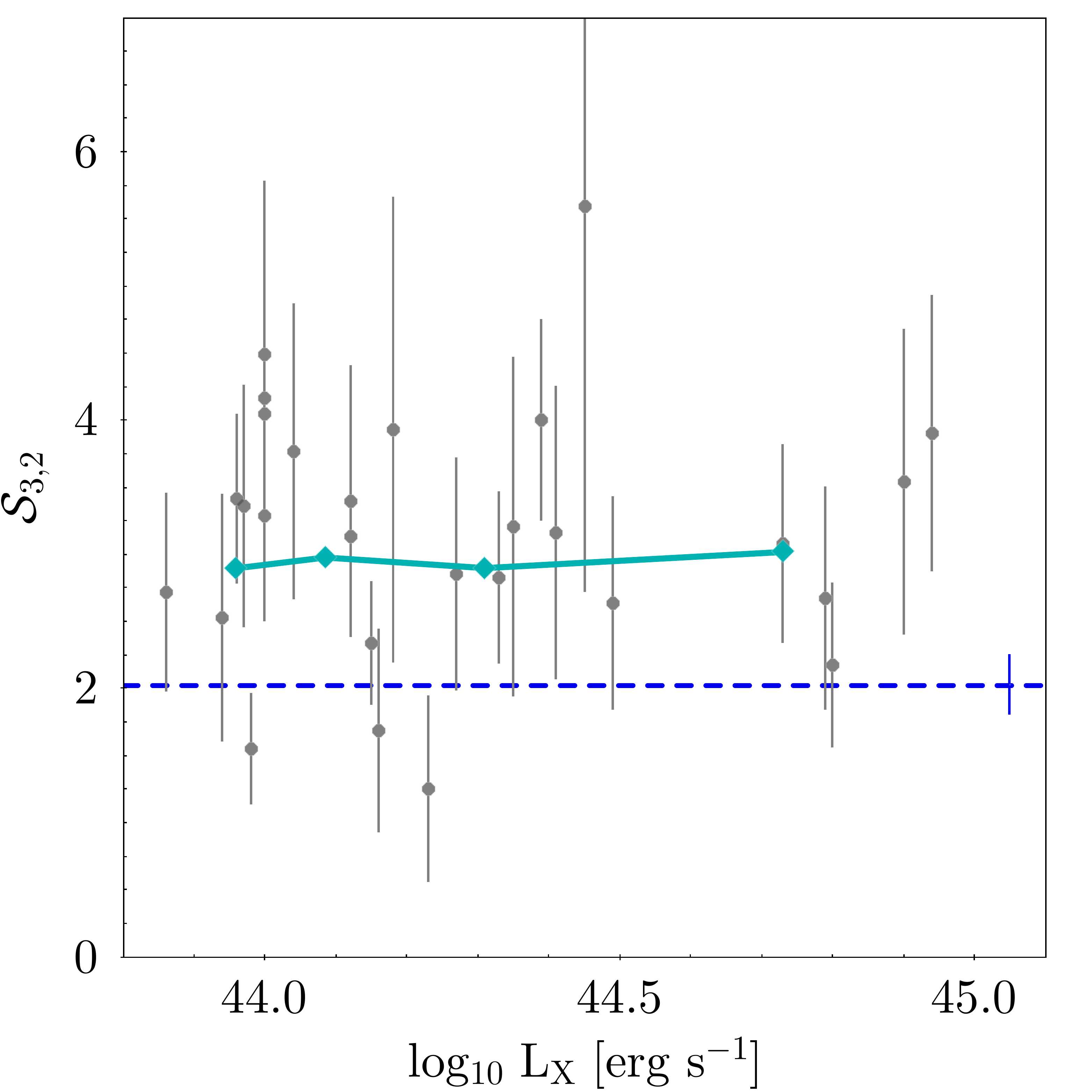}}
    \vspace{-5px}
            
  \subfloat{
   \label{fig:SFR21_Sp_Cl-sigma}
    \includegraphics[height=4.43cm]{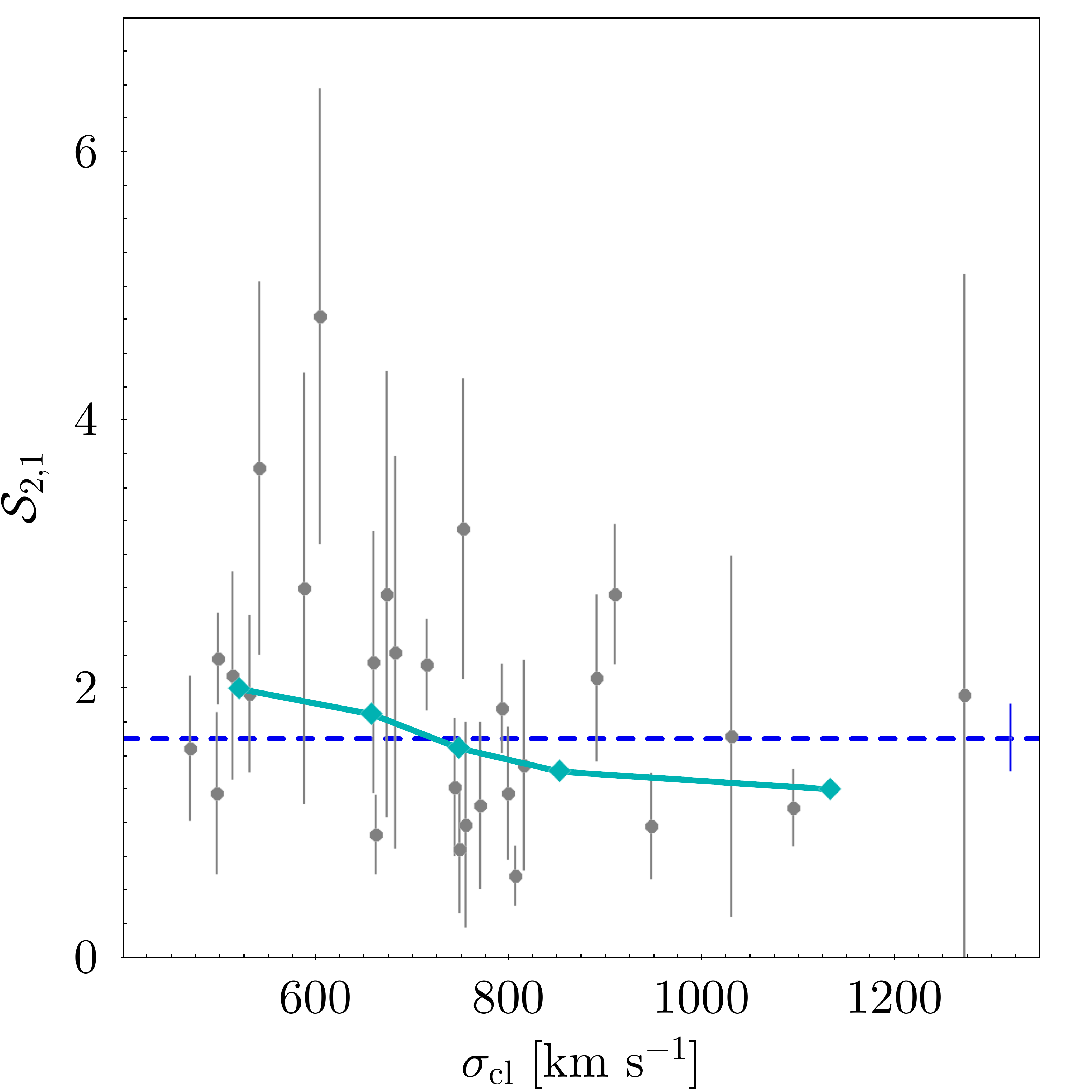}}
  \subfloat{
   \label{fig:SFR21_Sp_Cl-Lx}
    \includegraphics[trim={100 0 0 0}, clip,  height=4.43cm]{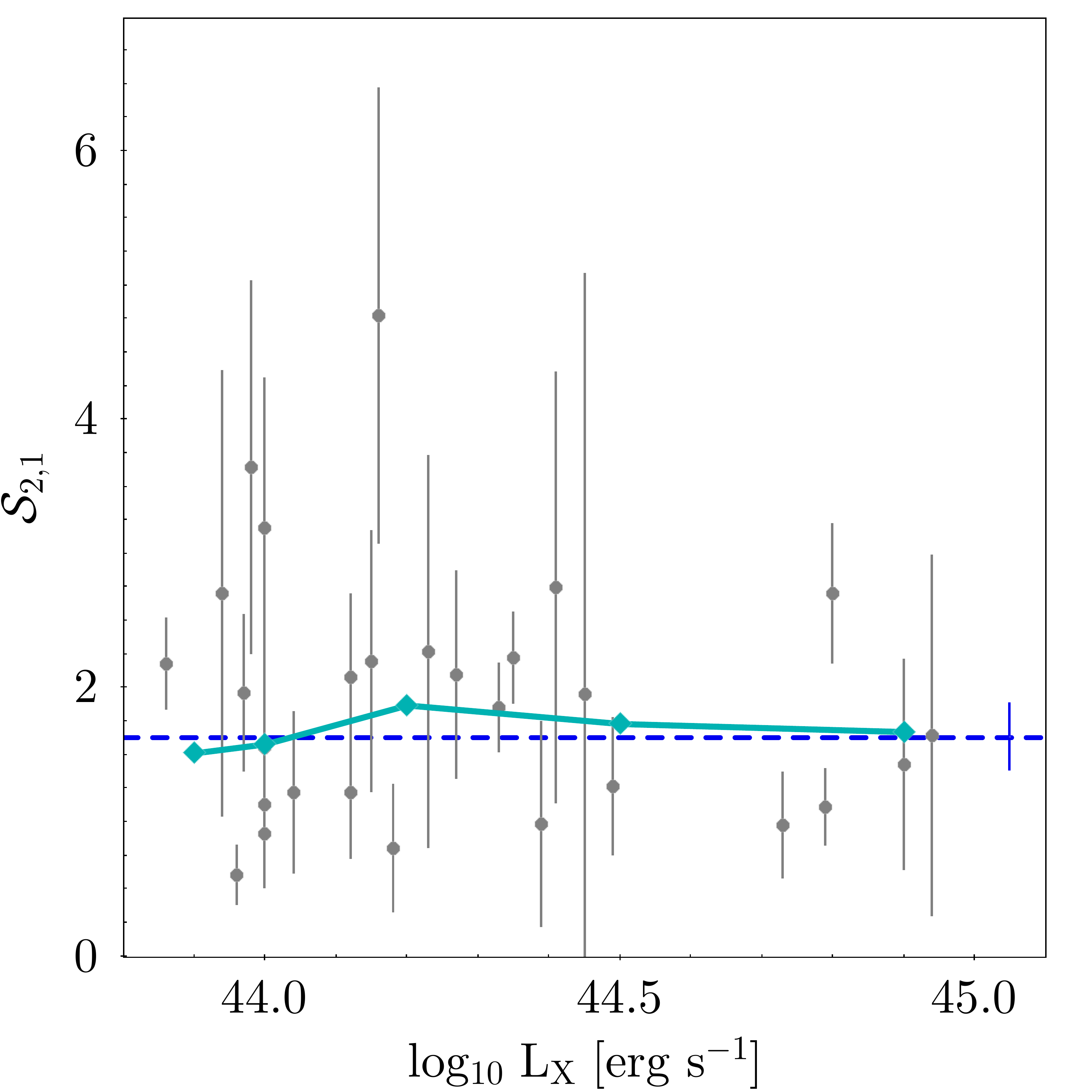}} 

\caption{Average quenching indices $S_{3,2}$ ({\it top row}) and $S_{2,1}$ ({\it bottom row}), for spiral galaxies in a sample of clusters (each grey point is a cluster, see text), as a function of cluster velocity dispersion ({\it left-hand panels}) and X-ray luminosity ({\it right-hand panels}). {\it Solid cyan lines:} index weighted mean; {\it dashed blue line:} index weighted mean for the field spiral sample. Error bars are calculated through bootstrapping.}
 \label{fig:SFR321_Sp_cluster-mass_ave}
\end{figure}

Possible differences in the quenching induced by the environment as a whole can be amplified by using quenching indexes. In this instance, we have selected clusters with both WINGS and OmegaWINGS data, with a minimum weighted number of 30 spirals (or approximately more than 20 spirals). We thus are left with a total of 28 clusters, for whose spiral galaxies we calculate the error-weighted mean $S_{3,2}$ and $S_{2,1}$, in bins of $\sigma_{\textrm{cl}}$ and L$_{\rm X}$; we also estimate the uncertainties by bootstrapping.

The results are shown in \autoref{fig:SFR321_Sp_cluster-mass_ave}. In the top panels, we present the relation between $S_{3,2}$ and the two cluster mass proxies. We do see that $S_{3,2}$ grows with increasing cluster velocity dispersion; a flat trend with X-ray luminosity is also observed. If we compare these relations with the average value found for galaxies in the field (dashed blue lines in \autoref{fig:SFR321_Sp_cluster-mass_ave}), we conclude that, in the cluster halo mass range probed by our sample, the environment boosts the quenching by a factor ranging from 1.36 to 1.73, for low- and high-mass clusters (gauged by $\sigma_{\textrm{cl}}$), respectively. This cannot be due to differences in the galaxy stellar mass (i.e., mass quenching), given that the mass distribution of spirals is very similar in clusters and in the field (see \autoref{fig:Mass_distr_types_sample}, right panel, in \autoref{sec:Stellar-mass_distribution}). 

On the other hand, a flat trend is observed for $S_{2,1}$ versus L$_{\rm X}$, with values very close to those encountered in the field. A different result is found when considering $S_{2,1}$ as a function of growing $\sigma_{\rm cl}$: in this case the average index decreases, from marginally above to marginally below the value for field galaxies. This result seems to contradict the evidence showing that a number of physical mechanisms halt, or at least hamper, star formation processes in clusters, leading to significant differences with respect to field galaxies.

Various interpretations can be offered to explain this result. One is stochasticity, due to the fact that the recent SFR is characterised by the presence of nebular emission lines that are representative of a very small range of stellar ages ($\sim 10^7$ yr), and whose intensities are also very sensitive to stellar age within this very range. Having kept the SFR of stellar populations with emission lines constant in the last 20 Myr (see \autoref{sec:Sinopsis}), we cannot detect SFR variations within such a short period. For example, a low SFR at age 4 Myr is somewhat equivalent to a higher SFR at slightly older ages. At older epochs, our observations sample and average star-forming processes over much longer timescales. The loss of time resolution in the star formation process is amplified by the fact that we only observe the central regions of all galaxies, while star formation is actually patchy and spatially discontinuous.

Secondly, not all interaction processes happening in clusters will lead directly to quenching. Ram pressure, for example, besides eventually causing star formation to stop through gas removal via hydrodynamical interactions, can be responsible for inducing moderate bursts of star formation at the beginning of the interaction, as observed in, e.g., \cite{Poggianti2016, Vulcani2018}. This effect is also seen in simulations \citep[e.g.,][]{fujita1999,roediger2005}, according to which the intensity of the burst depends on several factors: the relative velocity between the galaxy and the intracluster gas, the gas density, the galaxy inclination relative to its velocity vector. The combination of all these elements may result in the absence of a clear sign of decline in the most recent SFR of cluster spirals.

Last but not least, ram pressure is not the only mechanism that can enhance star formation: as proposed by \cite{Park-Hwang2009,Hwang2018}, and as we have also seen in \autoref{subsec:galaxy-neighbours}, the morphology of the most nearby galaxy can influence the present star formation level, both quenching and enhancing it, hence affecting the probability of observing one or the other effect. 

\subsection{Quenching in the cluster and field environments}
\label{subsec:quenching}
Galaxies in the local universe are on a path that naturally brings them to become passive objects. Several works \cite[see, e.g.,][and references therein, for a review on the topic]{Madau&Dickinson2014} have shown that the star-formation density of the universe peaked at around $z \approx 2$ (approximately 10.2 Gyr ago), and has declined exponentially since then. Evidence suggests that both galaxy mass and environment (more significantly for lower mass galaxies) have a crucial role in galaxy quenching \citep[e.g.,][]{Pintos-Castro2019}. Using WINGS data, \cite{Guglielmo2015} found that, for $z > 0.1$, SFHs are similar for galaxies with the same mass, regardless of morphology. For $z < 0.1$, however, late-type galaxies have, on average, higher SFR than early-types. 

Here, we try to disentangle the effects of galaxy mass and environment on quenching. To this end, we calculate the quenching indices as in \autoref{tab:quench-index-Morph}, but this time separating by stellar mass. 

The $S_{4,3}$ index, which reflects the early build-up of stellar mass, is much less dependent on stellar mass for spirals than for ellipticals or S0s, regardless of environment.

\begin{figure}
 \centering
 
  \subfloat{
    \includegraphics[trim={0 0 20 0}, clip, height=4.5cm]{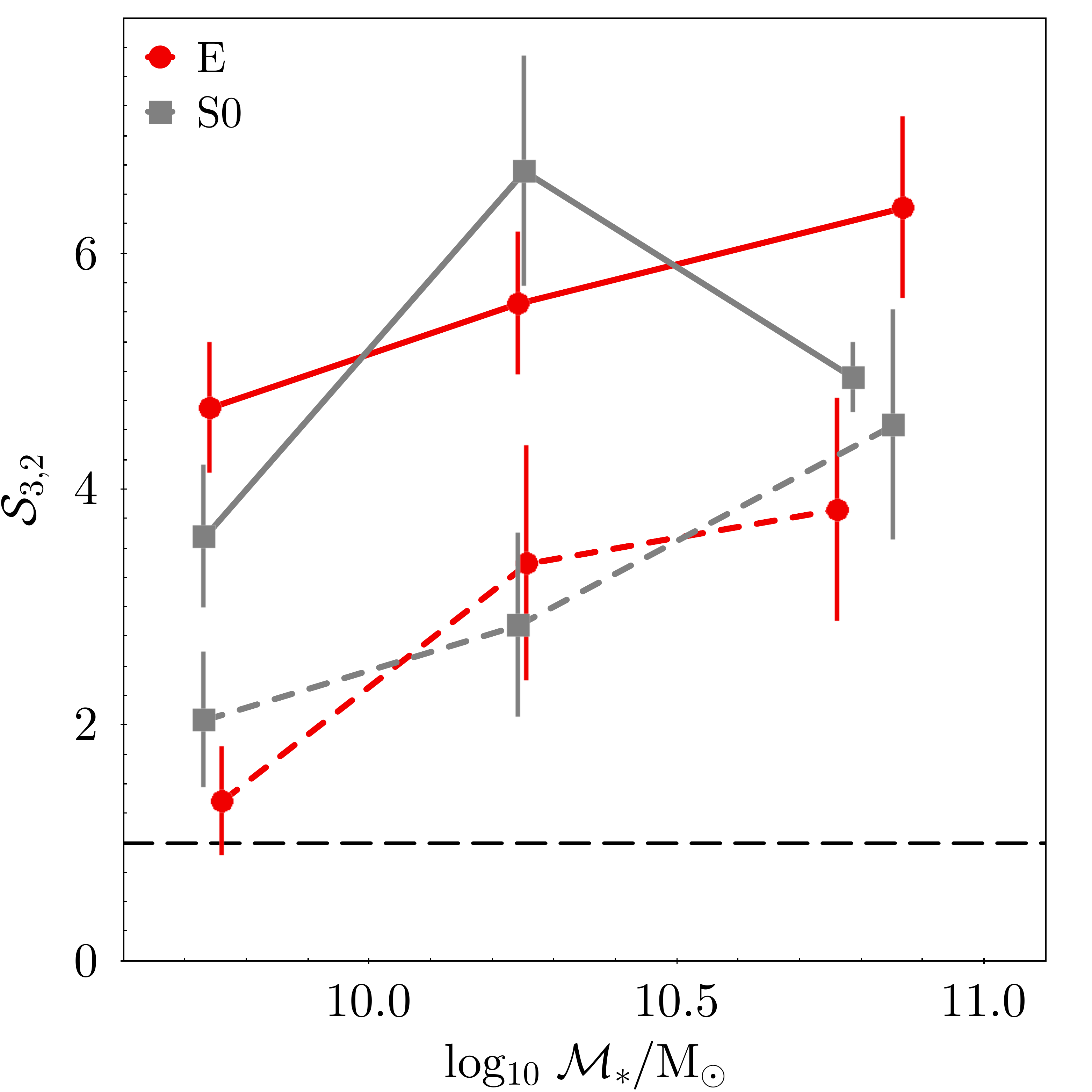}}
  \subfloat{
    \includegraphics[trim={80 0 0 0}, clip, height=4.5cm]{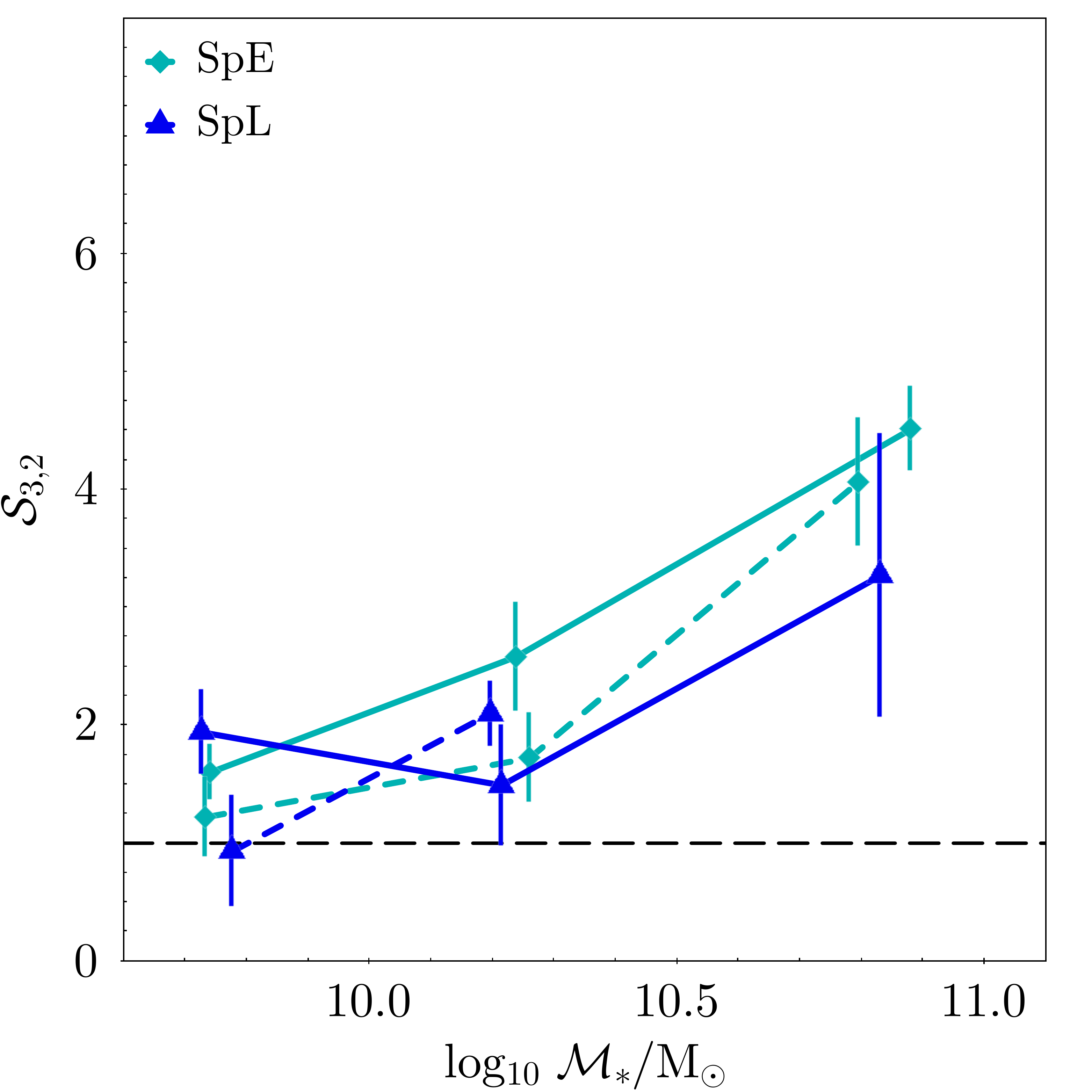}}

 \caption{Quenching index $S_{3,2}$ for early- ({\it left}) and late- ({\it right}) type galaxies in the cluster ({\it continuous lines}) and field ({\it dashed lines}) samples, divided into equal bins of stellar mass. Quenching occurs if index lies above the long-dashed black line.}
 \label{fig:quenchingindex}
\end{figure}

The $S_{3,2}$ index is presented in \autoref{fig:quenchingindex}, with continuous lines for cluster galaxies, and dashed lines for field ones. The $S_{3,2}$ index of field galaxies increases monotonically with mass for all morphologies, with values that are almost always lower than for their cluster counterparts, at fixed morphology and stellar mass. The strongest differences between environments are found for Es and S0s, whose index values, at fixed mass, take generally higher values than for spirals, even though the confidence interval around average values can be large. Differences induced by the environment are more subtle for spiral galaxies; for them, mass-quenching dominates over the environment. Once the mass effect is taken into account, what is left can be attributed to the physical processes happening in clusters. The fact that the differences between galaxies in clusters and in field are within the confidence interval, suggests that the time scales needed for quenching to be effective are larger than the $\sim 5.7\times 10^8$ years probed by our second age bin. At older ages, morphological transformations are more effective, and we likely observe strongly quenched spirals as S0s.

\begin{figure}
\centering
  \subfloat{
    \includegraphics[trim={0 0 20 0}, clip, height=4.5cm]{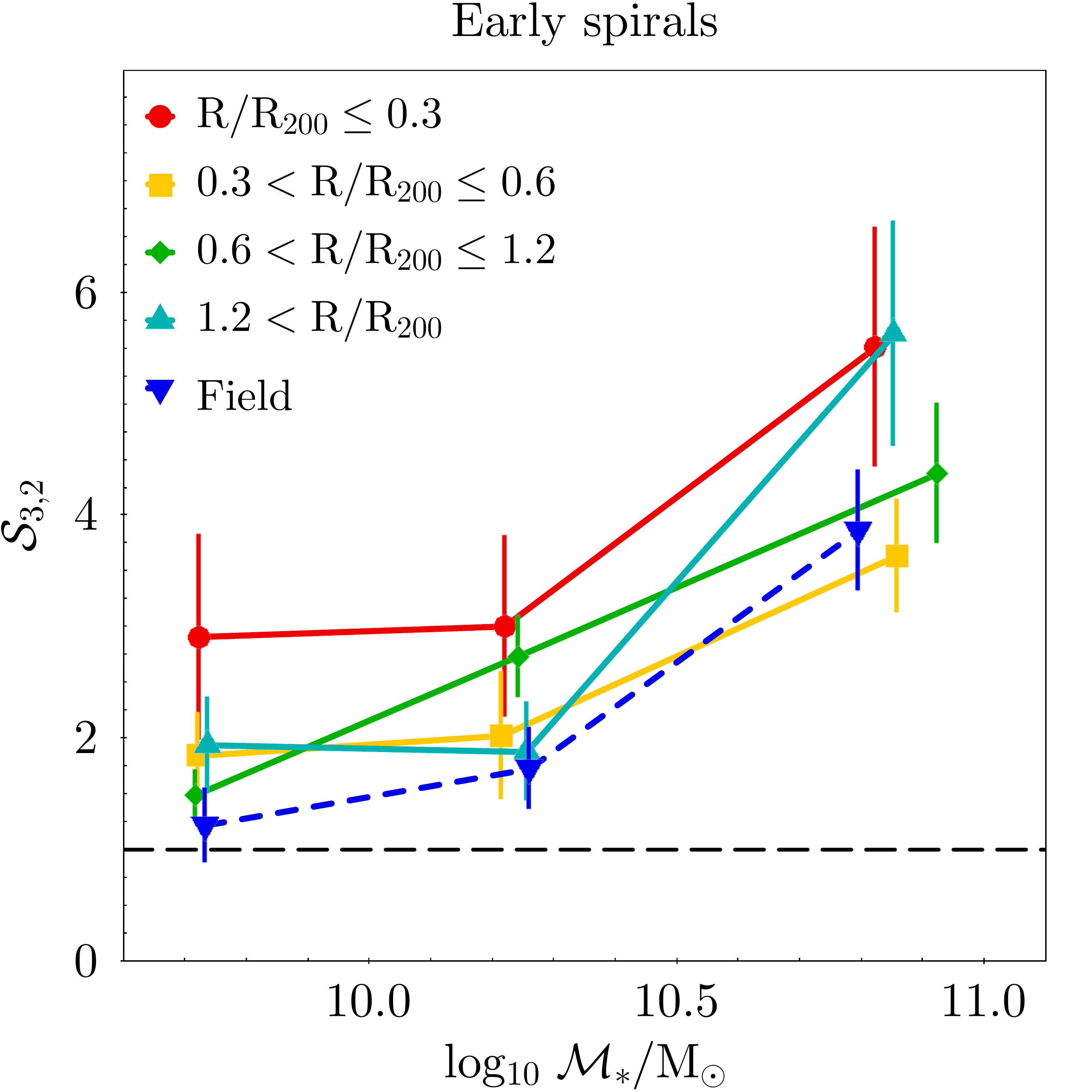}}

 \caption{Quenching index $S_{3,2}$ for early spirals in the cluster ({\it continuous lines}) and field ({\it dashed lines}) samples, divided into bins of clustercentric projected distance and stellar mass. The long-dashed black horizontal line separates quenching (above) from enhancement (below).}
 \label{fig:quenchingindex_rad}
\end{figure}

In spite of the tenuous environmental effects, spirals can help disentangle the influence of environment and mass, since they represent the population that has spent the shortest amount of time in the cluster environment. We hence analyse differences based on their projected location. In \autoref{fig:quenchingindex_rad}, we show the values of $S_{3,2}$ for SpEs as a function of mass, in the 4 bins of clustercentric projected distance that we have used, and for the field as well. On average, the largest differences with the field occur for galaxies closest to the cluster centre, but a clear tendency with clustercentric distance is not observed. This is somewhat expected: the location of a spiral galaxy with respect to the cluster centre is a transient property; it does not directly correlate with the time spent within the cluster, and hence under its ``quenching influence''. 

The $S_{2,1}$ index presents the largest differences between spirals and early-types, and this is a consequence of the fact that the percentage of bona fide quenched galaxies is much lower for spirals than for early-types. Overall, the results obtained for $S_{2,1}$ in spiral galaxies seem to be dominated by stochastic variations, due to the very nature of the SFR$_1$. However, in general, its values are always lower than those of $S_{3,2}$ at any given mass.

\subsection{Disentangling local density and radial distance}
\label{subsec:dissentangle}

In \autoref{subsec:SFH_proj-dist} and \autoref{subsec:SFH_LD}, we analysed how two parameterisations of the cluster environment properties, namely the projected distance to the cluster centre and the (projected) local density, may correlate with the SFHs of galaxies, depending on the morphological type. The two parameters are themselves correlated for $R/R_{200} < 1.8$ (see \autoref{fig:LD-R_R-neigh}, top row),  as also found by, e.g., \citealp{Oh2018}, with a larger scatter for larger projected distances. In this section, we make an attempt at disentangling their importance, by simultaneously analysing their effects on the stellar populations through the quenching indices.

We studied the mean quenching indices $S_{4,3}$, $S_{3,2}$, and $S_{2,1}$, for the cluster and field galaxy samples, separated into morphological classes. First of all, we noted no significant differences in the oldest index for any morphology. Cluster and field galaxies have similar quenching factors, regardless of LD or projected distance. 

On the other hand, $S_{3,2}$ is significantly larger in clusters than in the field for all morphological classes. Given that the mass distribution between field and cluster is similar for S0s and early spirals (see \autoref{sec:Stellar-mass_distribution}), it seems fair to conclude that, at least for these morphological classes, the environment is responsible for quenching.

Regarding the $S_{2,1}$ index, early-types show the largest differences with field galaxies, and they are more quenched than late-types as well. The $S_{2,1}$ index of elliptical galaxies  presents the clearest dependence with LD, with higher values of quenching happening at higher densities. This is noted for S0s as well, although to a lesser degree.  
On the other hand, spirals have quenching indexes very similar to their field counterparts and show at most mild trends for which none of the two environment parameterisations dominates over the other.

The highest SFR$_1$ for early-types is seen in those located in low-density regions and in the cluster outskirts, and decreases for log$_{10} \ \Sigma_{10}/$Mpc$^2\ \geqslant 1.0$ and projected distance $R/R_{200}\ \leqslant 1.5$. As for spiral galaxies, especially late spirals, higher values of SFR$_1$ may be found at all projected distances and for LD log$_{10} \ \Sigma_{10}/$Mpc$^2 \leqslant 2$. Since these galaxies are the least numerous, this might be due to contamination effects (i.e., the fact that the distances we measure are projected quantities). If star-forming late spirals are truly located close to the cluster centre, they may have momentarily increased their SFR due to ram pressure, which is more efficient in the inner cluster region ($R/R_{200} < 0.5$; \citealp{Zinger2018}). Conversely, for SFR$_2$, there is no clear trend for any morphological type or environment tracer. The same is true for older ages. 

The quenching effect is apparently detectable on early spirals when they are observed closer than one virial radius from the cluster centre and at log$_{10} \ \Sigma_{10}/$Mpc$^2 \gtrsim 1.8$, although they may keep some gas to form stars at smaller distances. 

\section{Discussion}
\label{sec:discussion}
The analysis we have performed in this paper aims at determining the relative importance of stellar mass, morphology, and environment when trying to detect and quantify the environmentally--driven changes that are responsible for the observed properties of galaxies in clusters, especially focusing on the stellar population properties. One of the questions that are often brought up is which, between the local and the large-scale environment, is most important for the fate of these galaxies. This is not an easy question to answer, since the characteristics of the local environment are somewhat influenced by those of the large-scale one: for example, more massive clusters are both richer in ICM and in galaxies, and this consequently raises the probability that an infalling galaxy will find itself embedded in a denser and hotter medium, with a higher likelihood of galaxy-galaxy interactions as well. Furthermore, recent environmentally--induced changes are more easily visible in later morphological types.

When we consider the morphological fractions of galaxies in local clusters, we do find evidence of a mild effect as a function of total cluster mass, as parameterized either by $L_{\rm{X}}$ or $\sigma_{\rm{cl}}$, with more massive clusters hosting a larger fraction of S0 galaxies, at the expense of the spiral population (see \autoref{fig:Fraction_types_Sigma_cl} and \autoref{fig:Fraction_types_Lum-X_cl} in \autoref{sec:Fraction_types_cluster}). 
Of course, this effect can be an indirect consequence of the fact that more massive clusters host a larger number of galaxies and make galaxy-galaxy interactions more frequent in inducing morphological transformations.
This finding, however, is somewhat different from what \cite{Poggianti2009} concluded, when analysing WINGS data, i.e., that morphological fractions were not significantly different in clusters with different masses. These discrepant results can be explained by the sample used in this work, which includes galaxies that are both fainter and farther from the cluster centres (up to $1 R_{200}$ and beyond).

Taking into account morphological type fractions is of crucial importance when looking for differences in the stellar populations, in order to properly understand and quantify pure environmental effects. An early-type galaxy located in a cluster is such, either because it was born like this, or because it has been transformed. This can happen if the galaxy has spent a significant amount of time in this dense environment. A broad way to study this effect is by looking at the rest-frame colours, at fixed mass: ellipticals and late spirals are almost indistinguishable from their field counterparts in this respect, while early spirals and S0s are found to be redder in clusters. The most straightforward interpretation of the two aforementioned results (morphological type fraction and colours) is that changes in the stellar populations happen faster than morphological transformation \citep{Balogh2004}. On the other hand, the fact that late spirals look alike (in colour and also in recent SFR) in clusters and the field probably means that most of them are relatively new to the cluster environment (as already proposed by \citealp{Cava2017}), and their overall properties have not been significantly affected yet. 

Given that dense environments will eventually quench star formation, differences between cluster and field galaxies should be traceable in their SFH (\autoref{fig:SFH_morphology}), and these effects should be stronger in those galaxies that have spent the longest time in clusters. Indeed, once the stellar mass is taken into account, the mean SFH of cluster galaxies is steeper (i.e., displaying faster quenching) than in the field for all morphologies, directly suggesting a higher quenching efficiency in clusters. This effect is more clearly seen with the quenching indices (see \autoref{tab:quench-index-Morph} and \autoref{tab:quench-indices-Sp-mass}):  $S_{4,3}$, which reflects the star formation conditions at the earliest epochs of galaxy life, shows, in general, the smallest differences when various morphologies and environments are considered, reflecting a common formation scenario in the early past. Only when separating by stellar mass, significant and important discrepancies are spotted, which is consistent with galaxy birth and growth in a downsizing scenario \citep{Cowie1996}. Furthermore, $S_{3,2}$ is on average higher for cluster early-types and decreases monotonically for late-types. This trend can be roughly viewed as a consequence of the time a galaxy has spent in the cluster environment, which we expect to be --on average-- increasingly larger when going from the earliest to the latest morphological types. The differences encountered for cluster and field ellipticals and S0s might be due to different formation channels for these galaxies. For S0 in particular, several studies point to the existence of two types of lenticular galaxies: those that are born as such, and those that have been transformed into S0 because of environmental effects \citep{Tapia2017, Fraser-McKelvie2018b}.

Concerning the star formation in Es, as highlighted in \autoref{fig:SFH_morphology}, the finding that the SFR$_1$ of elliptical galaxies is significantly higher in the field than in clusters is consistent with the results of \cite{Huang&Gu2009}, who found hints that ellipticals in lower-density environments may be more prone to hosting star formation. 

In the comparison between cluster and field populations, when galaxy stellar mass is taken into account, the differences in quenching are only ascribable to the environment, and they are seen in all morphological types. The mean SFH of cluster spirals (these objects most likely dominate the recently infalling population) only shows a mild dependence on total cluster mass. This effect is more easily spotted through the $S_{3,2}$ index, whose average values increase with larger cluster mass (i.e., cluster velocity dispersion; \autoref{fig:SFR321_Sp_cluster-mass_ave}). The $S_{2,1}$ index, where more recent effects (and on a much shorter timescale) should be visible, presents values that are remarkably similar for spirals in clusters and in the field. We believe this can be interpreted as a hint that local effects are the ones driving the ability to form stars, while correlations with large-scale ones are much more difficult to spot: large-scale effects are likely blurred by the collection of different mechanisms acting simultaneously, but with different time scales, intensities, and consequences. On the other hand, the fraction of star-forming spirals is always lower in clusters than in the field, with no correlation with total cluster mass (\autoref{fig:SF_fraction_Cl-mass}).

Both the redder colours and the higher fractions of quenched galaxies in clusters clearly show that a dense environment is hostile to star formation, and will eventually lead galaxies to stop forming stars. Nevertheless, the aforementioned results for $S_{2,1}$ in cluster and field spirals seem to be inconsistent with this picture. A likely explanation for the behaviour of $S_{2,1}$ may lie in the very short timescale probed by emission lines, which are the observable used to calculate the current SFR. This makes the SFR$_1$ calculation subject to the probability of observing a galaxy in a star-forming or in a quenched phase which, in turn, depends on currently ongoing interactions. For example, the presence of a nearby companion may both trigger and suppress a certain level of star formation activity (see also discussion below), and ram pressure has been demonstrated to be able to induce star formation as well. Conversely, SFR$_2$ probes a timescale more than 20 times longer, and hence it may be a better tracer of the cumulative effects of the mechanisms at play, which will eventually lead to quenching and we seem to be able to observe through $S_{3,2}$.

This picture is supported by the characteristics of the main-sequence relation observed for cluster spirals, which exhibits a larger scatter than the relation for spirals in the field (\autoref{fig:SFR-Mass_relation}). While the slope is similar in the two environments, cluster galaxies show both higher and lower star formation levels than field ones, something that can be attributed to the different interactions that cluster galaxies undergo, both enhancing and suppressing the SFR (note that this phenomenon is likely to be short-lived, since we only see it in SFR$_1$). The presence of a population of galaxies with significantly higher SFR, in particular, seems odd, given that the cluster environment is notoriously adverse to maintaining star formation. However, it may be explained as the product of the first stages of interaction with the cluster, during which ram pressure, before stripping away the interstellar medium from the galaxies, induces a star formation episode, as has been observed both in numerical simulations \citep[e.g.,][]{fujita1999,Kronberger2008,Kapferer2009} and in real data (e.g., \citealt{Vulcani2018}; \citealt{Roberts2020}; Fritz et al., in prep.). This should be the case as long as interactions do not affect the morphology so much that early spirals cease to be classified as such \citep{Lambas2012}.

In an effort to study the role played by the environment, we searched for possible differences in SFR$_1$ and SFR$_2$, at various projected clustercentric distances and local density values. We have done so for the different morphological types and separated by stellar mass as well, to better isolate the influence of each factor (\autoref{fig:Main-sequence}). Firstly, we found no significant variations in SFR$_1$ as a function of stellar mass, neither for ellipticals nor S0s, at all projected clustercentric distances. Secondly, S0 galaxies, which represent the dominant cluster population, display somewhat increasing SFRs as a function of increasing clustercentric distance, again independently of stellar mass. For ellipticals, however, the trend is not clear. The fact that the present SFR of early-types is independent of stellar mass is an indication that the star-formation mode/mechanism currently at play is different from that in spirals \citep[e.g.,][]{Silk2014}. Since early-type galaxies are gas-poor objects today, the most important factor to determine their SFR would be, not the amount of available gas, but rather external agents, like ram pressure or neighbour interactions, which may trigger star formation. This is further supported by the fact that an SFR-mass relation indeed exists for the SFR$_2$ of early-types, suggesting that processes that enhance and quench star formation are both at play in clusters, but that the former are much shorter-lived and less frequent. 

In the case of spirals, separated into stellar mass bins, SFR$_1$ increases with decreasing radius. The rise in SFR$_1$ from the outskirts is, indeed, mild, but common to all mass bins. This is observed until about $0.5 R_{200}$, after which SFR$_1$ decreases towards the innermost regions. (Note that, even though pronounced, the fall at high masses is characterised by large uncertainties). This decrement is more pronounced in more massive clusters, when $\sigma_{\rm{cl}}$ is taken as the cluster mass proxy. The result is not clear when $L_{\rm{X}}$ is considered. 

\autoref{fig:Main-sequence} also suggests that spirals closer in projection to the cluster centre are more prone to quenching in the high- and intermediate- mass bins, although this result needs to be taken with some caution because of the uncertainties, to which projection effects might also be contributing. The results for SFR$_2$ strongly resemble those for SFR$_1$ (again \autoref{fig:Main-sequence}): the only difference is that the SFR$_2$ of the most massive cluster spirals can be as high as in the field, with a maximum at around 0.5 $R_{200}$. This is consistent with ram pressure promoting star formation, and would naturally explain why at present the SFR of the most massive cluster galaxies is lower than in the field: their first star-forming episode would have consumed a significant fraction of their gas and, together with gas stripping, would more quickly lead to the halting of star formation, owing to gas depletion. 

A similar analysis was conducted by taking into account the local density (\autoref{fig:Main-sequence-LD}). If no distinction is made on morphological type, LD affects SFR$_1$, with higher densities resulting in lower SFR$_1$. When, however, morphological class is considered, we see that the relation between SFR$_1$ and LD is driven by Es and S0s, while for spirals SFR$_1$ is flat or even increases with growing LD. On the other hand, SFR$_2$ seems to be quite unaware of local density, for all morphologies and stellar masses. The only clear difference is seen in field galaxies, which display higher values of both SFR$_2$ and SFR$_1$.

The fact that SFR$_2$ shows no significant trend with LD can be interpreted as evidence that, at the age sampled by SFR$_2$, the local environment has already had a strong quenching effect on the SF activity, something that is also supported by a high $S_{3,2}$ quenching index. $S_{3,2}$ would be, of course, stronger for early-type galaxies, which have spent, on average, the longest time within the cluster.  

Exactly pinpointing the physical mechanisms responsible for this is, however, quite tricky: the inner regions of clusters are denser both in terms of the ICM and the number of galaxies, and interactions are thus potentially very effective. Recently, \cite{Vulcani2022}, exploiting the same dataset used in this work, have shown that the large majority of spirals are very likely to be affected by RPS at some point during their accretion into clusters, and this mechanism is more efficient in the inner cluster regions \citep[e.g.,][]{Zinger2018}. Moreover, it is in these same regions that rapid encounters with early-type galaxies are expected to be more frequent and, hence, more effective to accelerate the quenching process. 

Taken at face value, our results seem to imply that clusters quench SF more drastically in early- than in late-types, as differences with the field are stronger in the former. This is likely a direct consequence of the amount of time spent within a cluster. An alternative interpretation is that early-types are such because they have been more affected than spirals, and spirals are such because they have been less affected and, thus, able to keep their morphology, independently of their location and the time spent in the cluster. Furthermore, in some cases, cluster late-types display average SFR values higher than in the field (see, e.g., the SFR-mass relation for spirals). This, apparently counterintuitive, result can be understood when interactions within galaxy clusters are taken into account, as follows.

Several of the findings so far presented and discussed have a quite natural explanation in binary interactions. With the purpose of studying both the hydrodynamical and the gravitational/tidal effects of galaxy-galaxy interactions, \cite{Park-Hwang2009} and \cite{Park-Choi2009} analysed galactic properties in relation to the distance and the morphology of their closest neighbour. According to their results, a late-type galaxy located within a neighbour's virial radius is affected in different ways, depending on the neighbour's morphology: a late-type neighbour can provide gas and trigger star formation, while the extended hot gas halo of an early-type can act as a quenching agent, as also found by \cite{Patton2020} with simulations, and \cite{Cao2016} using observations.

We performed a similar analysis, looking for changes in the SFH, quenching indices, colour, and luminosity of galaxies, separated by morphological type, as a function of projected distance to and morphology (early- or late-type) of their closest neighbour. Unlike what we found for the clustercentric distance and the LD, a close encounter clearly affects the recent SFR: both early- and late-type galaxies display an enhancement in SFR$_1$ when they are located within the virial radius of a late-type closest neighbour. Similarly, the $S_{2,1}$ quenching index increases when a galaxy is within the virial radius of an early-type closest neighbour. These results are statistically more significant when spirals are considered, while for early-types, even though trends are observed, average values are within the confidence intervals. Somewhat counter-intuitively, the fraction of star-forming spirals steeply declines when located within 1 virial radius of a companion, independently of its morphology (see \autoref{fig:SF-active-Fract_neigh}), while, at the same time, SFR$_1$ is higher in spirals when they are within 1 virial radius of another spiral. This apparent contradiction might indicate that the enhanced star formation activity triggered by the hydrodynamic interaction with the late-type companion is short-lived: the interaction initially produces an increase of the SFR that quickly consumes part of the available gas, which in turn quenches the galaxy and decreases the star-forming fraction.

SFR$_2$ and  $S_{3,2}$ (as opposed to SFR$_1$ and $S_{2,1}$) do not seem to be affected by closest neighbour properties. This may actually be the consequence of three factors: (a) as demonstrated by \cite{Hwang2018}, the timescales of galaxy-galaxy encounters are very short; (b) SFR$_2$ averages the stellar population properties and galaxy interactions over more than 0.5 Gyr; (c) at earlier epochs, spiral galaxies were likely in less dense environments, hence experiencing less frequent encounters. In the case of early-type galaxies, the reduction in the average SFR with decreasing neighbour distance is likely a secondary effect of the MD relation: a denser environment is both more likely to host early-types and to be located in the innermost parts of a cluster. There, both the galaxy and the ICM densities reach their highest values, and are hence more likely to negatively affect the onset of star formation.

The influence of the closest neighbour is indeed modulated by the local environment, particularly by the LD (see \autoref{fig:SF_neigh_Sp_LD}): quenching effects are stronger at low-LD values, while at high LD the trends with distance and morphology of the neighbour are, on average, flat. We can think of two possible explanations for this: firstly, since on average higher densities occur at smaller clustercentric radii, galaxies have likely already lost most of their cold gas; secondly, encounters among galaxies there, although more frequent, are faster and hence shorter-lived.

Supporting this idea, \cite{Hwang2018}, using numerical simulations, found that high-speed multiple hydrodynamical interactions between a spiral and an early-type will eventually result in a loss of about 50\% of gas after six edge-on collisions, and about 90\% if the collisions are face-on. Interactions with the ICM will yield a similar gas loss, according to simulations \citep{Jachym2007}. This result is confirmed by \autoref{fig:SF-active-Fract_neigh_LD} (right panel): the star-forming fraction, at fixed neighbour distance, decreases more strongly at high than at low LD.

By the same token, while the average value of SFR$_1$ in spiral galaxies seems to be unaffected by either LD or by projected clustercentric distance, the fraction of star-forming spirals does show negative and positive correlations, respectively, with decreasing projected distance and local density (\autoref{fig:SFing_fraction}): eventually, dense environments (gauged both by number of galaxies and ICM density) will lead to the quenching of star formation, but on timescales larger than those probed by SFR$_1$. Mechanisms, such as ram pressure or binary interactions, can trigger a star-formation burst on short timescales, likely to be followed by a sudden quenching. The fact that we observe similar trends in the average SFR$_1$ and the quenched fraction, as a function of both LD and projected distance, is consistent with a short duration and high intensity of this active phase.
This is also reflected in higher SFR$_1$ for cluster than field spirals in the lowest mass bin, at both small cluster radii and high density (see \autoref{fig:Main-sequence} and \autoref{fig:Main-sequence-LD}).

This brings us to the question of whether it is position within the cluster or local density the most important factor affecting galaxy properties. We have found an indication that, at least concerning the star-forming fraction, the effect of clustercentric distance is marginally stronger (see \autoref{fig:SFing_fraction}). The two aspects are, however, not independent: the density of galaxies is higher in the cluster centre, where the ICM also reaches its peak density, so the two effects would add up. On the other hand, a galaxy might be located in a substructure of high galaxy density at a large galactocentric distance, hence in a region of lower ICM density; in this case, the effects of high galaxy number density will not be aided as much by the ICM. 

With the purpose of disentangling the two parameters, we analysed the quenching indices for different morphologies, simultaneously taking into account clustercentric distance and local density. 

As probed by $S_{3,2}$, projected distance seems to play a larger part in SF quenching for S0s, while SpEs seem to be equally affected by projected distance and LD, with the strongest quenching  occurring at the highest LD and the shortest distances. This may indicate, not only that the combined effect of the two produces the most potent quenching, but also that SF supression is more likely visible after about 0.5 Gyr, the average half-crossing time for galaxies in clusters. On the other hand, quenching at the most recent epochs (revealed by $S_{2,1}$) is significantly higher in clusters than in the field only for early-type galaxies; for late-types, quenching is on average very similar, if not indistinguishable, in clusters and the field. 

The latter result is coherently explained by the other findings gathered so far: the cluster environment will ultimately act as a quenching agent of star formation in recently infalling, star-forming (gas-rich), galaxies. However, there are several mechanisms, such as ram pressure and binary interactions, that might briefly boost star formation, hence partially compensating --on average-- for the quenching, and making star formation and quenching indicators similar to those in the field.

Since star formation is much rarer in ellipticals, the quenching index $S_{2,1}$ is highest for them among all morphologies, with a stronger dependence on LD than on clustercentric distance. The relations, however, show a quite width confidence interval. This is naturally explained by interactions with nearby late-type companions in cluster outskirts.

Differences in stellar content as a function of large-scale cluster properties suggest that the local environment (e.g., interactions among galaxies) has stronger effects than the large-scale one (e.g., cluster-galaxy interactions) in shaping present galaxy stellar properties, such as the stellar mass distribution \citep{Vulcani2012} and the SFR-stellar mass relation \citep{Calvi2018}. In fact, variations in stellar populations are more pronounced (even though sometimes subtle) relative to local properties, and their intensity/significance depends on morphology.

Clearly, different mechanisms can act at the same time, with different intensities and scale times. These include harassment, thermal evaporation, RPS, and starvation. They are correlated, either with the presence of nearby companions (i.e., the local density) or with the clustercentric distance (i.e., the local characteristics of the ICM), and affect both the content and the physical conditions of the gas and, hence, the ability of a galaxy to form stars. Our results indicate that the preferred processes are RPS (in the inner parts of clusters, preferably where $R/R_{200} < 0.5$), harassment (with repetitive galaxy encounters at low local densities), and tidal interactions (which quench star formation if the companion is an early-type, or raise the SFR when the closest neighbour is a spiral). Preprocessing, acting in the cluster outskirts, may be playing a part too. All these phenomena can act on galaxies of all stellar masses \citep{Bamford2009}, even on some located in the field, where there are both quenched spirals and early-types with some degree of star formation activity. 

\section{Summary and conclusions}
\label{sec:conclusions}
The buildup of the bulk of the baryonic mass in galaxies is known to depend at least on two intrinsic characteristics: their (final) total stellar mass, and their morphology. Yet, at the same time, both morphology and the recent SFH of galaxies are known to be strongly dependent on the environment they inhabit. Among the long-known observational evidences of the effects of a dense environment on the properties of galaxies are the so-called morphology-density relation, and the fact that quenched galaxies dominate the populations of local clusters. The processes leading to these observed characteristics are still a matter of investigation.

In this paper, we have undertaken an in-depth analysis of the stellar population properties of a mass-limited sample of cluster and field (non-cluster member) galaxies, with $\mathcal{M}_* > 3\times 10^9 \ \rm M_\odot$. We have mostly focused on investigating possible interdependencies between the SFH, the morphology, and both the local and large-scale environments.

To this end, we have used the WINGS/OmegaWINGS sample of cluster galaxies in the local universe ($0.02 < z < 0.09 $), leveraging galaxy properties, such as $B$ and $V$ photometry, redshift, cluster membership, morphology, and local density, as well as cluster properties like velocity dispersion and X-ray luminosity. We also used the spectrophotometric, non-parametric, code \texttt{SINOPSIS}, in order to derive fundamental properties of the stellar populations, such as masses, star formation rates, stellar ages, and star formation history. Concerning the latter, we have defined a ``quenching index'', $S_{i,j}$, as the ratio between the SFR in the two different age bins $i$ and $j$. The quenching index can detect possible effects of the environment on the ability of a galaxy to form stars.

The main results of this work can be summarised as follows:

\begin{enumerate}
       \item[1.] At fixed stellar mass and morphology, the star formation histories of cluster galaxies are steeper than those of field galaxies, indicating higher quenching efficiencies due to external agents. However, some mechanisms acting preferably in clusters, such as ram pressure or hydrodynamical interactions with nearby late-type galaxies, can instead favour the onset of star formation in recently acquired galaxies.
       
       \item[2.] When studying the properties of stellar populations, and in particular the star formation histories of cluster galaxies, morphology needs to be taken into account. The mode of star formation depends on the time spent within the cluster (on average, longer for early-types and shorter for late-types), and can in turn be roughly parameterized by morphology. Neglecting morphology might lead to underestimating other mechanisms at play, such as preprocessing, close neighbour interactions, or ram pressure.
       
       \item[3.] At low values of local density, the distance and morphology of the closest neighbour determine the properties of the current star-forming activity. In a high local density environment, though, we cannot rule out a possible influence, as the presence of multiple nearby galaxies could play a role.
       
       \item[4.] Quenching measured by the $S_{3,2}$ index on the recently accreted galaxy population (mainly spiral galaxies) is possibly stronger in more massive clusters (i.e., those with a higher velocity dispersion; see \autoref{fig:SFR321_Sp_cluster-mass_ave}), but this may result from the combination of local effects (see \autoref{fig:SF_neigh_Sp_LD} and \autoref{fig:Main-sequence}).
       
       \item[5.] A study of the effects of cluster environment on star-formation activity limited to assessing the fraction of passive galaxies, based on the presence/absence of emission lines, is bound to show only a partial picture, as star formation episodes in clusters can be both rapidly induced and quickly suppressed. A more complete portrayal arises by quantifying the star-formation activity still present in the fraction of active galaxies, and by looking at the full star formation history.
\end{enumerate}

We do find that cluster environment affects the ability of a galaxy to form stars, independently of morphological type. Still, it is difficult to disentangle the effects of neighbour proximity from those of the cluster as a whole. To try to shine more light on the issue, we have analysed whether the morphology of the closest (massive) neighbour may affect the properties of a given galaxy. We find that, indeed, very close late-type neighbours promote star formation, while very close early-type neighbours can quench it. Such binary interactions may likely be at the origin of, or at least contributing to, the Butcher-Oemler effect. The cumulative effects of repeated encounters with early-types, more frequent in the innermost cluster regions, will ultimately result in the shutting down of star formation. This result provides an observational context to the theoretical work of \cite{Hwang2018}. 

Going forward, it would be important to follow up surveys based on fibre spectroscopy with new ones, using integral field spectroscopy (in the same way the Mapping Nearby Galaxies at the Apache Point Observatory survey is a natural extension of the Sloan Digital Sky Survey). More complete spectroscopy surveys of cluster galaxies, such as the future WEAVE Wide-Field Cluster Survey, will allow the study not only of galaxies as a whole, but also of their stellar and gas dynamics, further helping to disentangle the different mechanisms at play.

\section*{Acknowledgements}

We are grateful to the anonymous referee for their suggestions and very careful review of the paper that greatly improved its presentation. J.F. and D.P.M acknowledge financial support from the UNAM-DGAPA-PAPIIT IN111620 grant, Mexico. R.A.G.L. acknowledges the financial support of DGAPA, UNAM, project IN108518, and of CONACyT, Mexico, project A1-S-8263. D.P.M. acknowledges support from a CONACyT scholarship.

\section*{Data Availability}

All data used for this work will be made publicly available through the Common Data Service \href{http://cds.u-strasbg.fr}{CDS}. 



\bibliographystyle{mnras}
\bibliography{example} 




\appendix

\section{Fraction of morphological types with cluster mass tracers}
\label{sec:Fraction_types_cluster}

One of the main goals of the WINGS survey was to study the fraction of morphological types in nearby clusters, as a function of cluster X-ray luminosity ($L_{\rm{X}}$) and velocity dispersion ($\sigma_{\rm{cl}}$; \citeauthor{Poggianti2009}, \citeyear{Poggianti2009}). Here, we extend the analysis to the clusters observed in OmegaWINGS, and thus include all cluster member galaxies within projected $R_{200}$. In the following, we gather the galaxies in three broad morphological classes: ellipticals, S0s, and spirals.

\begin{figure*}
 \centering
  \subfloat{
   \label{fig:fraction-E_Sigma_cl}
    \includegraphics[trim={0 0 0 0mm}, clip, width=0.35\textwidth]{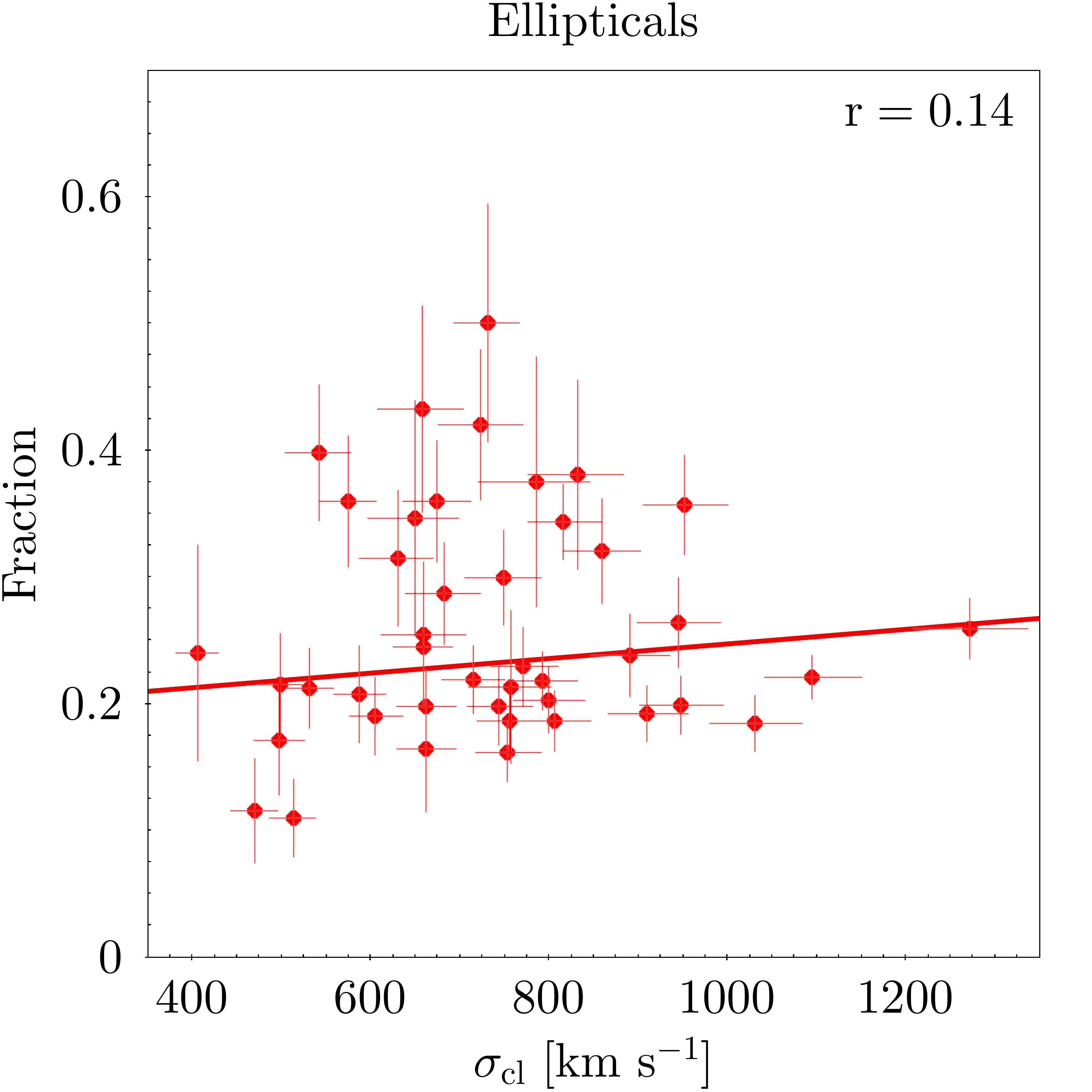}}
  \subfloat{
   \label{fig:fraction-S0_Sigma_cl}
    \includegraphics[trim={120 0 0 0mm}, clip, width=0.30\textwidth]{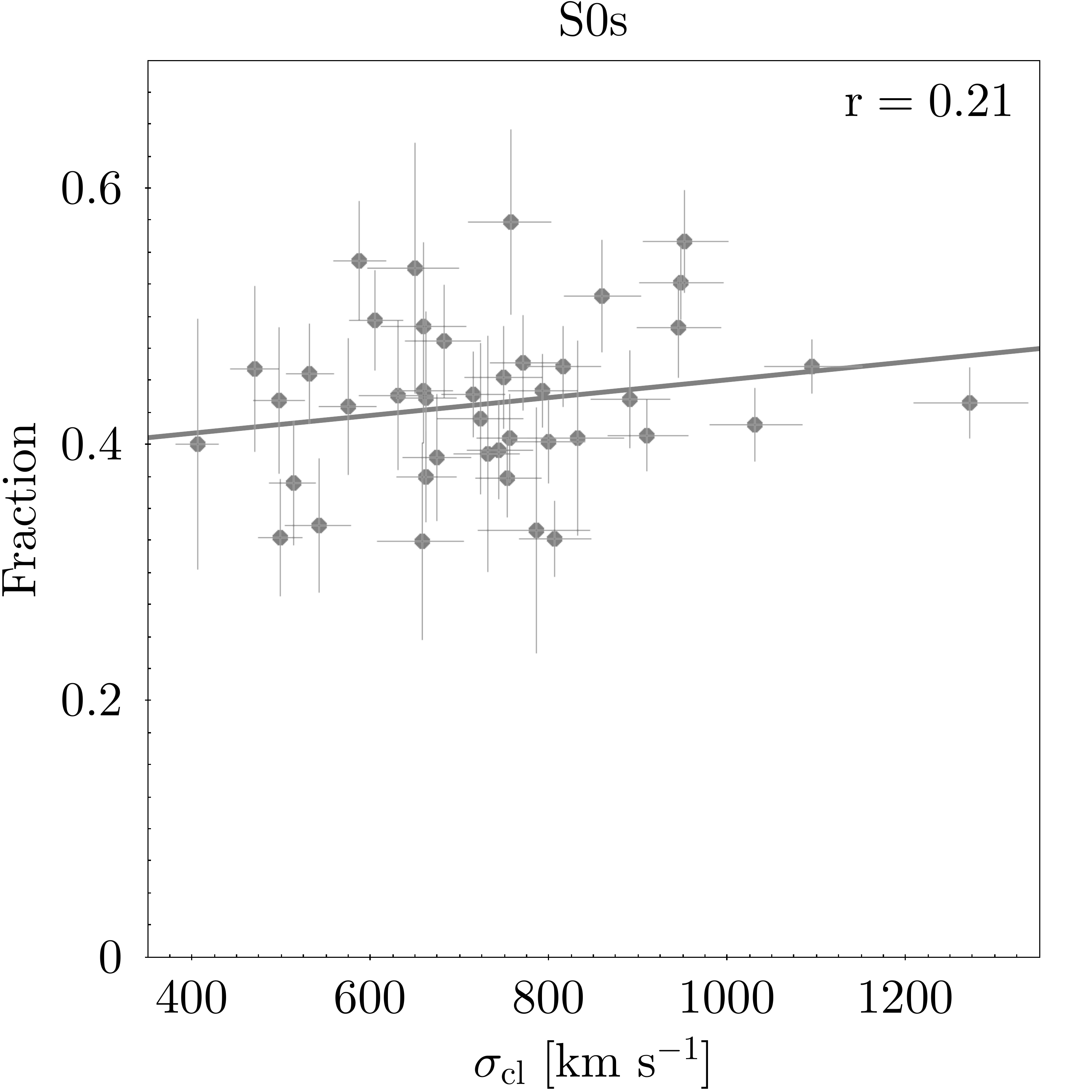}}
  \subfloat{
   \label{fig:fraction-S_Sigma_cl}
    \includegraphics[trim={120 0 0 0mm}, clip, width=0.303\textwidth]{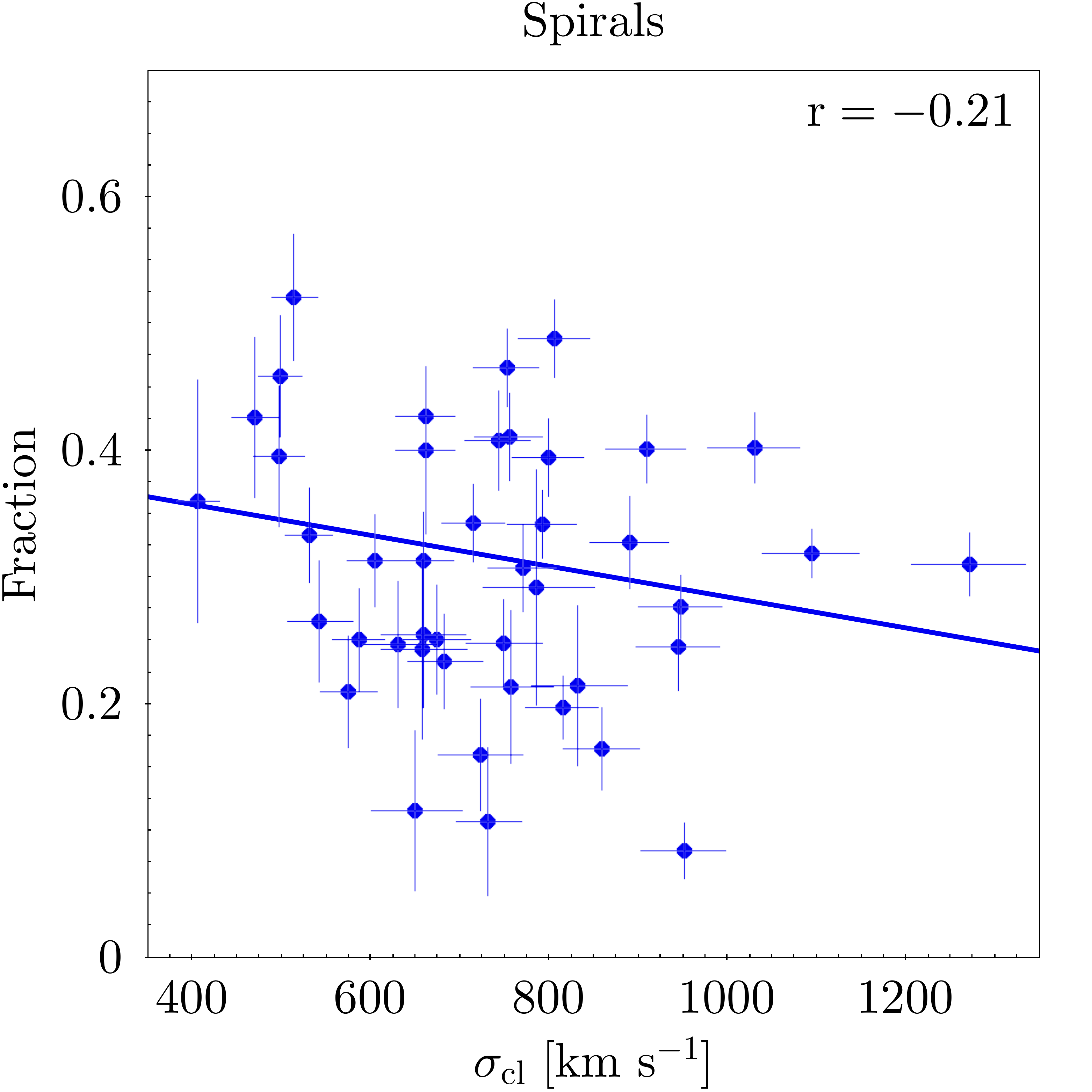}}      
 \caption{Fractions of galaxy morphological types in the WINGS and OmegaWINGS datasets, as a function of cluster velocity dispersion ($\sigma_{\rm{cl}}$): ellipticals (left), S0s (middle), and spirals (right). The error bars for the fractions are binomial. The least-square fit weighted by the errors is shown as a solid straight line in each panel. The $r$ value corresponds to Pearson's correlation coefficient.}
 \label{fig:Fraction_types_Sigma_cl} 
\end{figure*}

\begin{figure*}
 \centering
  \subfloat{
   \label{fig:fraction-E_LumX_cl}
    \includegraphics[trim={0 0 0 0mm}, clip, width=0.35\textwidth]{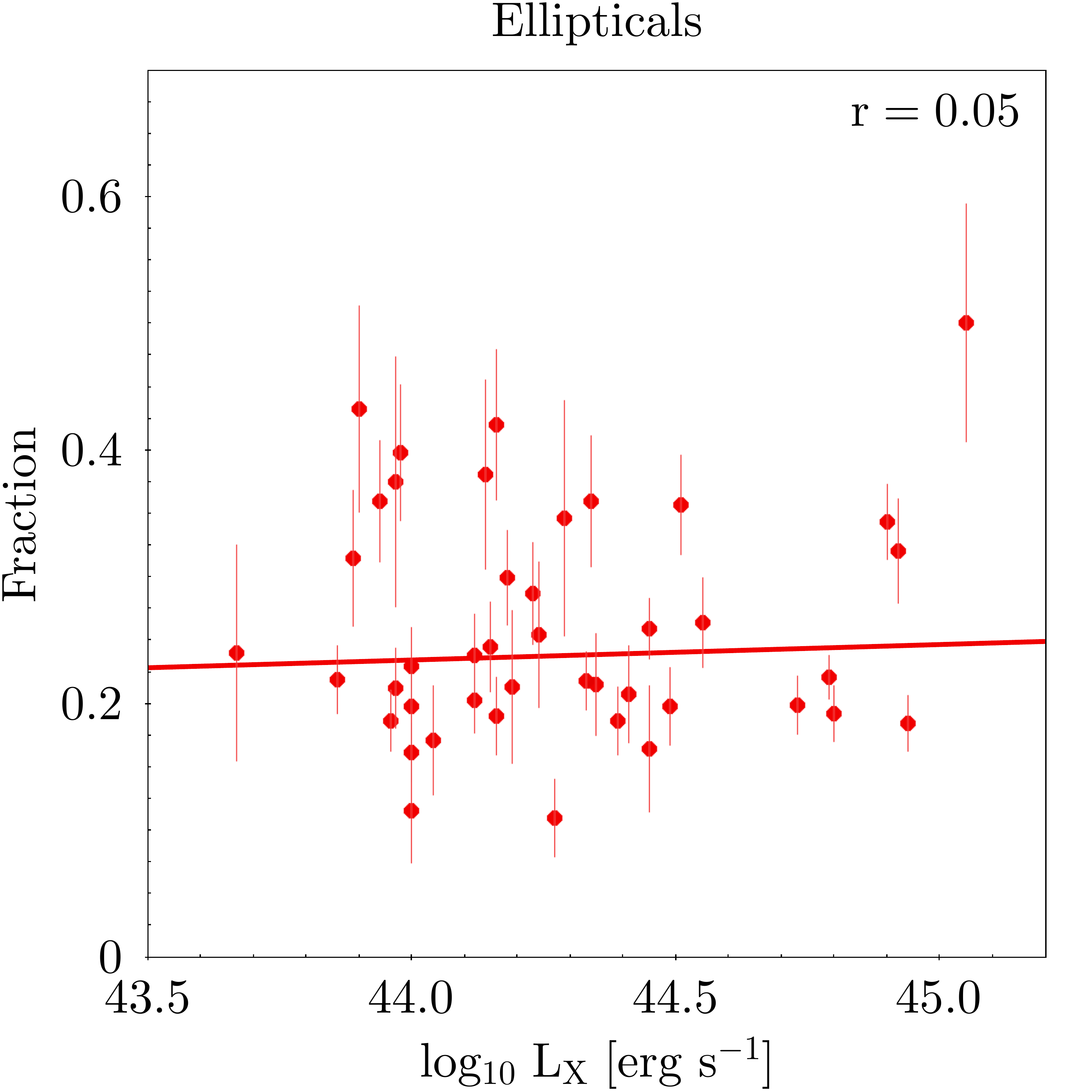}}
  \subfloat{
   \label{fig:fraction-S0_LumX_cl}
    \includegraphics[trim={120 0 0 0mm}, clip, width=0.30\textwidth]{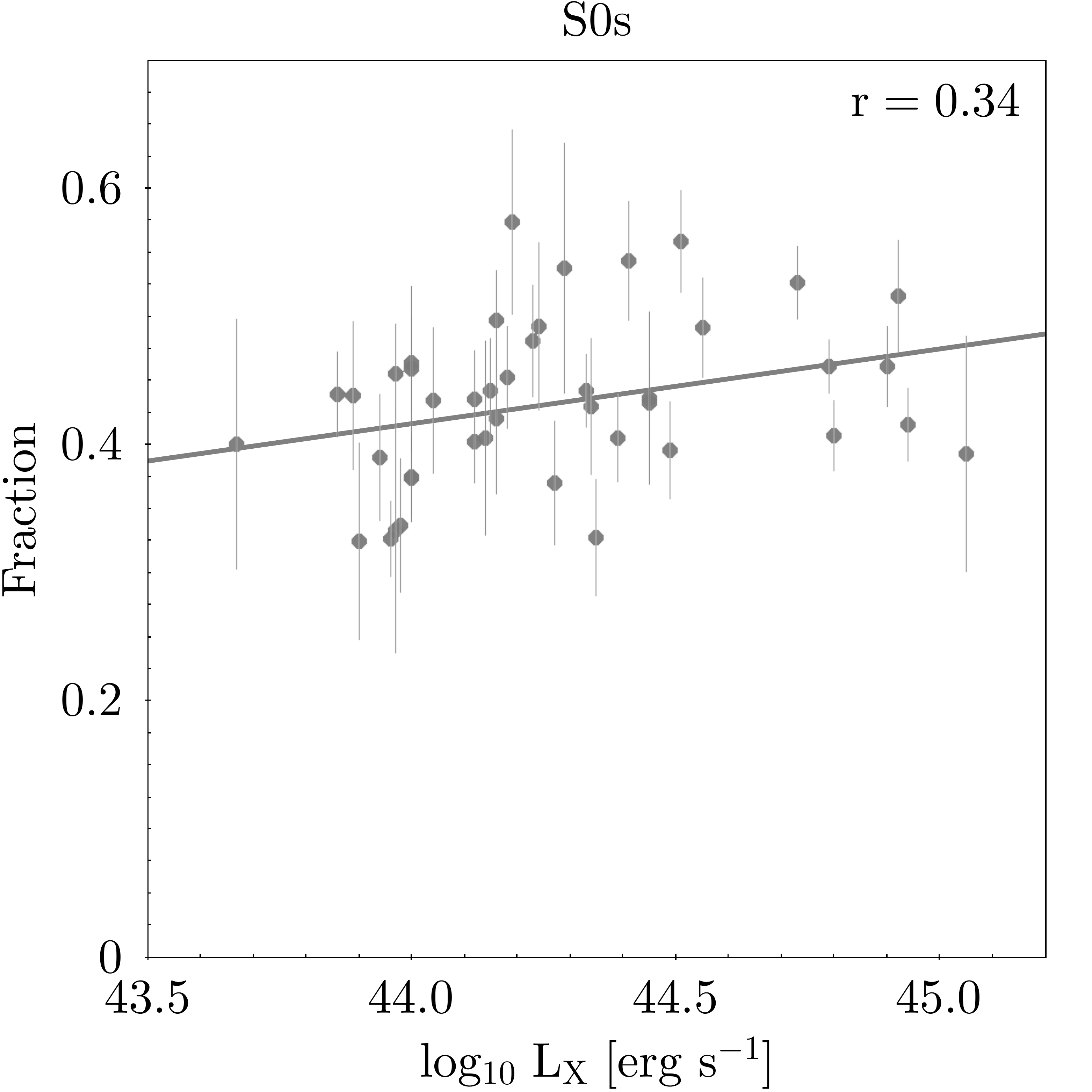}}
  \subfloat{
   \label{fig:fraction-S_LumX_cl}
    \includegraphics[trim={120 0 0 0mm}, clip, width=0.303\textwidth]{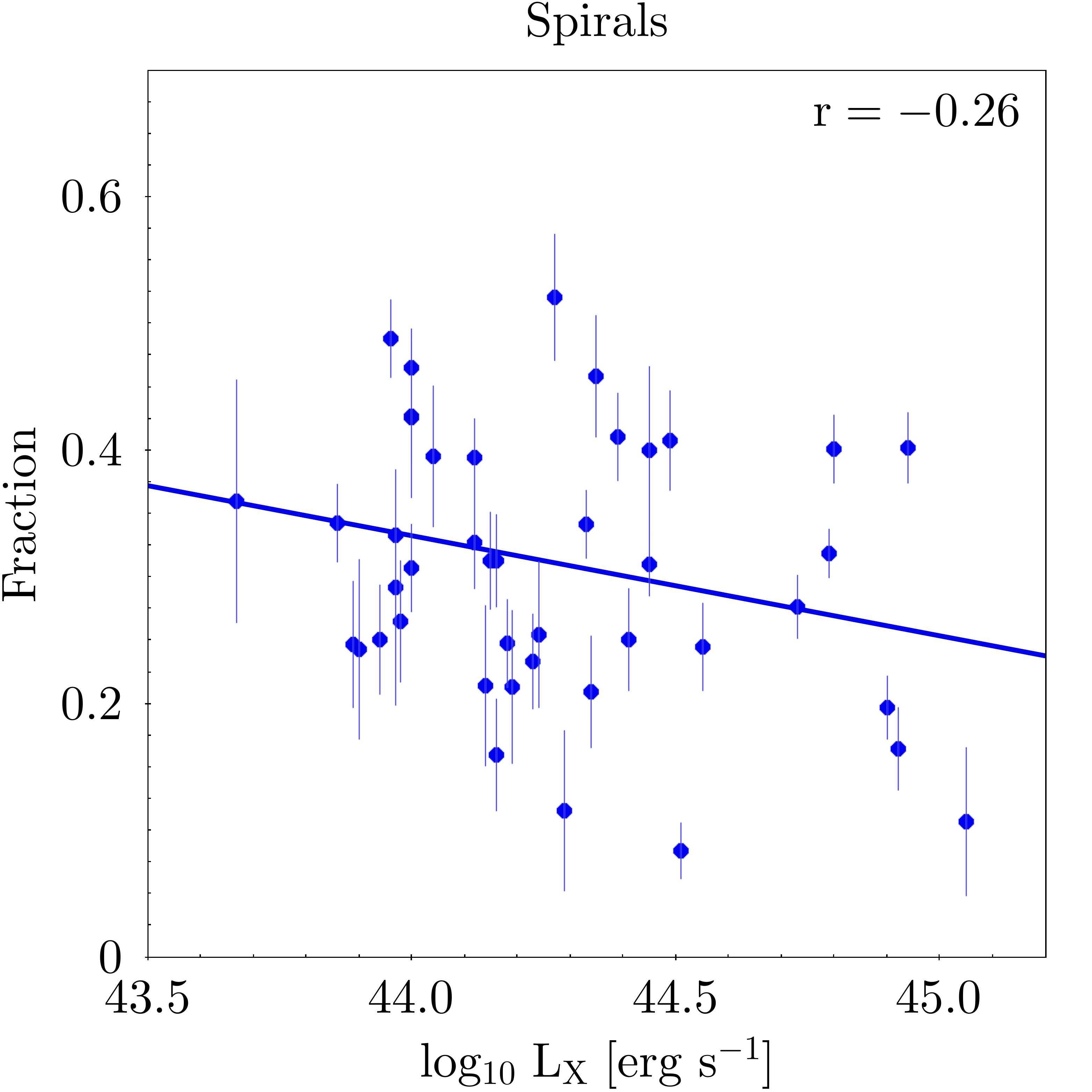}}
 \caption{Fractions of galaxy morphological types in the WINGS and OmegaWINGS datasets, as a function of cluster X-ray luminosity ($L_{\rm{X}}$). Panels as in \autoref{fig:Fraction_types_Sigma_cl}.} 
 \label{fig:Fraction_types_Lum-X_cl}
\end{figure*}

\autoref{fig:Fraction_types_Sigma_cl} shows the fractions of morphological types as a function of cluster velocity dispersion. Solid lines are the least-square fits, with a low correlation coefficient\footnote{Pearson's correlation coefficient is a measure of the statistical relation between two variables. It gives information about the magnitude of association, and the direction of the relation.} in all cases: 0.314 for ellipticals, 0.21 for S0s, and -0.21 for spirals. The trends of morphological fraction as a function of X-ray luminosity are shown in \autoref{fig:Fraction_types_Lum-X_cl}, together with the least-square fits. We also find low correlations: 0.34 for S0s, -0.26 for spirals, and 0.05 for ellipticals. Taking into account the binomial error bars in the fractions, and the errors in the $\sigma_{\rm cl}$ of clusters, these results indicate that the fractions of S0s and spirals may have a correlation with $\sigma_{\rm{cl}}$ and $L_{\rm{X}}$. While the S0 fraction increases with cluster mass, the opposite is true for spirals.

In \autoref{tab:morph_fractions}, we list the mean morphological fractions of clusters and the field. In comparison with the results obtained by \citet[][only for WINGS galaxies, i.e., galaxies within $0.6 R_{200}$ and with $M_V \leq -19.5$]{Poggianti2009}, our fraction of spiral galaxies is higher, at the expense of lower elliptical and S0 fractions. We attribute these results to our inclusion of OmegaWINGS data, that sample the galaxy population in the cluster outskirts, and hence (1) to the morphological evolution that spiral galaxies experiment as they are fall towards the cluster centres, and (2) to the fact that we only use cluster members without a cut in luminosity. The comparison between the two works could be telling that morphological transformations are quicker in less luminous (hence less massive) galaxies. Furthermore, sampling larger cluster areas allows the inclusion of regions where the transformation is ongoing or recent, while most early-types would dynamically be settled down towards the cluster centre.

\begin{table}
\caption{Morphological fractions of cluster members (up to $1\ R_{200}$) and field galaxies. Errors are binomial.}
\begin{center}
\begin{tabular}{ c c c } 
\hline
Type & Clusters & Field \\ \hline \hline 

E   & $26.0 \pm 4.3 \%$ & $14.7 \pm 0.9 \%$ \\ 
S0  & $43.5 \pm 4.9 \%$ & $25.8 \pm 1.1 \%$ \\ 
Sp  & $30.5 \pm 4.3 \%$ & $59.5 \pm 1.2 \%$ \\ 
SpE & $23.2 \pm 4.1 \%$ & $35.2 \pm 1.2 \%$ \\ 
SpL & $ 7.3 \pm 2.0 \%$ & $24.3 \pm 1.1 \%$ \\ \hline
\label{tab:morph_fractions}
\end{tabular}
\end{center}
\end{table}

\section{Stellar mass distributions}
\label{sec:Stellar-mass_distribution}
\begin{figure*}
\centering
   \subfloat{\label{fig:Mass-dist_Cl-Field_All}
    \includegraphics[trim={0 0 20 0mm}, clip, height=4.9cm]{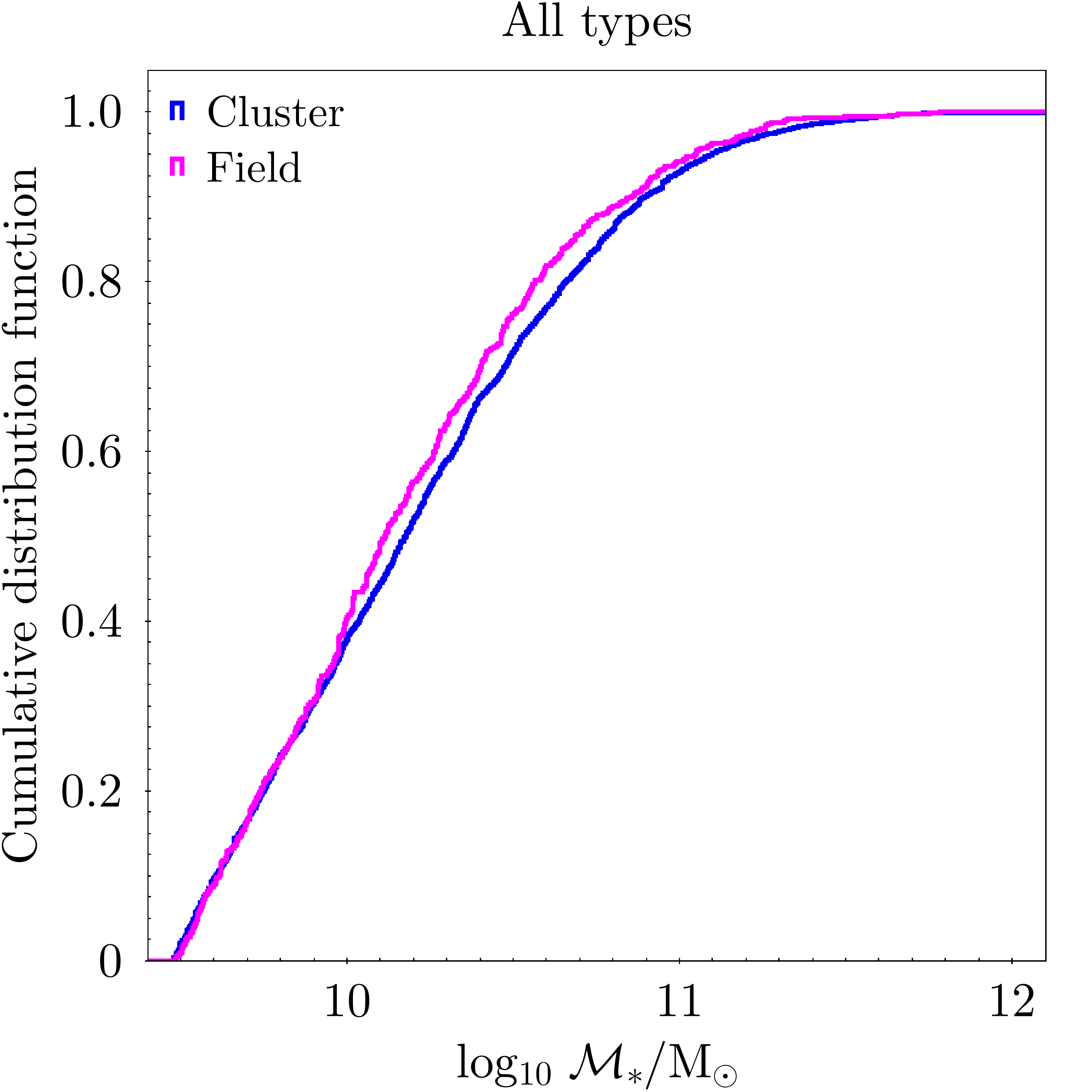}}
   \subfloat{\label{fig:Mass-dist_Cl-Field_E}
     \includegraphics[trim={100 0 20 0mm}, clip, height=4.9cm]{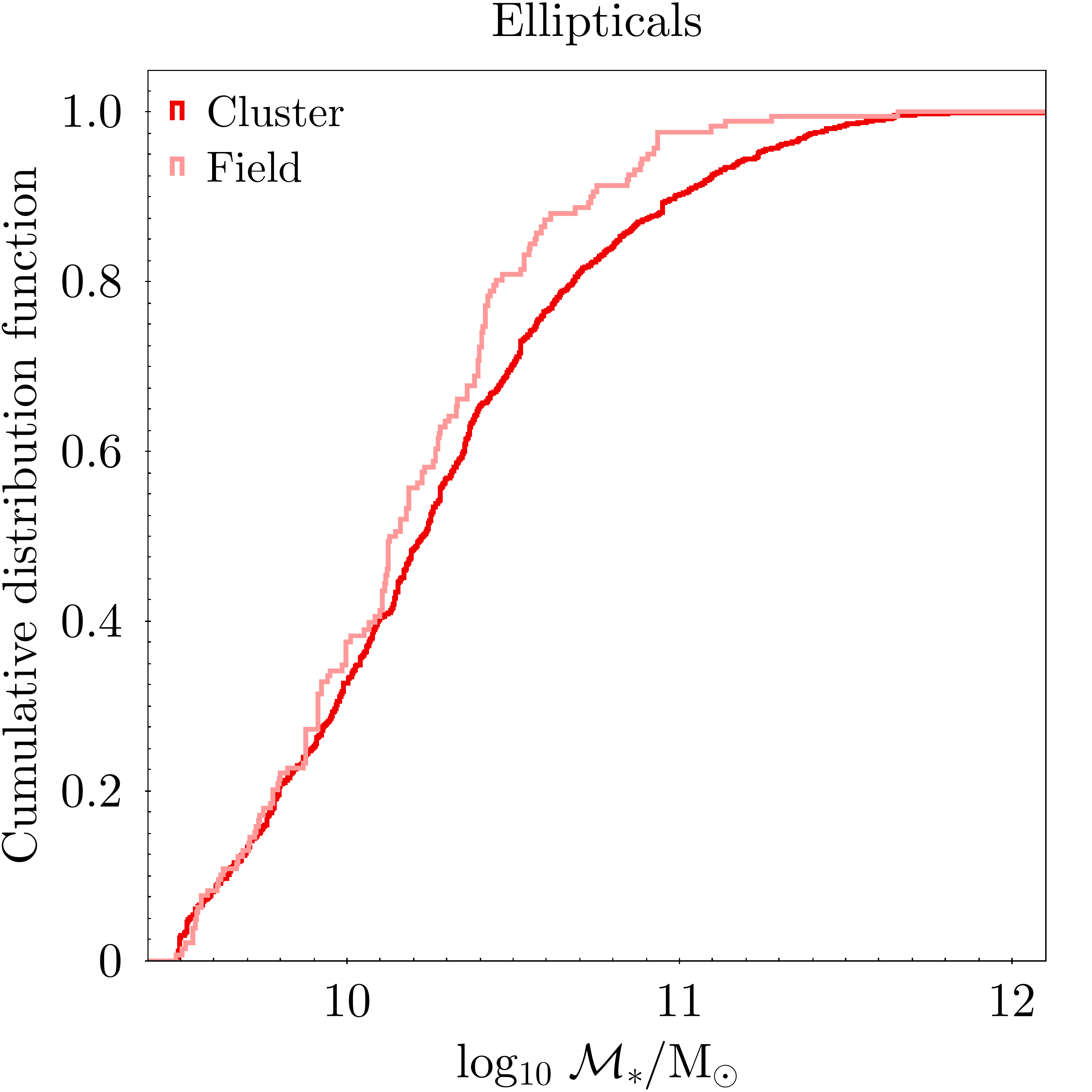}}
   \subfloat{\label{fig:Mass-dist_Cl-Field_S0}
     \includegraphics[trim={100 0 20 0mm}, clip, height=4.87cm]{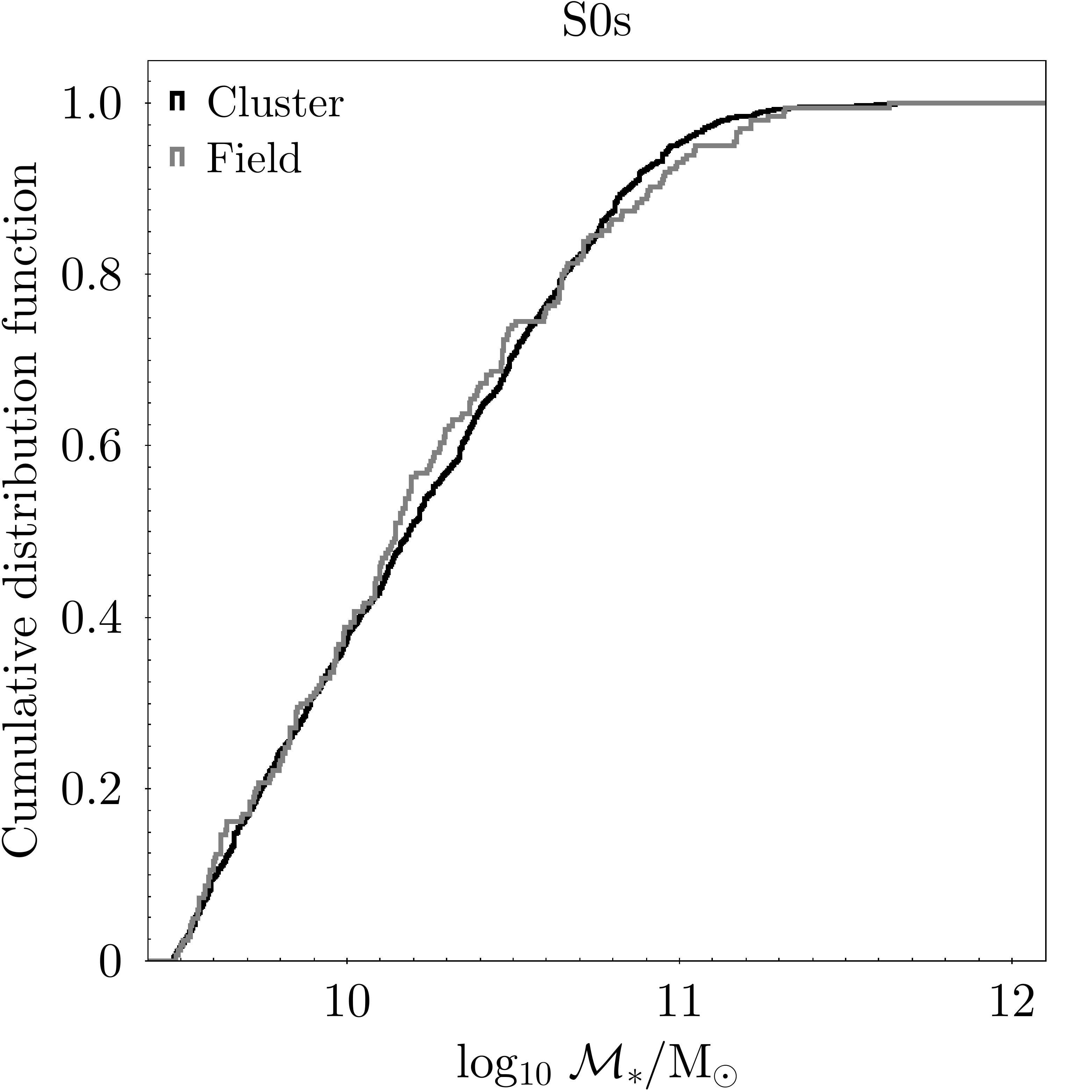}}
   \subfloat{\label{fig:Mass-dist_Cl-Field_Sp}
     \includegraphics[trim={100 0 20 0mm}, clip, height=4.9cm]{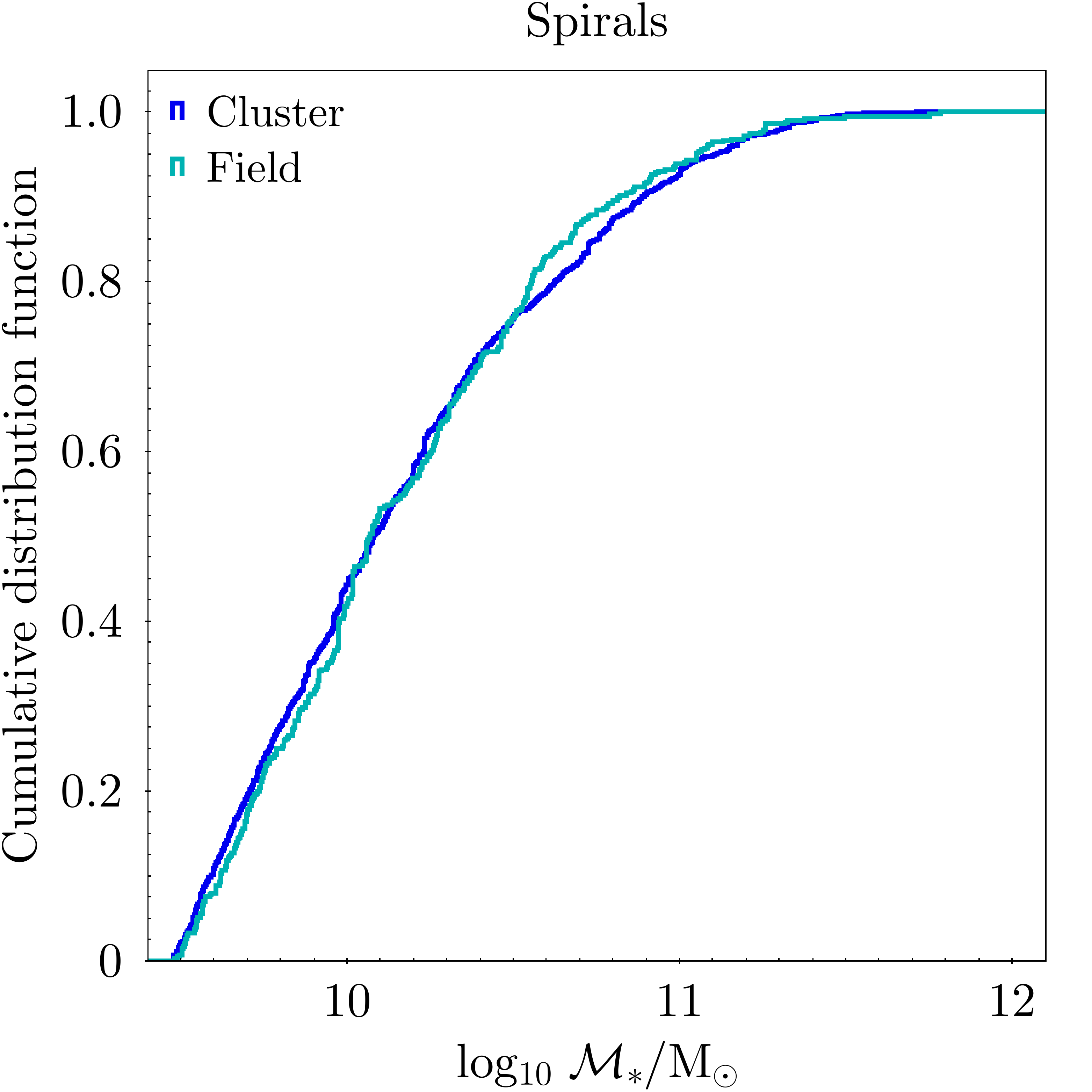}}

\caption{Cumulative distribution functions of weighted total stellar mass for the galaxy samples in clusters and in the field. {\it Left:}  all galaxies together; {\it centre left:} ellipticals; {\it centre right:} S0s; {\it right:} spirals.} 
 \label{fig:Mass_distr_types_sample}
\end{figure*}

In \autoref{subsec:Mass_distribution}, we present the total stellar mass distributions for early- and late-spiral galaxies in the final sample, both in clusters and in the field, as obtained with \texttt{SINOPSIS}. Here we show, in \autoref{fig:Mass_distr_types_sample}, the same distributions for elliptical, S0, and spiral galaxies, again comparing between clusters and field. The sample was selected with the criteria listed in \autoref{subsec:Sample_selection}. We apply a  Kolmogorov-Smirnoff test to the stellar mass distributions of cluster member and field galaxies, corrected for incompleteness. The results are listed in \autoref{tab:KS-test}. The null hypothesis, that cluster and field mass functions come from the same parent distribution, is rejected if the critical value $P < 0.05$. At face value, we find that cluster and field mass distributions are different for all types (taken together), ellipticals, and late spirals. Conversely, the distributions of S0s on the one hand, early spirals, and all spirals together on the other, are the same. 

\begin{table}
\caption{\label{tab:KS-test} $D$ and $P$ values of the K-S test for the stellar mass distributions of the cluster and field galaxy samples, according to morphological type.}
\begin{center}
\begin{tabular}{ c c c } 
\hline
Type & $D$ & $P$ \\ \hline \hline

All & $0.054$ & $0.007$ \\ 
E   & $0.130$ & $0.004$ \\ 
S0  & $0.058$ & $0.329$ \\ 
Sp  & $0.047$ & $0.239$ \\ 
SpE & $0.086$ & $0.017$ \\
SpL & $0.138$ & $0.007$ \\ \hline
\end{tabular}
\end{center}
\end{table}

For this study, the number of galaxies in the field is smaller than those in the cluster sample, and these differences are even higher for weighted numbers. We checked whether the results above for the K-S test could be affected by this effect. We made 1000 realisations of cluster sub-samples with the same number of galaxies as in the field sample (the probability of a galaxy to be chosen is given by its weight), and compared the cluster subsamples with the field, using again the K-S test, for each morphological type. We also counted the number of realisations for which both distributions are different. Results are shown in \autoref{tab:KS-test_subsamples}.

We find that the differences in the sizes of the cluster and the field samples do affect the results.
The probability of having different parent distributions in the cluster and field is 0.662 for all types together, 0.676 for ellipticals, and 0.825 for late spirals, while for S0s, all spirals together, and early spirals, this probability is very low. As a control test, we also compare the cluster subsamples with the whole cluster sample and find that, for all the morphologies analysed, they come from the same parent distribution with a probability larger than 0.98.

\begin{table}
\caption{\label{tab:KS-test_subsamples} $D$ and $P$ mean values of the K-S test for the stellar mass distributions of  cluster subsamples and the field sample. These means correspond to 1000 realisations of cluster subsamples with the same number of galaxies as the field sample, for each morphological type. The last column indicates the number of times, out of 1000, that $P < 0.05$.}
\begin{center}
\begin{tabular}{ c c c c} 
\hline
Type & $D$ & $P$ & $N$ \\ \hline \hline

All & $0.063$ & $0.060$ & 662\\ 
E   & $0.151$ & $0.054$ & 676\\ 
S0  & $0.067$ & $0.562$ & 14\\ 
Sp  & $0.062$ & $0.232$ & 73\\ 
SpE & $0.081$ & $0.217$ & 165\\
SpL & $0.147$ & $0.030$ & 825\\ \hline
\end{tabular}
\end{center}
\end{table}

\cite{Calvi2013} studied the mass function for WINGS galaxies, compared to the ``general field," through the PM2GC. Our results are generally consistent with that work: fixing the environment, the mass function changes with morphology, and fixing morphology, the mass function is independent of the environment. Nevertheless, for the present paper, this is not true in the case of ellipticals: for them, the mass distributions are overall statistically different in the cluster and field. This discrepancy is most likely attributable to the fact that our field elliptical sample is much smaller than theirs. Furthermore, the mass limits adopted in the two works are different, as well as the samples themselves (we use WINGS+OmegaWINGS for cluster and field galaxies, while they only include WINGS and PM2GC). What is, however, common to the two samples is that the most massive ellipticals are indeed found in clusters.

\clearpage
\clearpage
\onecolumn
\section{WINGS clusters}
\label{sec:Clusters_table}
Some basic properties of the WINGS cluster sample defined in \autoref{sec:Dataset} are listed in \autoref{tab:WINGS_clusters}: mean redshift ($z$) and velocity dispersion ($\sigma_{\rm{cl}}$) obtained from the data themselves \citep{Biviano2017, Gullieuszik2020}, virial radius ($R_{200}$), and X-ray luminosity (L$_{\rm X}$) taken from the ROSAT All-Sky Survey \citep{Ebeling1996, Ebeling1998, Ebeling2000}.

\begin{longtable}{l c r r l c c c}
\caption{Properties of the WINGS/OmegaWINGS cluster sample: redshift ($z$), velocity dispersion ($\sigma_{\rm{cl}}$ in km s$^{-1}$), virial radius ($R_{200}$ in Mpc), and X-ray luminosity (L$_{\rm X}$ in erg s$^{-1}$). WINGS and OmegaWINGS columns indicate if the cluster was observed in spectroscopy. The last column shows whether WINGS/OmegaWINGS spectra have sufficient signal-to-noise ratio to run the \texttt{SINOPSIS} code.}
\label{tab:WINGS_clusters} \\

\hline
Cluster & $z$ & $\sigma_{\textrm{cl}}$ & $R_{200}$ & log$_{10} \ \textrm{L}_{\rm X}$ & WINGS & OmegaWINGS & SINOPSIS \\ \hline 
\endfirsthead

\multicolumn{8}{c}{\tablename\ \thetable\ -- \textit{Continued from previous page.}} \\
\hline 
Cluster & $z$ & $\sigma_{\textrm{cl}}$ & $R_{200}$ & log$_{10} \ \textrm{L}_{\rm X}$ & WINGS & OmegaWINGS & SINOPSIS \\ \hline
\endhead 

\hline \multicolumn{8}{r}{\textit{Continued on next page.}} \\
\endfoot \hline \endlastfoot

A1069	& 0.06528 & 542	& 1.180 & 43.98 & \checkmark & \checkmark & \checkmark \\
A119	& 0.04436 & 952	& 2.250 & 44.51 & \checkmark & --    & \checkmark \\
A1291	& 0.05090 & 413	& 0.860 & 43.64 & \checkmark & --    & --    \\
A133	& 0.06030 & 623	& 1.292 & 44.55 & --    & --    & --    \\
A147	& 0.04470 & 387	& 0.808 & 43.73 & --    & --    & --    \\
A151	& 0.05327 & 771	& 1.670 & 44.00	& \checkmark & \checkmark & \checkmark \\
A160	& 0.04317 & 738	& 1.600 & 43.58 & \checkmark & --    & --    \\
A1631a	& 0.04644 & 715	& 1.390 & 43.86 & \checkmark & \checkmark & \checkmark \\
A1644	& 0.04691 & 945	& 1.890 & 44.55 & \checkmark & --    & \checkmark \\
A1668	& 0.06340 & 654	& 1.354 & 44.20	& --    & --    & --    \\
A168	& 0.04518 & 498	& 0.970 & 44.04	& --    & \checkmark & \checkmark \\
A1736	& 0.04610 & 918	& 1.916 & 44.37	& --    & --    & --    \\
A1795	& 0.06291 & 731	& 1.720 & 45.05 & \checkmark & --    & --    \\
A1831	& 0.06340 & 444	& 0.919 & 44.28 & \checkmark & --    & --    \\
A193	& 0.04852 & 758	& 1.580 & 44.19 & --    & \checkmark & \checkmark \\
A1983	& 0.04517 & 407	& 0.950 & 44.67 & \checkmark & --    & --    \\
A1991	& 0.05860 & 570	& 1.330 & 44.13 & \checkmark & --    & --    \\
A2107	& 0.04166 & 519	& 1.150 & 44.04 & \checkmark & --    & --    \\
A2124	& 0.06692 & 733	& 1.090 & 44.13 & \checkmark & --    & --    \\
A2149	& 0.06750 & 459	& 0.948 & 43.92 & --    & --    & --    \\
A2169	& 0.05780 & 524	& 1.088 & 43.65	& \checkmark & --    & --    \\
A2256	& 0.05810 & 1376& 2.856 & 44.85	& --    & --    & --    \\
A2271	& 0.05840 & 460	& 0.955 & 43.81	& --    & --    & --    \\
A2382	& 0.06442 & 807	& 1.730 & 43.96	& \checkmark & \checkmark & \checkmark \\
A2399	& 0.05793 & 662	& 1.550 & 44.00	& \checkmark & \checkmark & \checkmark \\
A2415	& 0.05791 & 683	& 1.190 & 44.23	& \checkmark & \checkmark & \checkmark \\
A2457	& 0.05889 & 605	& 1.310 & 44.16	& \checkmark & \checkmark & \checkmark \\
A2572a	& 0.03900 & 546	& 1.144 & 44.01	& \checkmark & --    & \checkmark \\
A2589	& 0.04217 & 1147& 2.750 & 44.27	& \checkmark & --    & --    \\
A2593	& 0.04188 & 523	& 1.210 & 44.06	& \checkmark & --    & --    \\
A2622	& 0.06100 & 732	& 1.517 & 44.03	& \checkmark & --    & --    \\
A2626	& 0.05509 & 650	& 1.480 & 44.29	& \checkmark & --    & \checkmark \\
A2657	& 0.04000 & 829	& 1.735 & 44.2	& --    & --    & --    \\
A2665	& 0.05620 & -- & -- & -- & -- & -- & --  \\
A2717	& 0.04989 & 470	& 1.170 & 44.00	& --    & \checkmark & \checkmark \\
A2734	& 0.06147 & 588	& 1.380 & 44.41	& --    & \checkmark & \checkmark \\
A311	& 0.06570 & -- & -- & -- & -- & -- & --  \\
A3128	& 0.06033 & 793	& 1.580 & 44.33	& \checkmark & \checkmark & \checkmark \\
A3158	& 0.05947 & 948	& 1.940 & 44.73	& \checkmark & \checkmark & \checkmark \\
A3164	& 0.06110 & 991	& 2.054 & -- & --  & --    & --    \\
A3266	& 0.05915 & 1095& 2.310 & 44.79 & \checkmark & \checkmark & \checkmark \\
A3376	& 0.04652 & 756	& 1.650 & 44.39 & \checkmark & \checkmark & \checkmark \\
A3395	& 0.05103 & 1272& 2.760 & 44.45 & \checkmark & \checkmark & \checkmark \\
A3490	& 0.06880 & 660	& 1.363 & 44.24 & \checkmark & --    & \checkmark \\
A3497	& 0.06800 & 724	& 1.496 & 44.16 & \checkmark & --    & \checkmark \\
A3528a	& 0.05441 & 891	& 1.880 & 44.12 & --    & \checkmark & \checkmark \\
A3528b	& 0.05350 & -- & -- & 44.30 & -- & --  & -- \\
A3530	& 0.05480 & 674	& 1.401 & 43.94 & --    & \checkmark & \checkmark \\
A3532	& 0.05536 & 662	& 1.550 & 44.45 & --    & \checkmark & \checkmark \\
A3556	& 0.04796 & 531	& 1.100 & 43.97	& \checkmark & \checkmark & \checkmark \\
A3558	& 0.04829 & 910	& 1.950 & 44.80	& --    & \checkmark & \checkmark \\
A3560	& 0.04917 & 799	& 1.790 & 44.12	& \checkmark & \checkmark & \checkmark \\
A3667	& 0.05528 & 1031& 2.220 & 44.94	& --    & \checkmark & \checkmark \\
A3716	& 0.04599 & 753	& 1.720 & 44.00	& --    & \checkmark & \checkmark \\
A376	& 0.04752 & 832	& 1.660 & 44.14	& \checkmark & --    & \checkmark \\
A3809	& 0.06245 & 499	& 1.040 & 44.35	& \checkmark & \checkmark & \checkmark \\
A3880	& 0.05794 & 514	& 1.200 & 44.27	& --    & \checkmark & \checkmark \\
A4059	& 0.04877 & 744	& 1.580 & 44.49	& --    & \checkmark & \checkmark \\
A500	& 0.06802 & 660	& 1.800 & 44.15	& \checkmark & \checkmark & \checkmark \\
A548b	& 0.04410 & 842	& 1.759 & 43.48	& --    & --    & --    \\
A602	& 0.06210 & 834	& 1.728 & 44.05	& --    & --    & --    \\
A671	& 0.04939 & 730	& 1.490 & 43.95	& \checkmark & --    & --    \\
A754	& 0.05445 & 816	& 1.660 & 44.90	& \checkmark & \checkmark & \checkmark \\
A780	& 0.05650 & -- & -- & 44.82 & -- & -- & --    \\
A85	    & 0.05568 & 859	& 2.020 & 44.92	& --    & \checkmark & \checkmark \\
A957x	& 0.04496 & 631	& 1.420 & 43.89	& \checkmark & \checkmark & \checkmark \\
A970	& 0.05872 & 749	& 1.630 & 44.18	& \checkmark & \checkmark & \checkmark \\
IIZW108	& 0.04889 & 575	& 1.199 & 44.34	& \checkmark & \checkmark & \checkmark \\
MKW3s	& 0.04470 & 604	& 1.580 & 44.43	& \checkmark & --    & --    \\
Rx0058	& 0.04840 & 696	& 1.451 & 43.64	& \checkmark & --    & \checkmark \\
Rx1022	& 0.05480 & 582	& 1.210 & 43.54	& \checkmark & --    & --    \\
Rx1740	& 0.04410 & 540	& 1.128 & 43.7	& \checkmark & --    & --    \\
Z1261	& 0.06440 & -- & -- & -- & -- & -- & --  \\
Z2844	& 0.05027 & 425	& 0.880 & 43.76	& \checkmark & --    & --    \\
Z8338	& 0.04953 & 658	& 1.350 & 43.9	& \checkmark & --    & \checkmark \\
Z8852	& 0.04077 & 786	& 1.630 & 43.97	& \checkmark & --    & \checkmark  \\
\end{longtable}


\bsp	
\label{lastpage}
\end{document}